\documentclass[5p,twocolumn,authoryear]{elsarticle}

\usepackage{natbib}

\usepackage{graphicx}
\usepackage{amsmath}
\usepackage{amssymb}

\usepackage{longtable}
\usepackage{dcolumn}

\newcolumntype{d}[1]{D{.}{.}{#1}}
\newcolumntype{p}[1]{D{p}{\mbox{$\,\pm\,$}}{#1}}

\newcommand{\skytel}{S\&T }
\newcommand{\aj}{Astron. J. }
\newcommand{\aap}{Astron. Astrophys. }
\newcommand{\jrasc}{Journal of the RAS of Canada }
\newcommand{\baas}{Bulletin of the AAS }
\newcommand{\planss}{Planetary and Space Science }
\newcommand{\pasp}{Publ. Astron. Soc. Pacific }

\journal{Icarus}

\begin{document}

\begin{frontmatter}

\title{Combining asteroid models derived by lightcurve inversion with asteroidal occultation silhouettes}

\author[durech]{Josef \v{D}urech\corref{cor}} 
\ead{durech@sirrah.troja.mff.cuni.cz}
\author[kaasalainen]{Mikko Kaasalainen}
\author[herald]{David Herald}
\author[dunham]{David Dunham}
\author[timerson]{Brad Timerson}
\author[durech]{Josef Hanu\v{s}}
\author[frappa]{Eric Frappa}
\author[talbot]{John Talbot}
\author[hayamizu]{Tsutomu Hayamizu}
\author[warner]{Brian D. Warner}
\author[pilcher]{Frederick Pilcher}
\author[galad1,galad2]{Adri\'an Gal\'ad}

\address[durech]{Astronomical Institute, Faculty of Mathematics and Physics, Charles University in Prague, 
			V Hole\v{s}ovi\v{c}k\'ach 2, CZ-18000 Prague, Czech Republic}
\address[kaasalainen]{Department of Mathematics, Tampere University of Technology, 
			P.O. Box 553, 33101 Tampere, Finland}
\address[herald]{3 Lupin Pl, Murrumbateman, NSW, Australia}
\address[dunham]{International Occultation Timing Association (IOTA) and KinetX, Inc., 7913 Kara Ct., Greenbelt, MD 20770, USA}
\address[timerson]{International Occultation Timing Association (IOTA), 623 Bell Rd., Newark, NY, USA}
\address[frappa]{1 bis cours Jovin Bouchard 42000 Saint-Etienne, France}
\address[talbot]{Occultation Section, Royal Astronomical Society of New Zealand, P.O. Box 3181,  Wellington,  New Zealand}
\address[hayamizu]{Japan Occultation Information Network (JOIN), Sendai Space Hall, 2133-6 Nagatoshi, Kagoshima pref, Japan}
\address[warner]{Palmer Divide Observatory, 17955 Bakers Farm Rd., Colorado Springs, CO 80908, USA}
\address[pilcher]{4438 Organ Mesa Loop, Las Cruces, NM 88011, USA}
\address[galad1]{Modra Observatory, FMFI Comenius University, 842 48 Bratislava, Slovakia}
\address[galad2]{Ond\v{r}ejov Observatory, AV \v{C}R, 251 65 Ond\v{r}ejov, Czech Republic}

\cortext[cor]{Corresponding author}

%Asteroid models from lightcurves and occultations}

\begin{abstract}

Asteroid sizes can be directly measured by observing occultations of stars by asteroids. When there are enough 
observations across the path of the shadow, the asteroid's projected silhouette can be reconstructed. Asteroid shape models 
derived from photometry by the lightcurve inversion method enable us to predict the orientation of an asteroid for the
time of occultation. By scaling the shape model to fit the occultation chords, we can determine the asteroid size with
a relative accuracy of typically $\sim 10\%$. We combine shape and spin state models of 
44 asteroids (14 of them are new or updated models) 
with the available
occultation data to derive asteroid effective diameters. In many cases, occultations allow us to reject one of two possible
pole solutions that were derived from photometry. We show that by combining results obtained from lightcurve inversion with
occultation timings, we can obtain unique physical models of asteroids.

\end{abstract}

\begin{keyword}
Asteroids\sep Occultations\sep Photometry
\end{keyword}

\end{frontmatter}

\section{Introduction}
Occultations of stars by asteroids are relatively frequent events systematically observed 
mainly by amateur astronomers.  The projected silhouette of the occulting asteroid
can be derived from accurate timings of the disappearance and reappearance of the star 
measured by several observers placed across the path of the shadow.
Apart from a precise measurement of the relative astrometric position of the star and the asteroid at the time of the event, 
the main scientific value of a well observed occultation is 
the direct and relatively accurate measurement of the asteroid's dimension.
Asteroid sizes derived from occultations 
can serve as an independent check of values obtained by, e.g., thermal
radiometric observations. 
From the diameter and the absolute magnitude of an asteroid, we can derive its geometric albedo \citep{She.Ted:06}.
The number of observed occultations has increased dramatically with
the availability of high-accuracy star catalogues based on the Hipparcos data \citep{Dun.ea:02}. 
Nowadays, occultations can be predicted well in advance with
sufficient accuracy. A further dramatic increase of the number of accurate predictions
will come with the stellar catalogue based on the astrometric mission Gaia \citep{Tan.Del:07}.

Over 1700 occultations have been observed 
so far. The majority of them 
were observed by less than three observers and they
do not provide any reliable silhouette estimation.
These occultations can yield only a lower limit of the asteroid size.
However, there are several hundred occultation events with a 
sufficient number of well defined
chords that provide enough data
to reconstruct more or less accurately the projected silhouette.

Until now, the usual final result of all reduced occultations was an ellipse fitting
the projected silhouette and approximating the asteroid's dimensions. In a few cases, 
the occultation data were processed together with optical lightcurves in order to derive 
the asteroid's triaxial ellipsoid shape model 
\citep{Dru.Coc:89, Dun.ea:90, Sat.ea:93, Sat.ea:00}.

The reconstruction of a 3D shape model from several occultation 2D projections is 
possible in principle but still unrealized in practice due to the lack of data.
On the other hand, there are more than one hundred asteroid shape models 
that have been derived  
in the past decade by the lightcurve inversion method of \cite{Kaa.Tor:01} and \cite{Kaa.ea:01}.
These models 
are archived in the Database of Asteroid Models from Inversion Techniques 
(DAMIT)\footnote{\tt http://astro.troja.mff.cuni.cz/projects/asteroids3D}
\citep{Dur.ea:10}.
 
The reliability of models derived by the lightcurve inversion technique has been
confirmed by their comparison with models obtained by radar or with direct spacecraft imaging
and by laboratory experiments \citep{KaaS.ea:05}.
Convex models are inevitably only approximations of real shapes and cannot provide us with shape details. However, 
the derived rotation periods and spin directions have very good agreement 
with independent results \citep{Kaa.ea:01}. Using the model, the asteroid's 
orientation can be computed for the epoch of the occultation and the observed 
projection can be compared with the predicted silhouette of the model. The basic 
result of such combination of occultation data and shape models is that the models
from lightcurve inversion can be calibrated to absolute dimensions
of the asteroid \citep{Tim.ea:09}. Moreover, if we include the occultation profile into the 
optimization procedure, we can refine the model and reveal shape details that are 
in principle unobtainable from lightcurves only. 
Lightcurve inversion provides unique solutions only for convex shapes, so occultations can help in detecting concavities.

For this paper, we have selected more than forty such asteroids, that have lightcurve inversion models in DAMIT and also 
occultation observations that
can be used for scaling the model. We also present ten new or updated models that we also scale using the occultation data.
Although some multi-chord occultation data are of sufficient quality to enable us to carry out the full multi-data inversion with
fitting not only the size but also the shape to the occultation data,
we leave this for a forthcoming paper.
We describe the data reduction and optimization process in Sect.~\ref{data_reduction} and we report our results in Sect.~\ref{results}.
   
\section{Data reduction and optimization}
\label{data_reduction}

\subsection{Occultations}
During an occultation, the asteroid's shadow moves on the Earth's surface and observers 
within the occultation path measure the times of disappearance and reappearance of
the occulted star. The negative reports from observers outside of the path 
put limits on the dimensions of asteroid's cross-section. The  duration of a typical event
is from seconds to tens of seconds. If the observations are made visually, the
reaction times of individual observers can involve significant systematic errors.

The occultation data we use were taken mainly from the NASA PDS database \citep{Dun.Her:09} that contains data up to 2008. 
More recent occultation data were
provided directly by the International Occultation Timing Association (IOTA)\footnote{{\tt http://www.lunar-occultations.com/iota/iotandx.htm}}, 
by the Asteroidal Occultation Observers in Europe\footnote{\tt http://www.euraster.net}, 
by the Occultation Section of the Royal Astronomical Society of New Zealand\footnote{\tt http://occsec.wellington.net.nz/planet/plnreslt.htm}, 
and by Japanese observers of asteroidal occultations\footnote{\tt http://uchukan.satsumasendai.jp/data/occult-e/occult-e.html}.
The data were processed in the standard
way described, for example, by \citet{Mil.Ell:79} or \citet{Was.ea:79}. Here we follow
the notation used by \citet{Kaa:03}.

Each observed time of disappearance and reappearance defines a point on the limb of
the asteroid. We define the fundamental plane that passes through the center
of the Earth and is instantaneously perpendicular to the line connecting the 
star (assuming to be at infinity) and the center of the asteroid. 
Then the coordinate system $(\xi, \eta)$ on
the fundamental plane is defined by two unit vectors
\begin{eqnarray}
\hat{\vec{s}}_\xi & = & (-\sin\delta \cos\alpha, -\sin\delta \sin\alpha, \cos\delta), \\
\hat{\vec{s}}_\eta & = & (\sin\alpha, -\cos\alpha, 0), \nonumber 
\end{eqnarray}
where $\alpha$ and $\delta$ are the right ascension and declination of the
occulted star. For prediction purposes, it is important to use apparent coordinates of the star. However, for our
purposes, the difference between the apparent and the mean position can be neglected and we use J2000.0 coordinates.

The geocentric coordinates of an observer are projected
onto the fundamental plane
\begin{equation}
\label{eq_2}
(\xi, \eta) = [ \hat{\vec{s}}_\xi \cdot (\vec{x} + \Delta\vec{v} \, \Delta t),
                \hat{\vec{s}}_\eta \cdot (\vec{x} + \Delta\vec{v} \, \Delta t)],
\end{equation}
where $\vec{x}$ is the observer's position on the Earth in the sidereal
equatorial frame, $\vec{x} = (R \cos\varphi \cos\theta, R \cos\varphi \sin\theta, R \sin\varphi)$,
where $R$ is the observer's distance from the Earth's center, $\varphi$ is the geocentric latitude
and $\theta$ is the local sidereal time, $\Delta \vec{v}$ denotes the differential
space velocity $\vec{v}_\mathrm{Earth} - \vec{v}_\mathrm{asteroid}$ in the equatorial 
frame ($\Delta \vec{v}$ is assumed to be constant during the occultation), $\Delta t$ is the time 
of disappearance/reappearance measured from some epoch. If the $\Delta t$ 
times for individual observations do not differ much (compared to the asteroid's rotation period)
the points $(\xi, \eta)$ can be taken as a `snapshot' 
of the asteroid's projection at the time $\Delta t$. This approximation is not needed in the analysis but we used it because the 
asteroids rotated by only a small number of degrees during the observations in most cases. Moreover, as we only scaled the models without actually 
modifying their shapes, this simplification is fully appropriate.

The assumption of a linear differential space velocity made in Eq.~(\ref{eq_2}) is appropriate if (i) the 
reference time is set so that $\Delta t \lesssim 10\,$min, and (ii) all the observations were made from the same
region of the world, which means that the individual timings $\Delta t$ do not differ more than $\sim 10\,$min.
This was fulfilled for all occultations we present in Sect.~\ref{results}. The maximum relative difference between projections
computed according to Eq.~(\ref{eq_2}) and those computed according to a more precise quadratic expression
\begin{align}
(\xi, \eta) =  & \left[ \hat{\vec{s}}_\xi \cdot \left(\vec{x} + \Delta\vec{v} \, \Delta t + \frac{1}{2} \Delta\dot{\vec{v}} \, (\Delta t)^2 \right), \right.  \nonumber \\
               & \ \left. \hat{\vec{s}}_\eta \cdot \left(\vec{x} + \Delta\vec{v} \, \Delta t + \frac{1}{2} \Delta\dot{\vec{v}} \, (\Delta t)^2 \right) \right],
\end{align}
is only $\sim 1\,$\%, which is well below the accuracy we need.

\subsection{Shape models from lightcurve inversion}

The shape models of the asteroids used in this work were taken from DAMIT or derived
from new observations (see Sect.~\ref{results}).
 All models were derived 
by the lightcurve inversion method 
described in \citet{Kaa.Tor:01} and \citet{Kaa.ea:01}. They are convex polyhedrons
(the convexity is the property of almost all DAMIT models, but nonconvex models can be used as well)
with triangular facets defined by radius vectors $\vec{r}$ of the surface points.
Their orientation in space at any given time can be computed from the ecliptic 
longitude and latitude 
$(\lambda_\mathrm{p}, \beta_\mathrm{p})$ of the spin axis direction 
(the $z$ axis in the asteroid Cartesian coordinate frame), the sidereal rotation period $P$
and the initial rotation angle $\varphi_0$ given for some epoch. 
The detailed description of the rotation matrices of the transformation from the asteroid body
frame to the ecliptic frame is given by \cite{Dur.ea:10}.
Thus, for a given time
$\Delta t$ of an occultation, we can project an asteroid onto the fundamental plane.
The radius vectors $\vec{r}$ are transformed to the equatorial coordinate system
$\vec{r}_\mathrm{eq}$. Then the projected 
coordinates $(\xi_\mathrm{mod}, \eta_\mathrm{mod})$ of the polyhedron vertices are
\begin{equation}
(\xi_\mathrm{mod}, \eta_\mathrm{mod}) = (\hat{\vec{s}}_\xi \cdot \vec{r}_\mathrm{eq}, 
                   \hat{\vec{s}}_\eta \cdot \vec{r}_\mathrm{eq})
                   + (\xi_0, \eta_0),
\end{equation}
where $(\xi_0, \eta_0)$ is some offset depending on the angular distance between 
the star and the asteroid. The silhouette of a convex model is the convex hull of projected vertices.
 
\subsection{Scaling the model}
Although the lightcurve inversion models represent the global shape of 
asteroids well, they are not very detailed and 
usually do not reveal nonconvex features \citep{Dur.Kaa:03}.
The complementary
occultation data can add such details in the models.
However, only a limited number of observed occultations clearly reveal a distinct nonconvex feature
(see \cite{Tim.ea:09}, for example). 
For most occultations, the errors of individual chords are too large or the number of chords is too low to enable us
to create a reliable nonconvex model. For these reasons, we fix the shape and only scale it as $c\cdot\vec{r}$
to give the best fit to the occultation silhouette. 
Also the rotation parameters
$(\lambda_\mathrm{p}, \beta_\mathrm{p}, \varphi_0, P)$ are fixed on the values determined by lightcurve inversion.
Thus the only free parameters are the scale $c$ and the offset distances $\eta_0, \xi_0$.

By changing $c$, $\eta_0$, and $\xi_0$, we minimize the $\chi^2$ measure 
\begin{equation}
\label{eq_chisq}
\chi^2 = \sum_{j=1}^N \frac{[(\xi_j, \eta_j)_\mathrm{occ} - (\xi_j, \eta_j)_\mathrm{model}]^2}{\sigma_j^2},
\end{equation}
 where $(\xi_j, \eta_j)_\mathrm{occ}$ are projections of 
observed timings 
onto the fundamental plane, $(\xi_j, \eta_j)_\mathrm{model}$ are 
intersections between the projected asteroid's limb and the line going through the 
 point $(\xi_j, \eta_j)_\mathrm{occ}$ with the direction
of occultation chord $\hat{\vec{s}}_v$ (or $-\hat{\vec{s}}_v$), and $\sigma_j$ are errors of $(\xi_j, \eta_j)_\mathrm{occ}$. 
The $\hat{\vec{s}}_v$ is a unit vector in the direction of the shadow movement
on the fundamental plane. 
We use the JPL Horizons\footnote{\tt http://ssd.jpl.nasa.gov/?horizons}
ephemeris system to compute $\hat{\vec{s}}_v$ of individual observers with respect to the asteroid.

The main source of error affecting the occultation silhouette points $(\xi_j, \eta_j)_\mathrm{occ}$ is 
the timing of dis- and re-appearance of the star. Errors in timing propagate
to errors in the position on the projection plane that are aligned in the direction of $\hat{\vec{s}}_v$. This is the reason why
we minimize the difference
between the model's silhouette and occultation silhouette along $\hat{\vec{s}}_v$ in Eq.~(\ref{eq_chisq}). 

The errors $\sigma_j$ are not always reported or even worse, are underestimated. So we often have to choose realistic values. 
This introduces a subjective aspect into the modelling process. Fortunately, the results derived in the next section are not too 
sensitive to the particular choice of weighting of individual observations. The uncertainty of timings becomes more important 
for short occultations -- then the uncertainty of the size is dominated by timing errors, not by errors introduced 
by too simple shape model. 

Timings determined visually are affected by observers' reaction times. When the star is relatively bright and easy to see, 
visual reaction times are less than $\sim 1.5\,$s. For fainter stars, reaction times of 2 to 3 seconds are possible. 
Observers estimate their reaction times and correct the timing for this effect. However, systematic errors introduced
by this effect are often present, as can be seen when a visual chord is compared with a nearby chord measured electronically. 
In many cases, the lengths of visual chords are correct, but they are significantly shifted in time with respect to the other chords.
For this reason, we allowed visual chords to `float' along the relative velocity vector $\hat{\vec{s}}_v$. This
shift in the fundamental plane corresponds to the shift $\Delta\tau$ in timings. For visual observations, we modified Eq.~(\ref{eq_2})
to
\begin{equation}
\label{eq_shift}
(\xi, \eta) = [ \hat{\vec{s}}_\xi \cdot (\vec{x} + \Delta\vec{v} \, (\Delta t + \Delta\tau)),
                \hat{\vec{s}}_\eta \cdot (\vec{x} + \Delta\vec{v} \, (\Delta t + \Delta\tau))],
\end{equation}
where $\Delta\tau$ are free parameters that are optimized to get the lowest $\chi^2$. To avoid too large shifts in time, we 
include the penalty function 
\begin{equation}
\chi^2_\mathrm{new} = \chi^2 + \sum_i \left( \frac{\Delta\tau_i}{\gamma} \right)^2,
\end{equation}
where $\gamma$ is chosen subjectively to trade-off between the goodness of the fit and the magnitude of the time shift. Usually, 
$\Delta\tau_i$ values of the order of tenths of second are sufficient. In some cases $\Delta\tau_i$ are of the order of seconds.

In principle, also some chords observed with video or CCD can be misplaced with respect to other chords due to bad absolute timing. Shifting such
chords according to Eq.~(\ref{eq_shift}) would improve the fit but one has to be sure that the discrepancy with other chords is caused by bad timing, not 
by some real features of the shape. We applied the shift only to chords observed visually.

\section{Results}
\label{results}

In this section, we briefly describe all asteroids for which we scaled the shape models to fit the occultation data. We list all
occultations used in our analysis in Table~\ref{table1}. For each occultation event, we list the total number of chords 
$N_\mathrm{total}$ we used 
(that can be lower than the number of observed chords because we sometimes reject clearly erroneous chords), 
the number of chords $N_\mathrm{phot}$ measured photoelectrically, using video, or CCD, the angle $\Delta\phi$ of which the asteroid 
rotated during the occultation, the mean duration of the event $t_\mathrm{mean}$, and the reference to the paper where the occultation results were published. 
For each occultation observed visually, the average absolute time shift $|\Delta \tau|$ of visual chords (Eq.~\ref{eq_shift}) is listed.
Observations that report
only one timing (disappearance or reappearance) are counted as individual chords.

Table~\ref{table_observers} lists all observers that participated in observations of recent occultations that were not included in the database 
of \cite{Dun.Her:09}. We list all observers, including those that reported `misses'. If some observer provided more than one chord, the number 
of chords is given in parentheses.

In Table~\ref{table2}, we list the derived equivalent diameter $D$ for each model
(which is the diameter of a sphere with the same volume as the shape model) and their estimated errors, 
the ecliptic latitude $\lambda_\mathrm{p}$ and longitude $\beta_\mathrm{p}$ of the pole direction, the sidereal rotation period $P$, 
the diameter 
$D_\mathrm{IRAS}$ derived from the IRAS infrared measurements by \cite{Ted.ea:04}, the rms residual of the fit with visual chords allowed to move
along the relative velocity vector (rms$_1$) and with visual chords fixed (rms$_2$),
and the reference
to the paper where the original shape model was published. If one of the pole solutions is preferred, the corresponding values are typeset in bold. We estimated 
the uncertainty of the equivalent size as twice the rms residual of the fit. In cases where the errors in timing are the main
source of uncertainty of the diameter (6~Hebe, for example), we estimated the uncertainty of the size by propagating the errors of timings.

For asteroids that have known mass, we could compute the density. However, such density estimates would be biased, beacause the volume of a convex model is 
larger than the real volume of the asteroid. Therefore we do not provide any density estimations.

\begin{figure}[t]
\begin{center}
\includegraphics[width=\columnwidth]{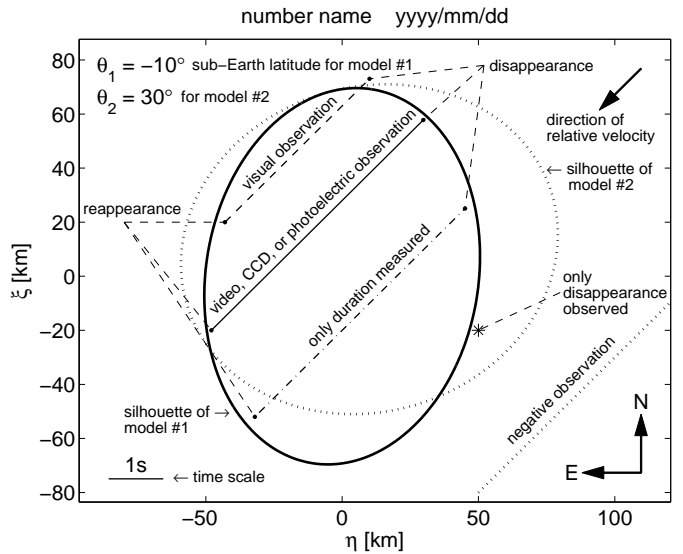}
\caption{Explanation of line types and symbols used in Figs.~\ref{Pallas_fig}--\ref{Varsavia_fig}.}
\label{fig_example}
\end{center}
\end{figure}

\begin{figure}
\begin{center}
\includegraphics[width=0.32\columnwidth]{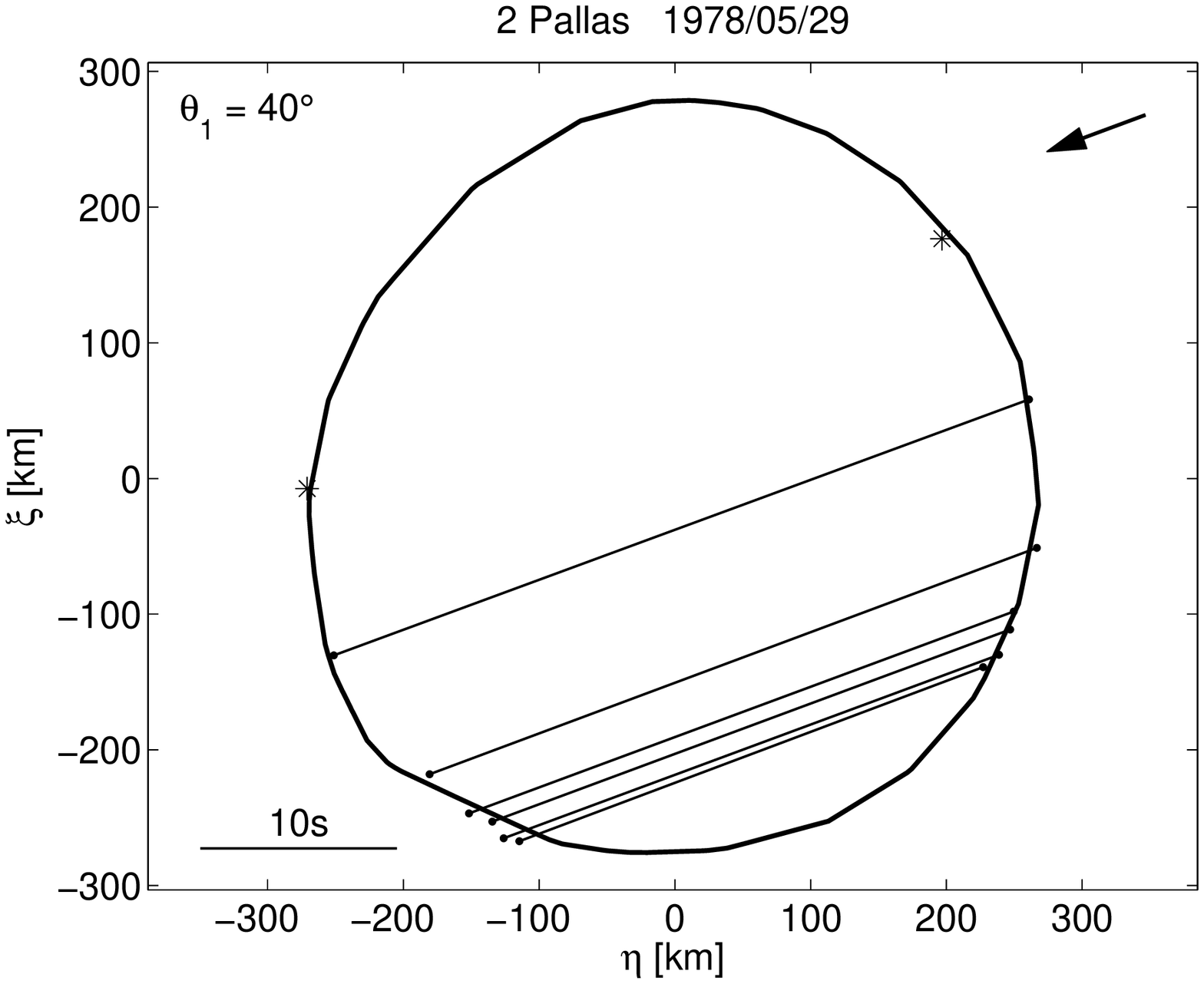}
\includegraphics[width=0.32\columnwidth]{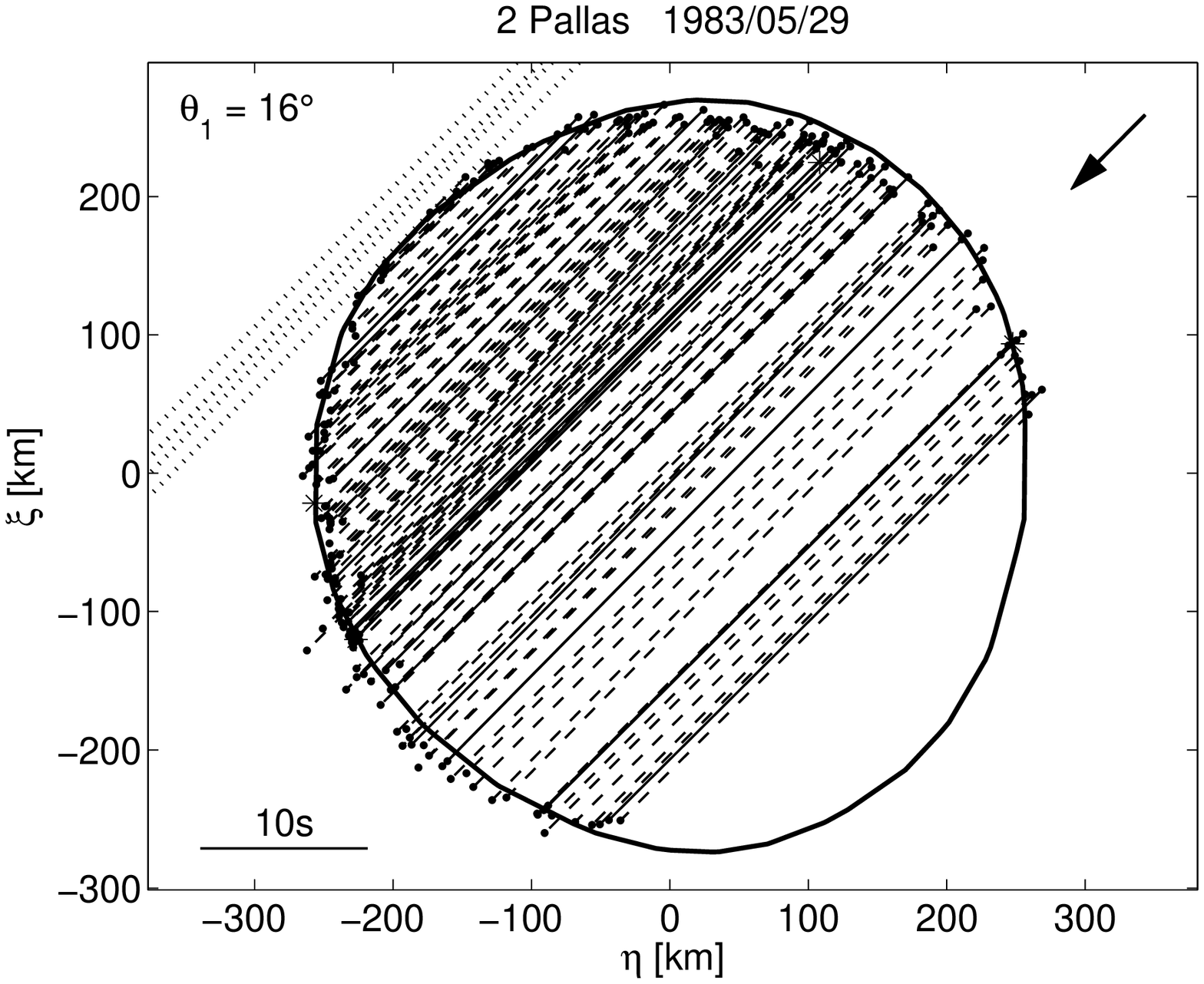}
\includegraphics[width=0.32\columnwidth]{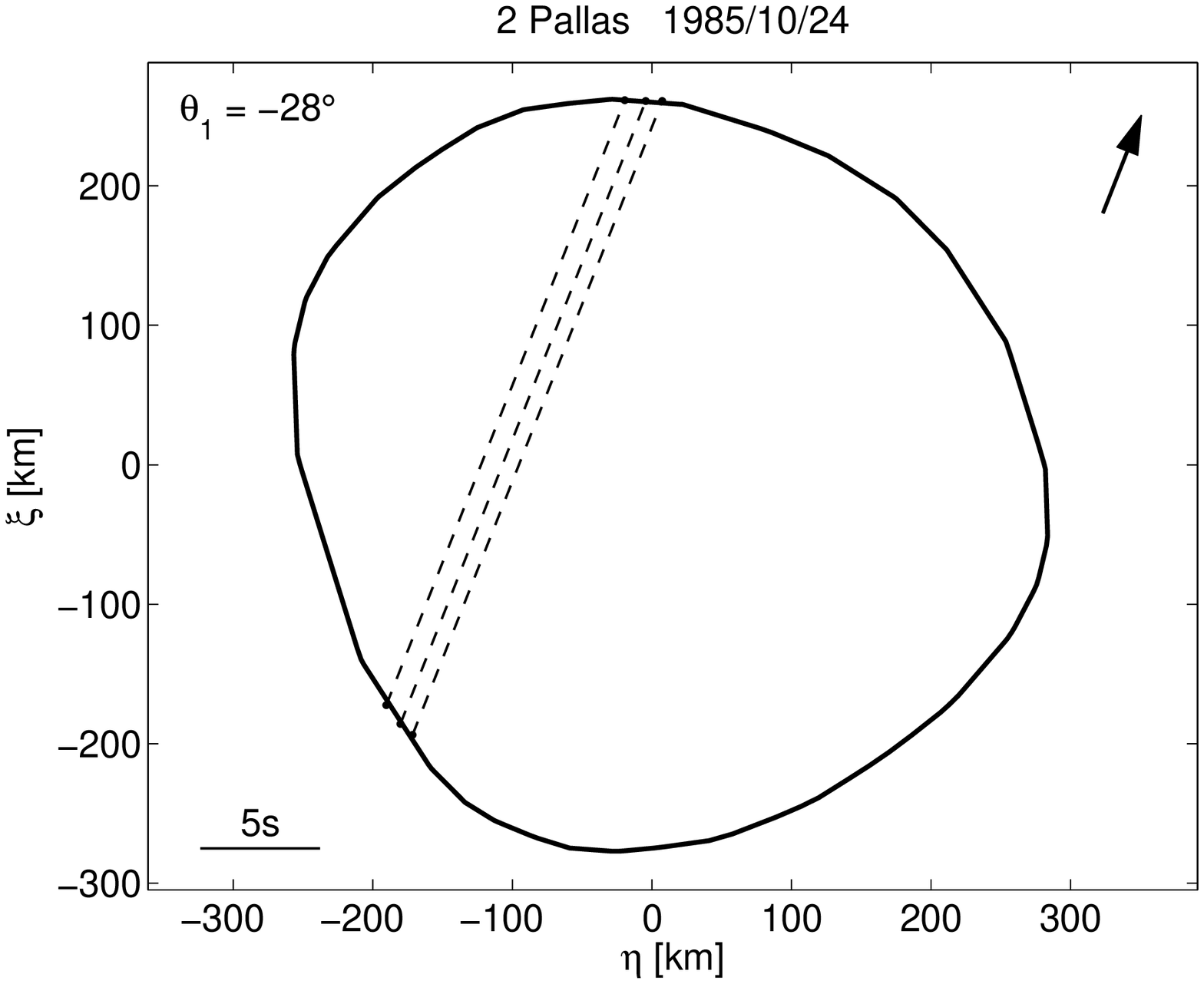}
\includegraphics[width=0.32\columnwidth]{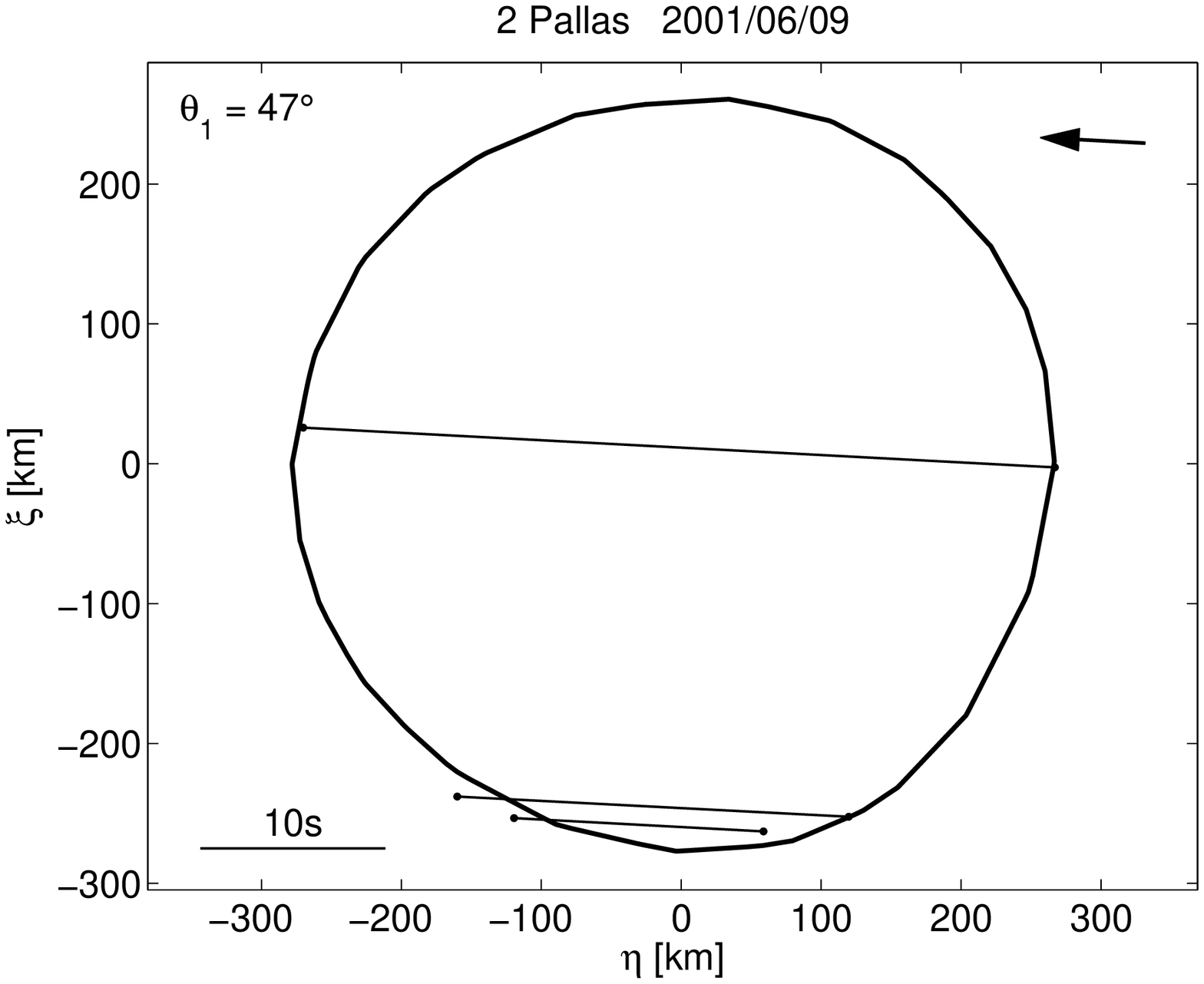}
\includegraphics[width=0.32\columnwidth]{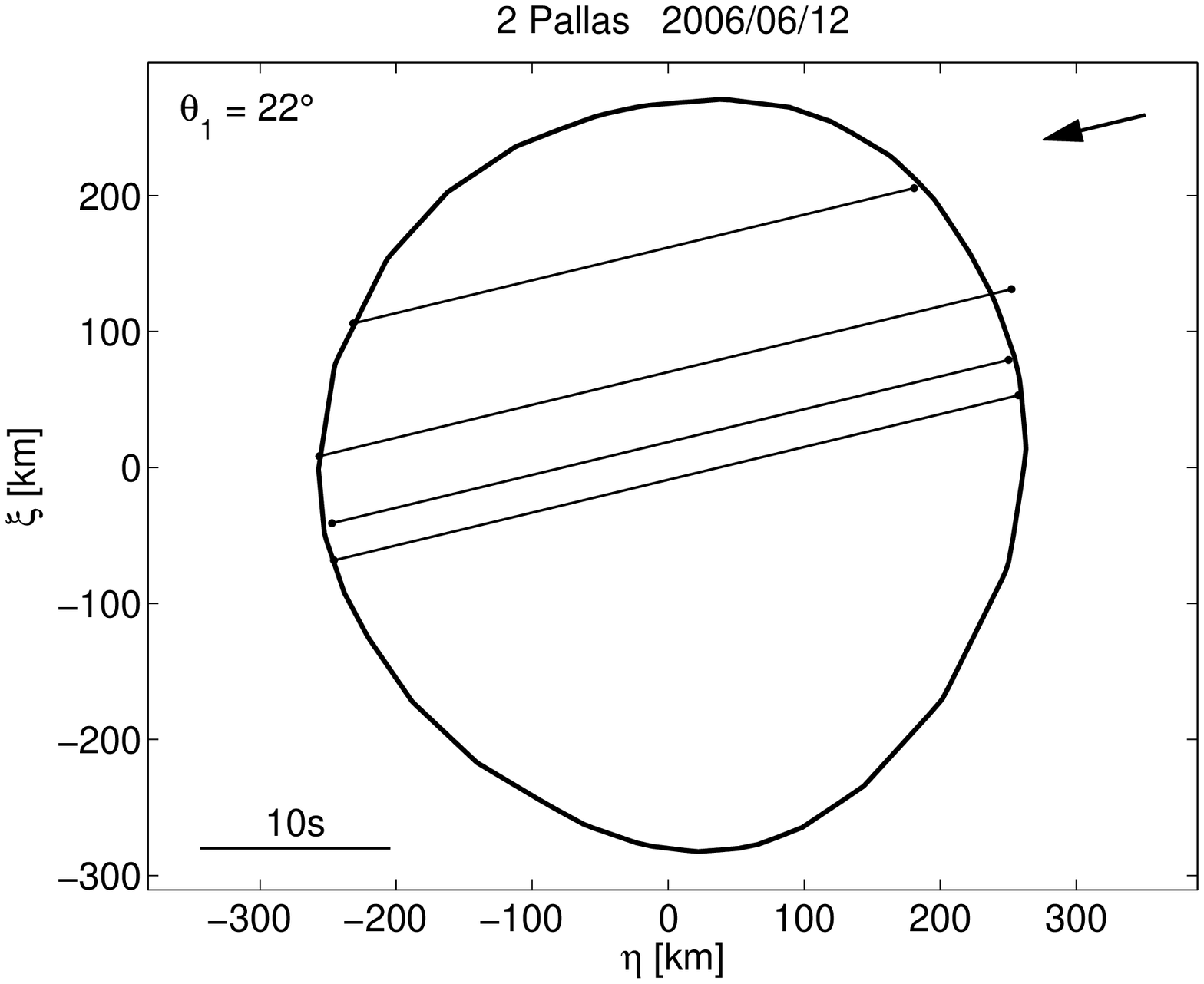}
\caption{(2) Pallas}
\label{Pallas_fig}
\end{center}
\end{figure}

\begin{figure}
\begin{center}
\includegraphics[width=0.32\columnwidth]{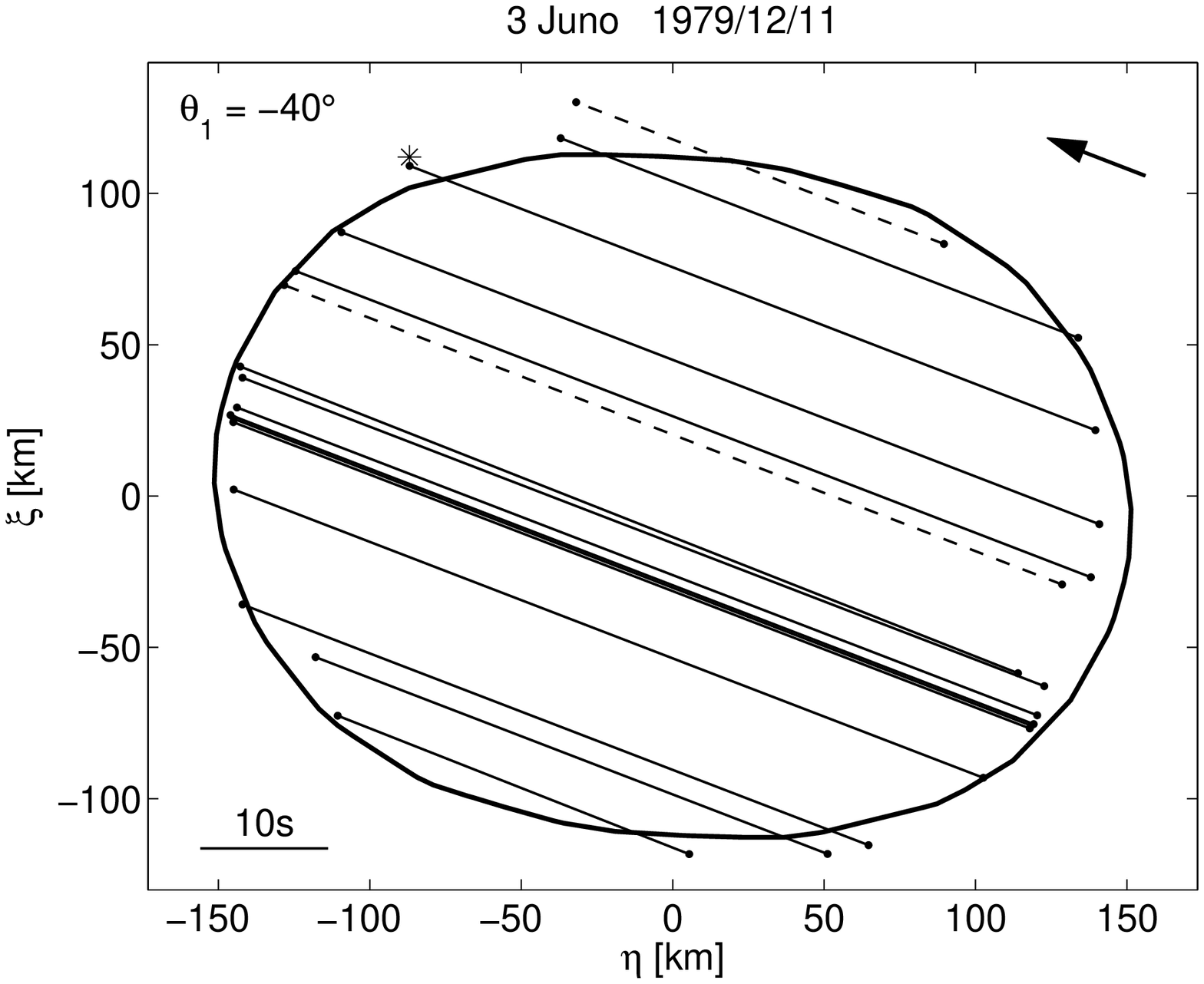}
\includegraphics[width=0.32\columnwidth]{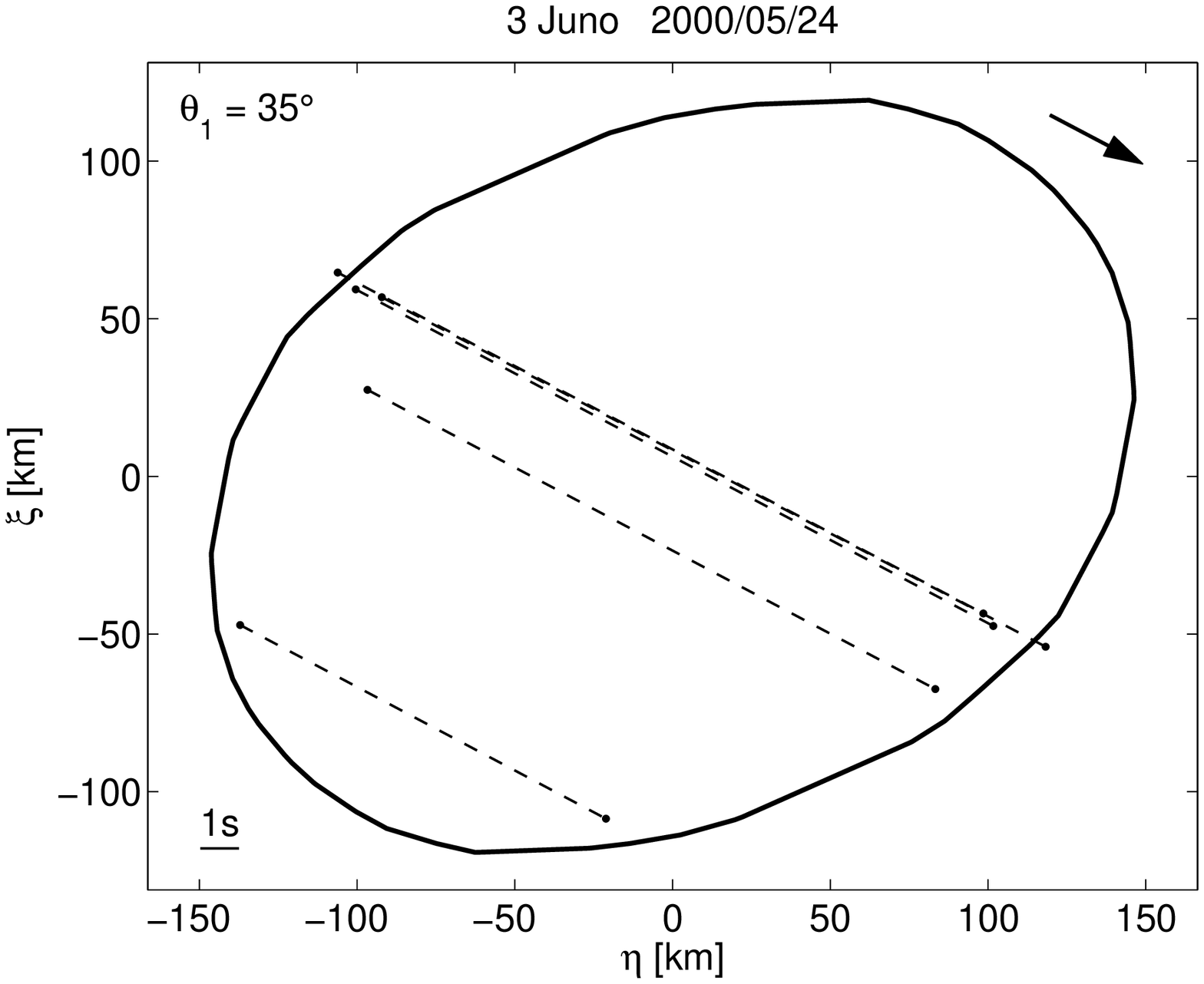}
\caption{(3) Juno}
\label{Juno_fig}
\end{center}
\end{figure}

\begin{figure}
\begin{center}
\includegraphics[width=0.32\columnwidth]{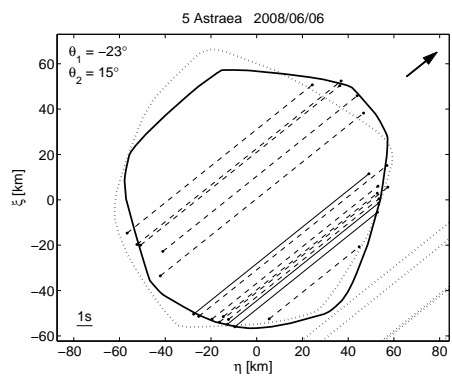}
\caption{(5) Astraea. The solid contour corresponds to the model with the pole direction $(126^\circ, 40^\circ)$, 
the dotted one to the pole $(310^\circ, 44^\circ)$.}
\label{Astraea_fig}
\end{center}
\end{figure}

\begin{figure}
\begin{center}
\includegraphics[width=0.32\columnwidth]{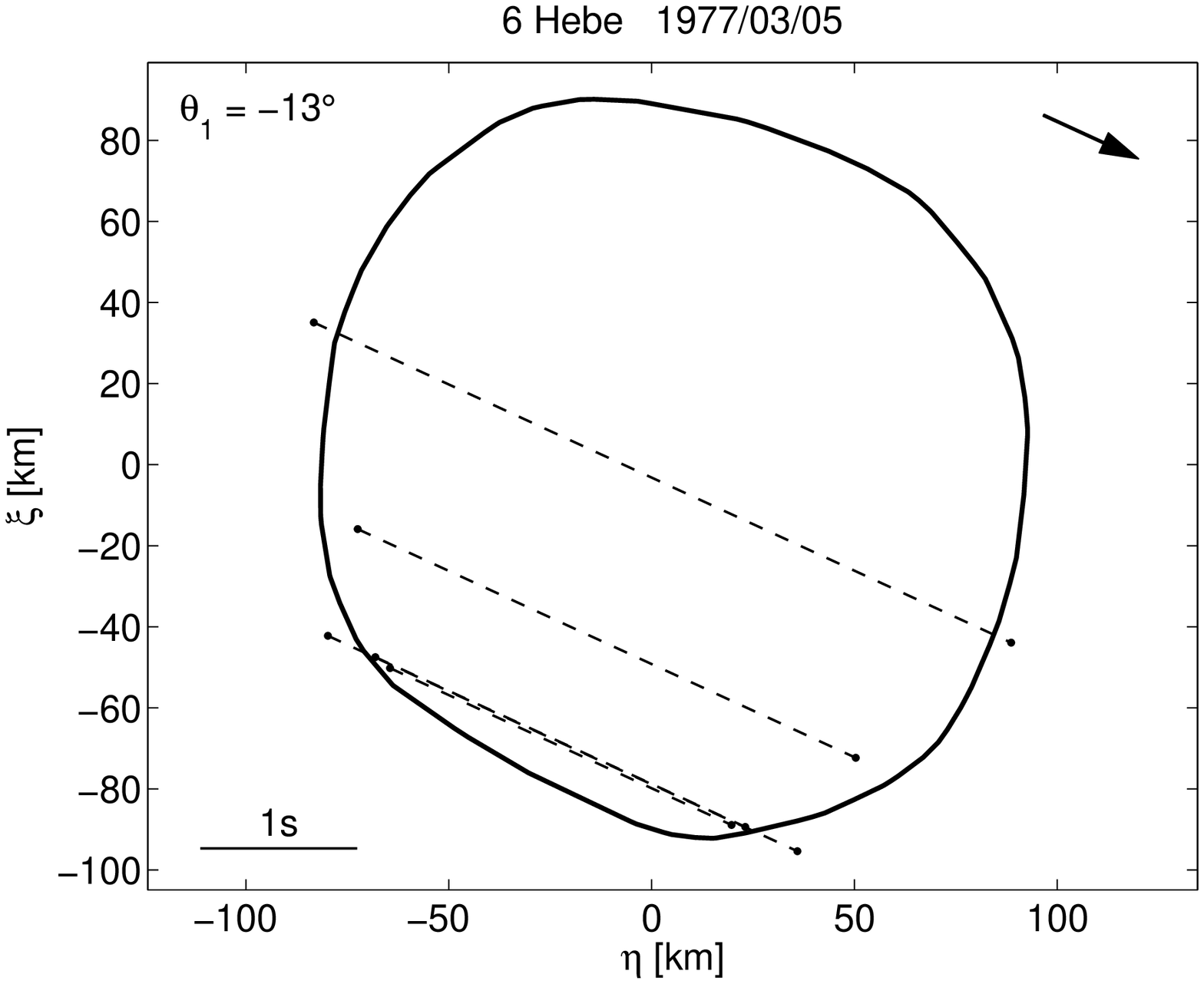}
\caption{(6) Hebe.}
\label{Hebe_fig}
\end{center}
\end{figure}

\begin{figure}
\begin{center}
\includegraphics[width=0.32\columnwidth]{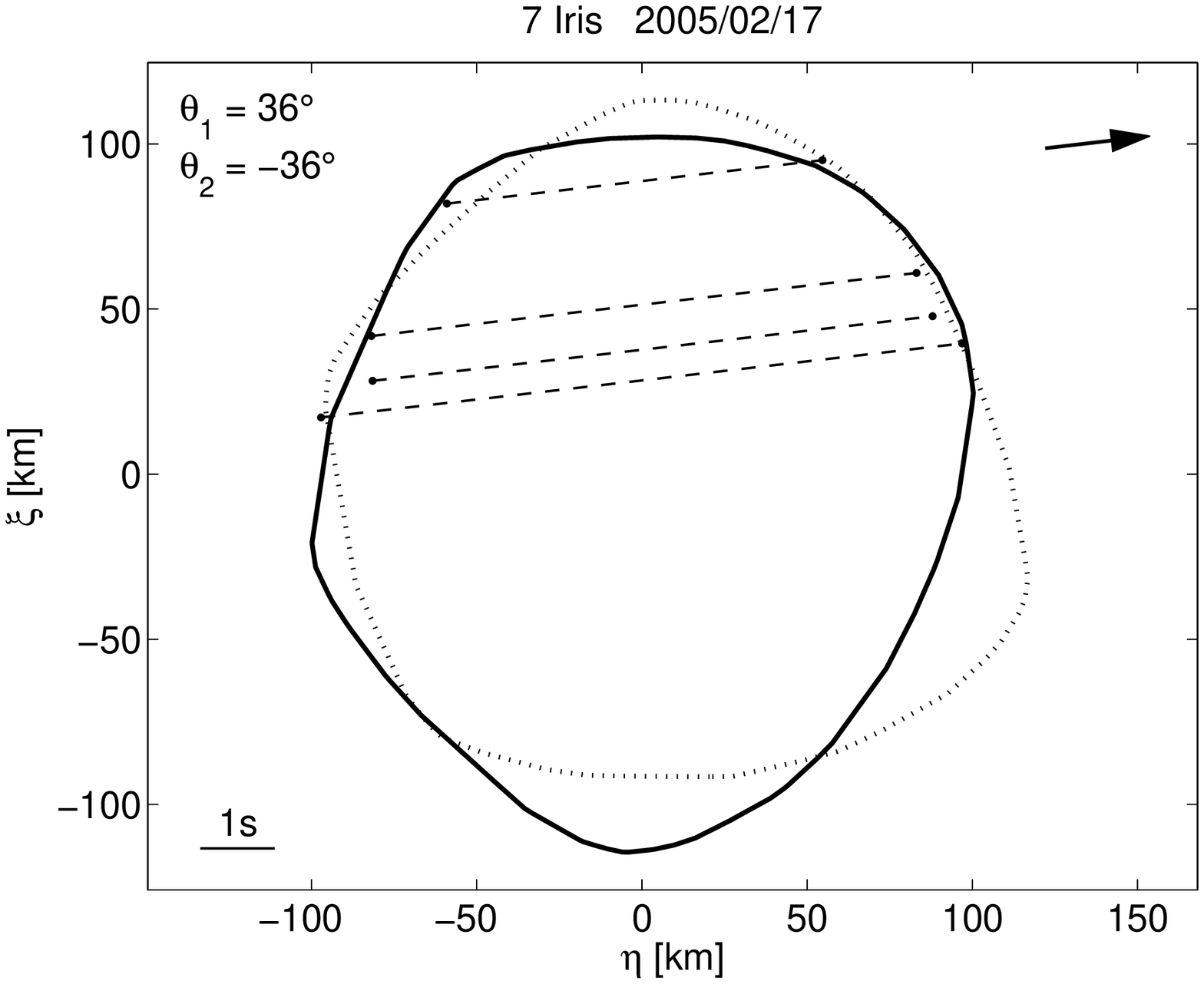}
\includegraphics[width=0.32\columnwidth]{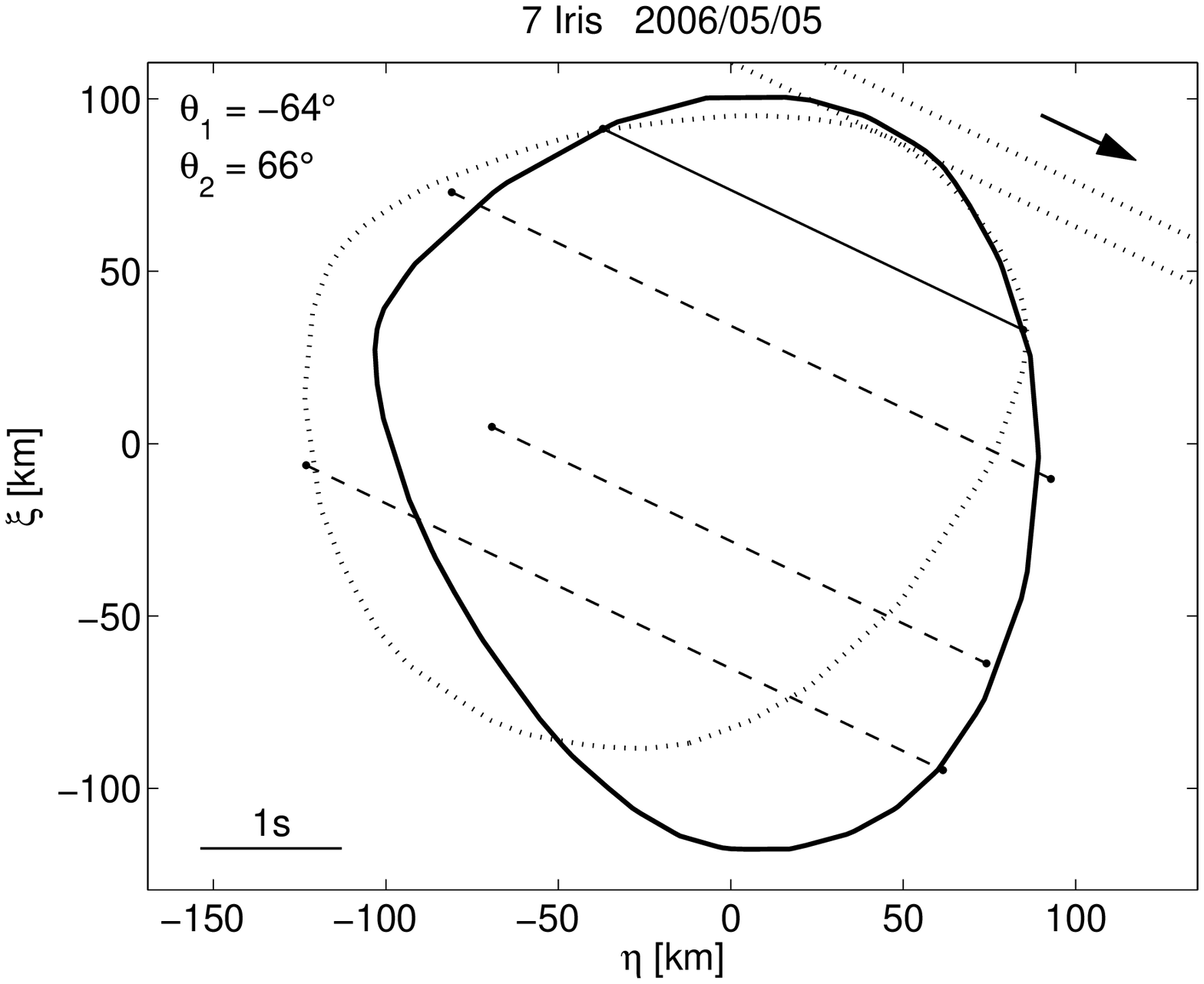}
\caption{(7) Iris. The solid contour corresponds to the pole $(20^\circ, 14^\circ)$ and the dotted one to the 
pole $(199^\circ, -2^\circ)$.}
\label{Iris_fig}
\end{center}
\end{figure}

\begin{figure}
\begin{center}
\includegraphics[width=0.32\columnwidth]{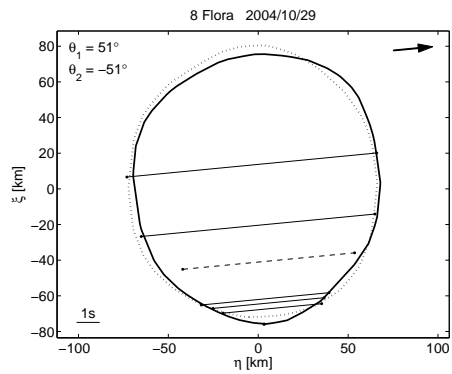}
\caption{(8) Flora. The solid contour corresponds to the pole $(335^\circ, -5^\circ)$, the dotted one to $(155^\circ, 6^\circ)$.}
\label{Flora_fig}
\end{center}
\end{figure}

\begin{figure}
\begin{center}
\includegraphics[width=0.32\columnwidth]{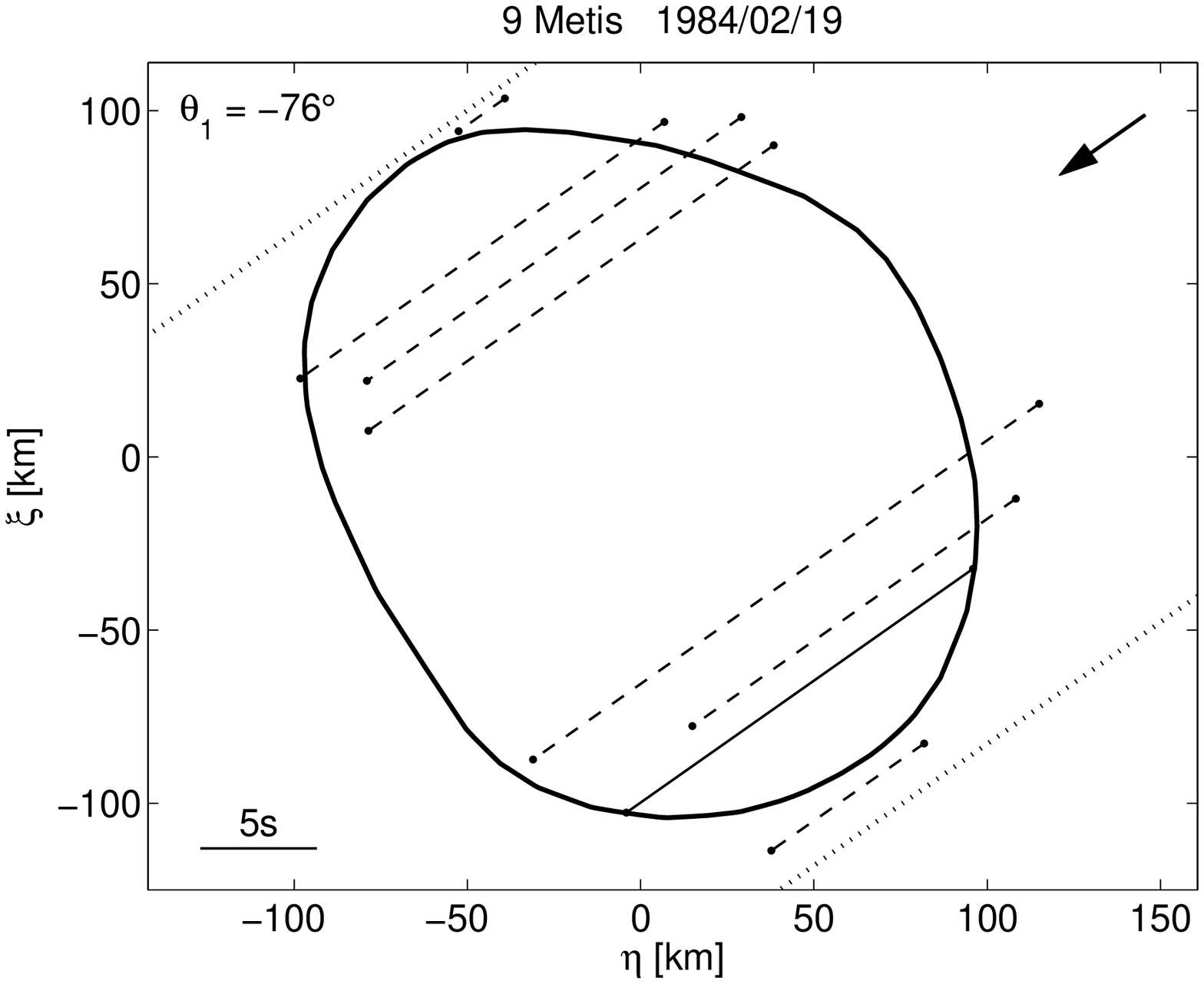}
\includegraphics[width=0.32\columnwidth]{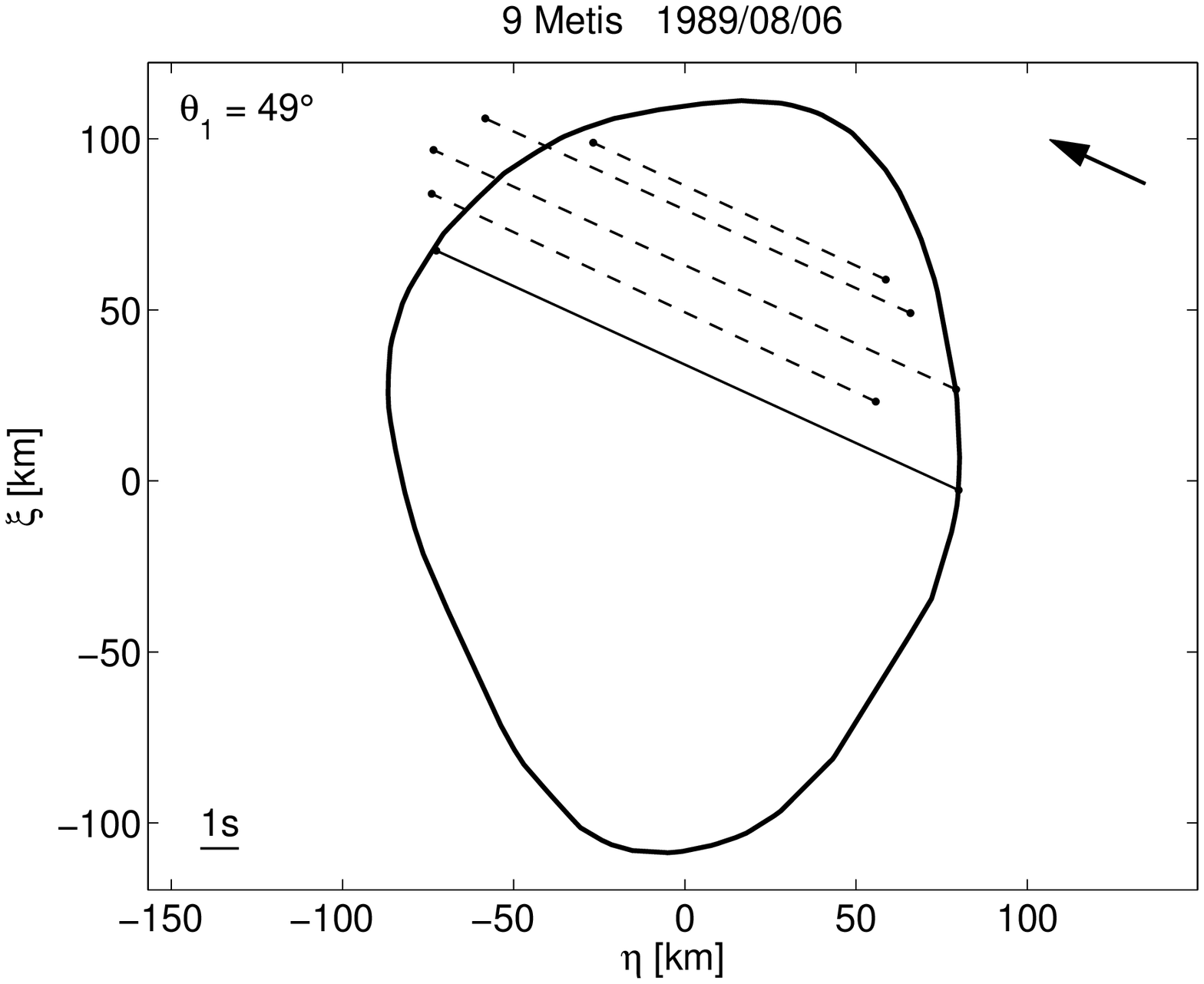}
\includegraphics[width=0.32\columnwidth]{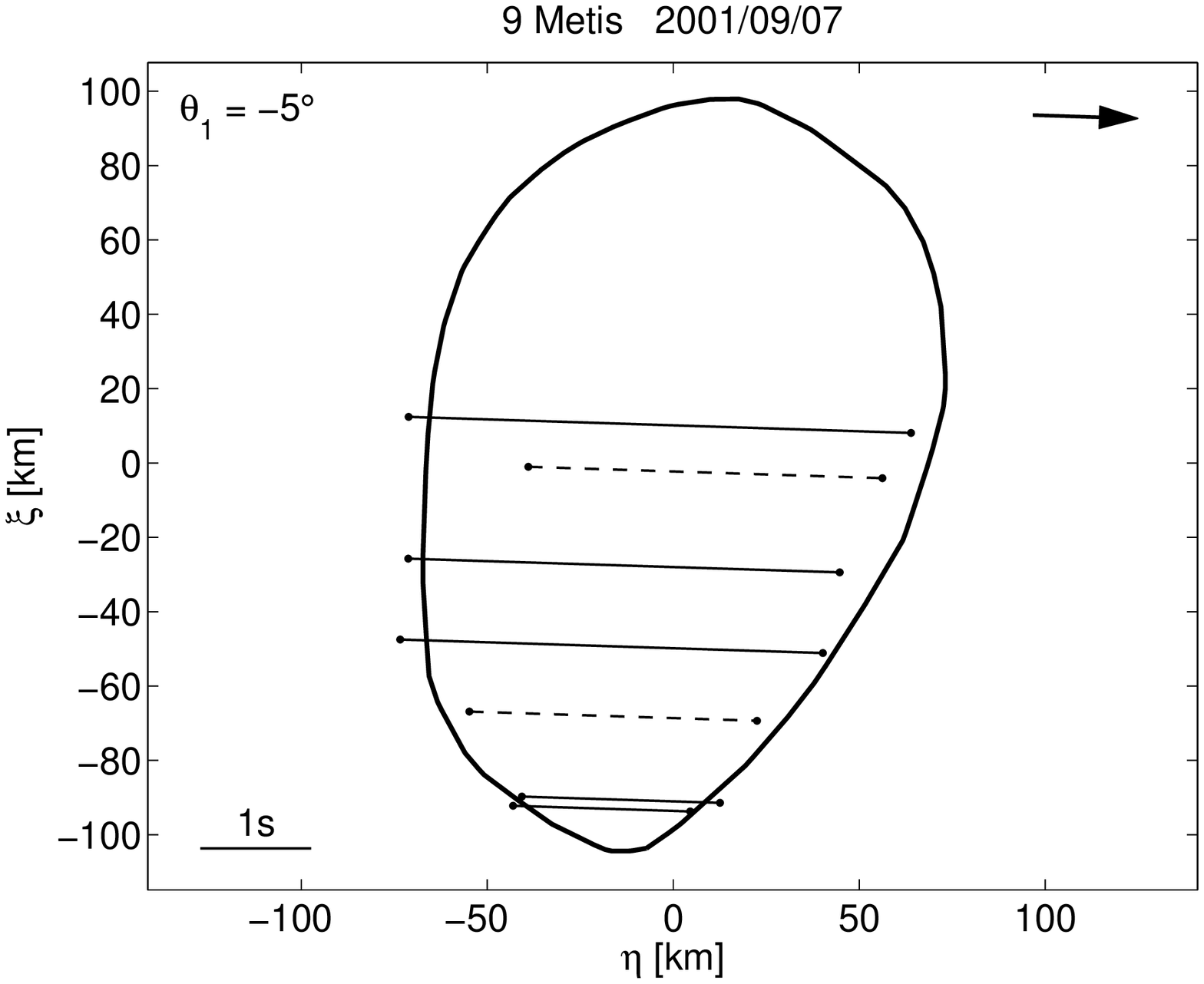}
\includegraphics[width=0.32\columnwidth]{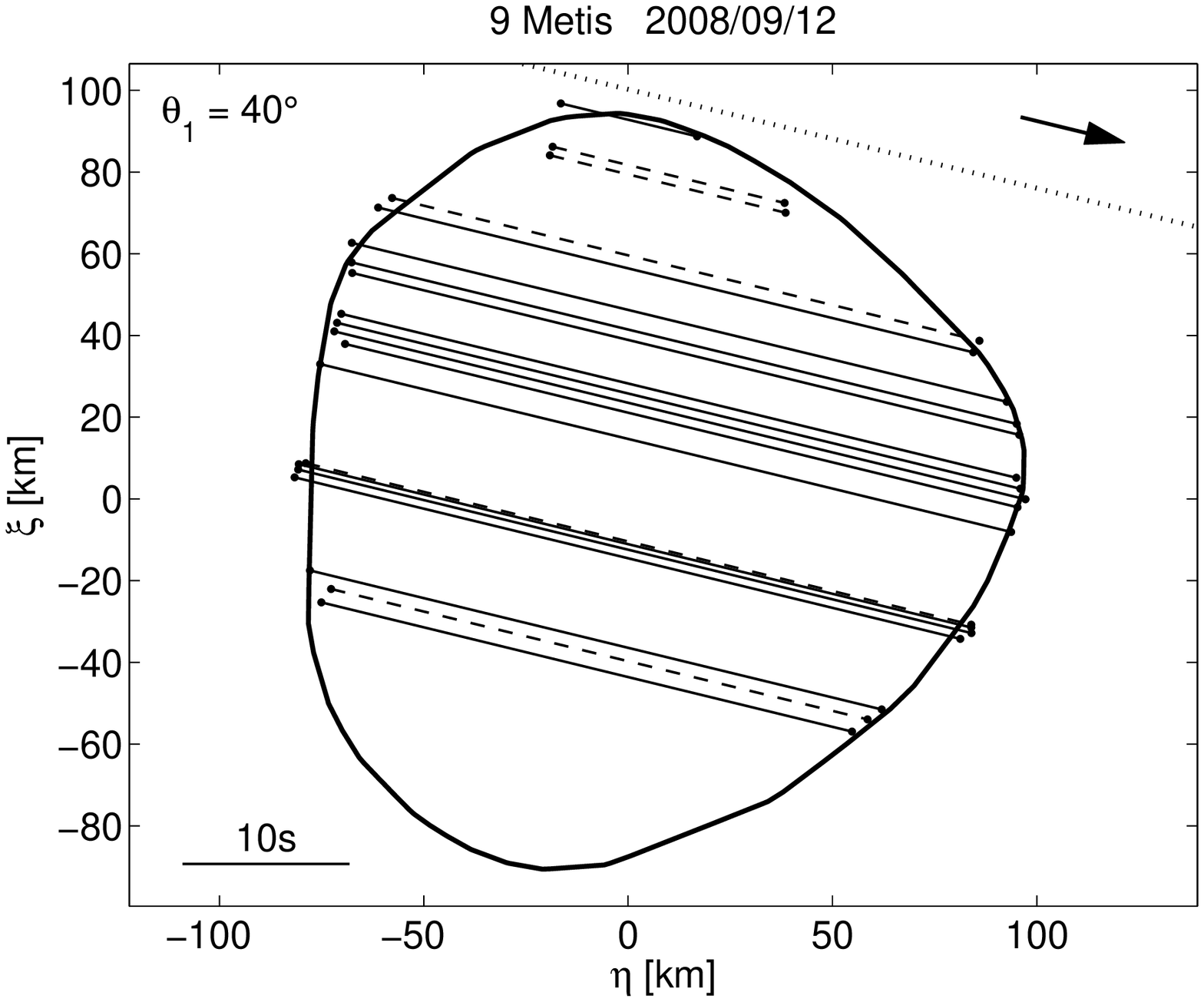}
\caption{(9) Metis.}
\label{Metis_fig}
\end{center}
\end{figure}

\begin{figure}
\begin{center}
\includegraphics[width=0.32\columnwidth]{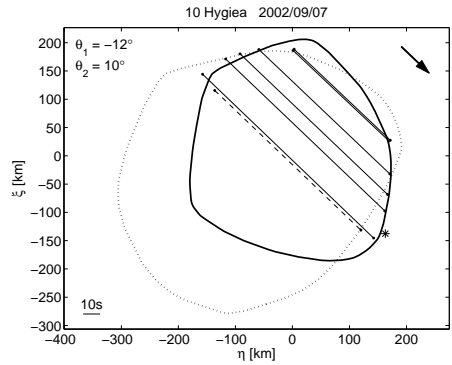}
\caption{(10) Hygiea. The solid contour corresponds to the pole $(122^\circ, -44^\circ)$, the dotted one to $(312^\circ, -42^\circ)$.}
\label{Hygiea_fig}
\end{center}
\end{figure}

\begin{figure}
\begin{center}
\includegraphics[width=0.32\columnwidth]{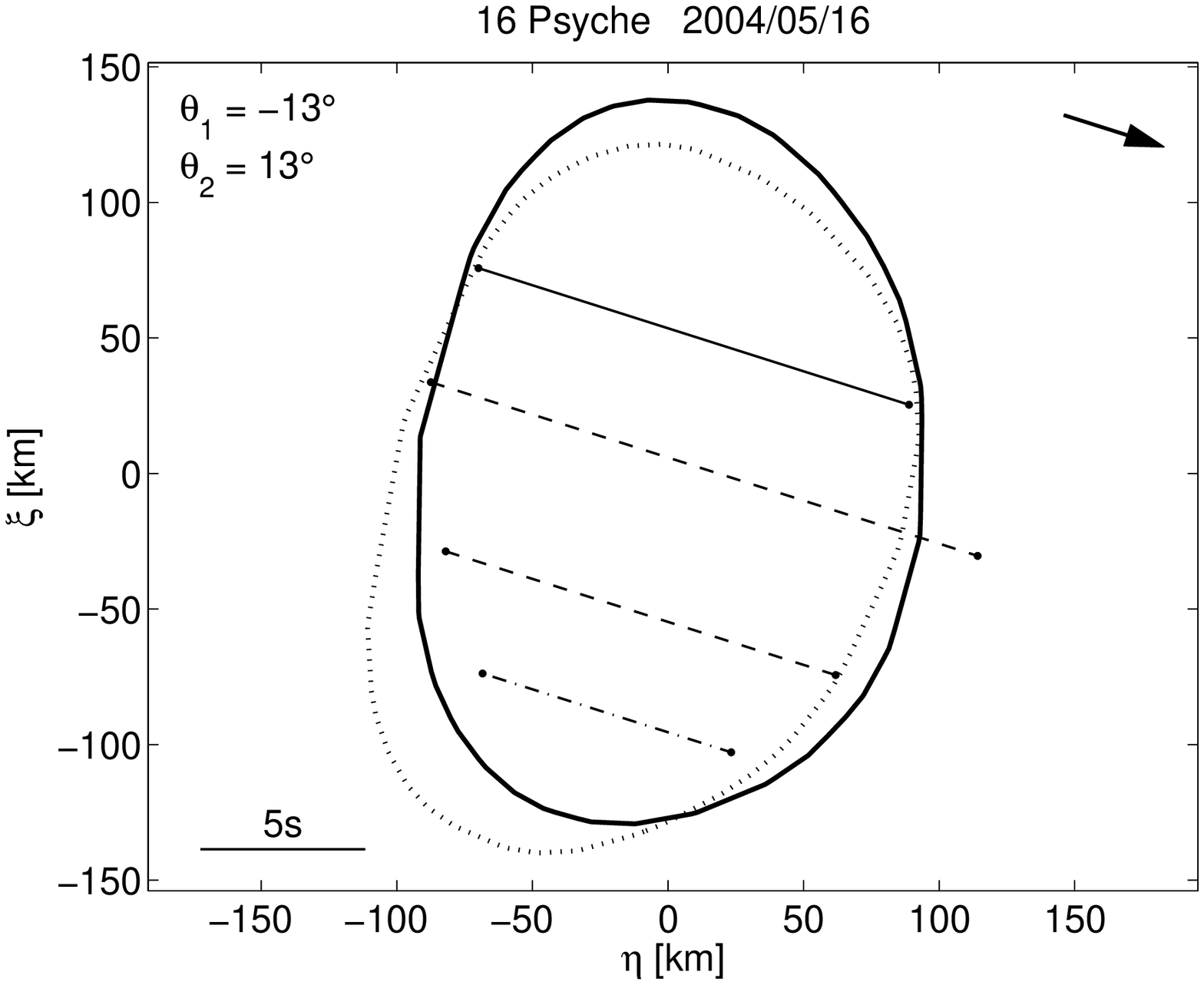}
\includegraphics[width=0.32\columnwidth]{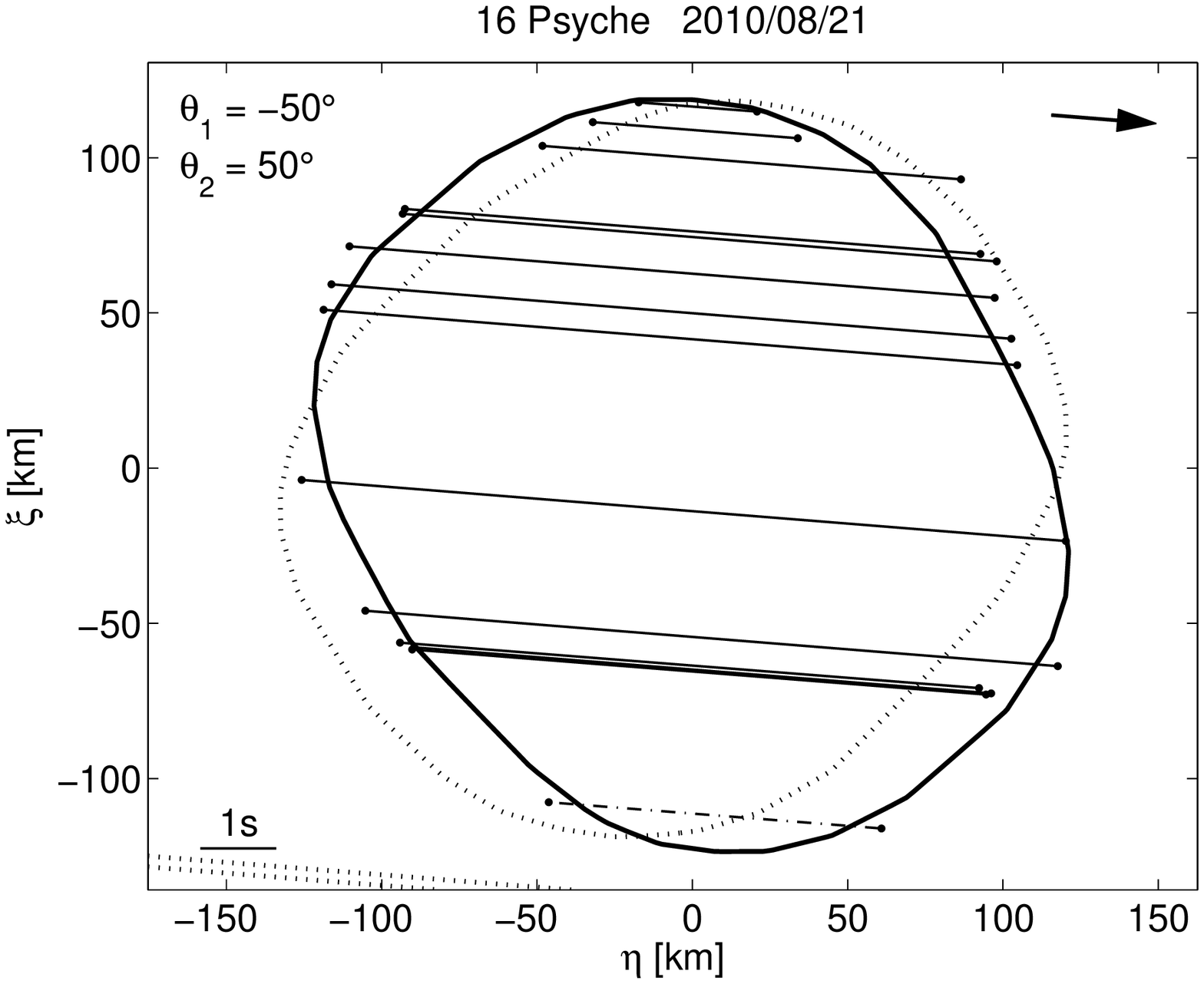}
\caption{(16) Psyche. The solid contour corresponds to the pole $(33^\circ, -7^\circ)$, the dotted one to $(213^\circ, 1^\circ)$.}
\label{Psyche_fig}
\end{center}
\end{figure}

\begin{figure}
\begin{center}
\includegraphics[width=0.32\columnwidth]{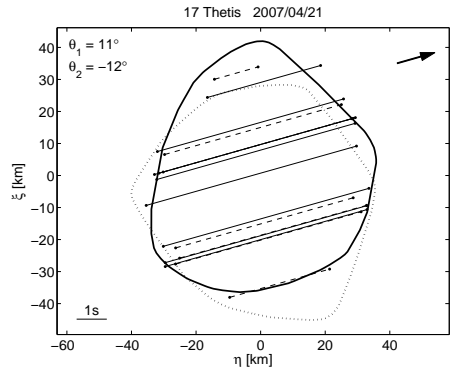}
\caption{(17) Thetis. The solid contour corresponds to the pole $(236^\circ, 20^\circ)$, the dotted one to $(55^\circ, 10^\circ)$.}
\label{Thetis_fig}
\end{center}
\end{figure}

\begin{figure}
\begin{center}
\includegraphics[width=0.32\columnwidth]{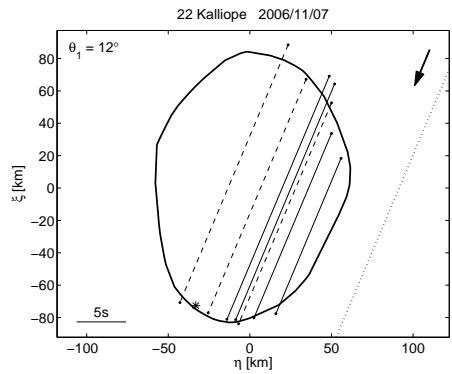}
\caption{(22) Kalliope.}
\label{Kalliope_fig}
\end{center}
\end{figure}

\begin{figure}
\begin{center}
\includegraphics[width=0.32\columnwidth]{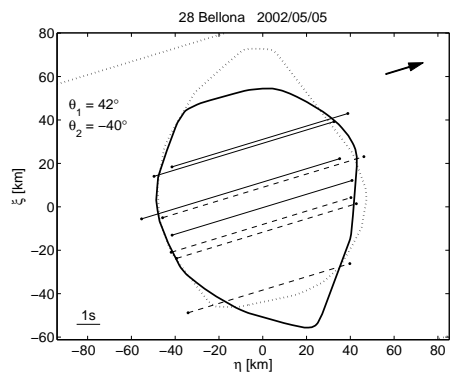}
\caption{(28) Bellona. The solid contour corresponds to the pole $(282^\circ, 6^\circ)$, the dotted one to $(102^\circ, -8^\circ)$.}
\label{Bellona_fig}
\end{center}
\end{figure}

\begin{figure}
\begin{center}
\includegraphics[width=0.32\columnwidth]{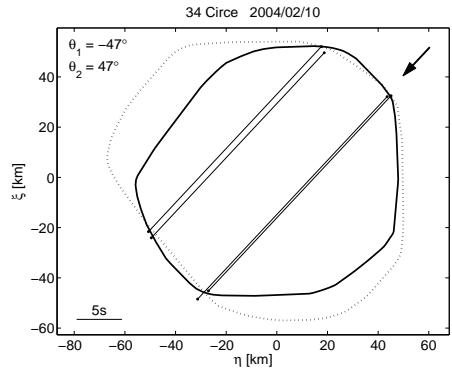}
\caption{(34) Circe. The solid profile corresponds to the pole $(94^\circ, 35^\circ)$, the dotted one to $(275^\circ, 51^\circ)$.}
\label{Circe_fig}
\end{center}
\end{figure}

\begin{figure}
\begin{center}
\includegraphics[width=0.32\columnwidth]{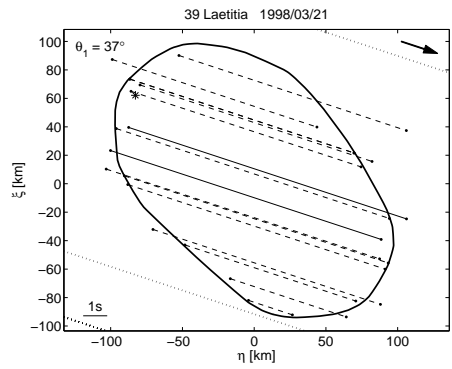}
\caption{(39) Laetitia.}
\label{Laetitia_fig}
\end{center}
\end{figure}

\begin{figure}
\begin{center}
\includegraphics[width=0.32\columnwidth]{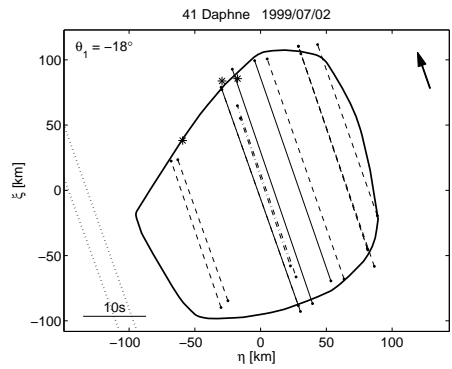}
\caption{(41) Daphne.}
\label{Daphne_fig}
\end{center}
\end{figure}

\begin{figure}
\begin{center}
\includegraphics[width=0.32\columnwidth]{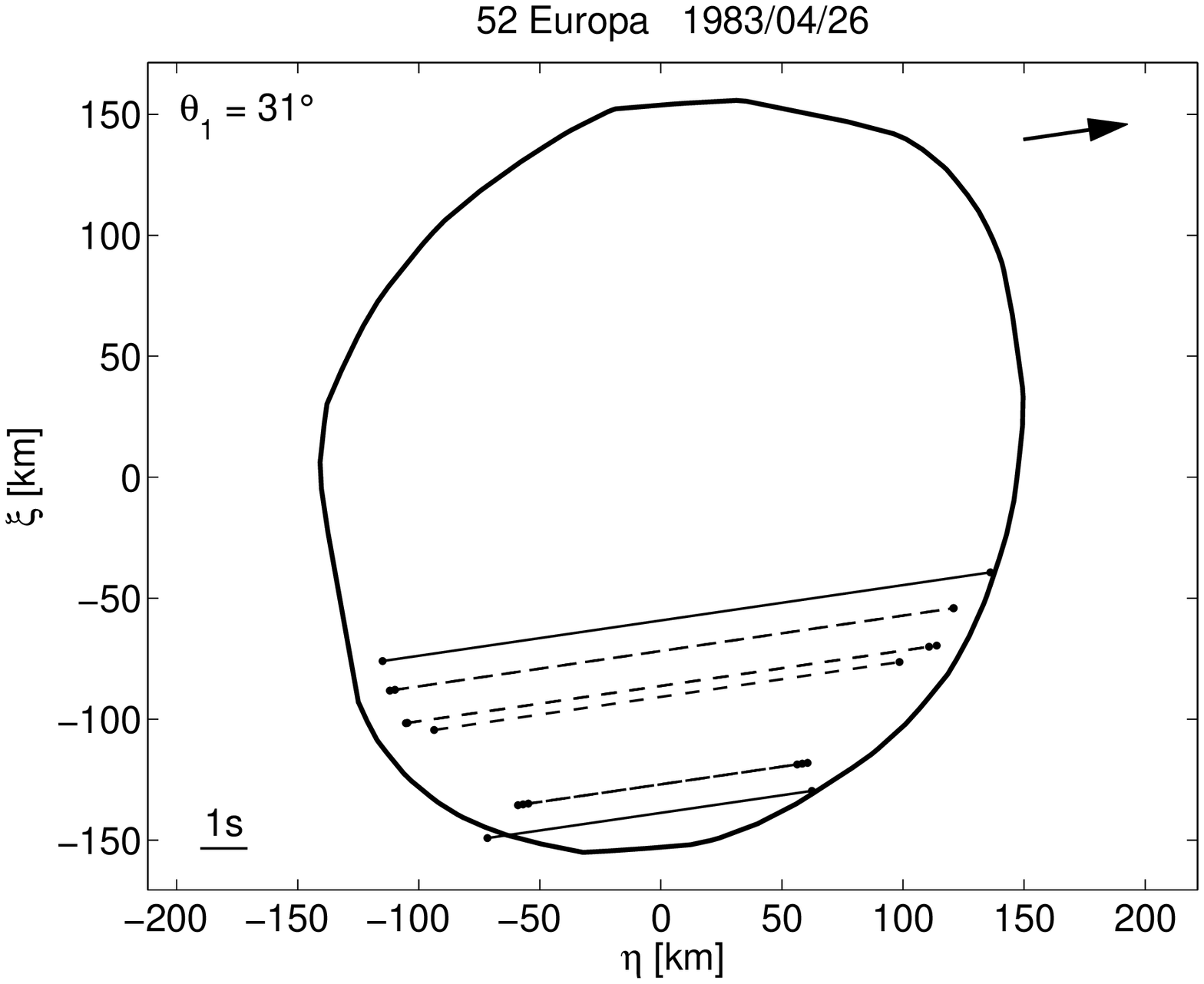}
\includegraphics[width=0.32\columnwidth]{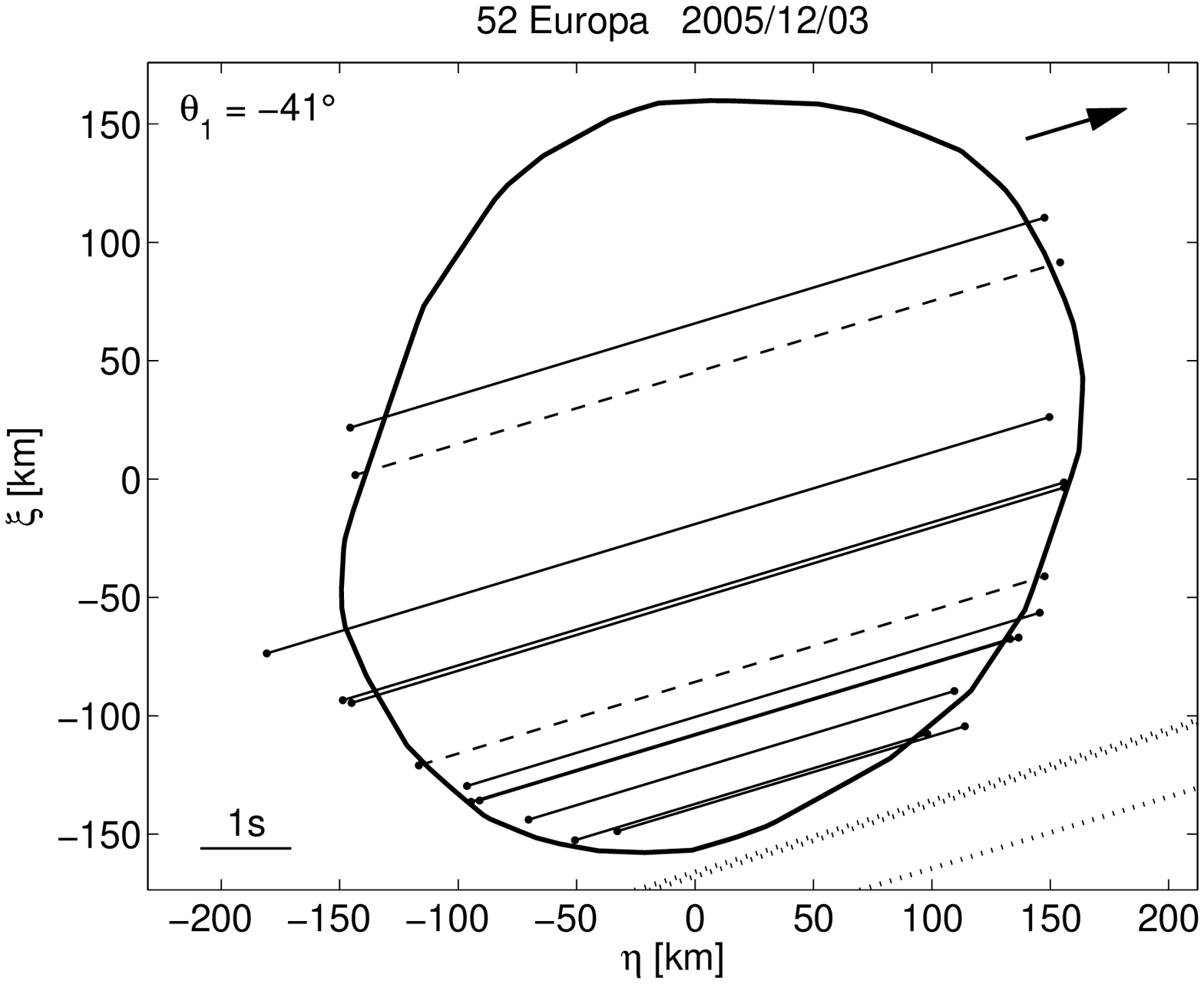}
\caption{(52) Europa.}
\label{Europa_fig}
\end{center}
\end{figure}

\begin{figure}
\begin{center}
\includegraphics[width=0.32\columnwidth]{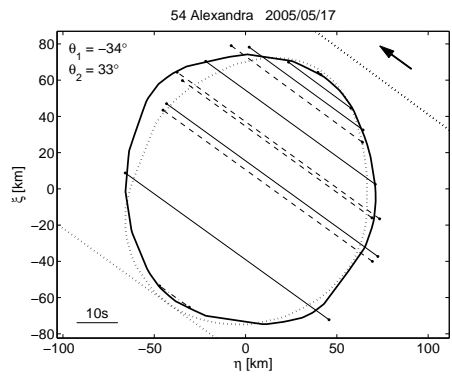}
\caption{(54) Alexandra. The solid profile corresponds to the pole $(318^\circ, 23^\circ)$, the dotted one to $(156^\circ, 13^\circ)$.}
\label{Alexandra_fig}
\end{center}
\end{figure}

\begin{figure}
\begin{center}
\includegraphics[width=0.32\columnwidth]{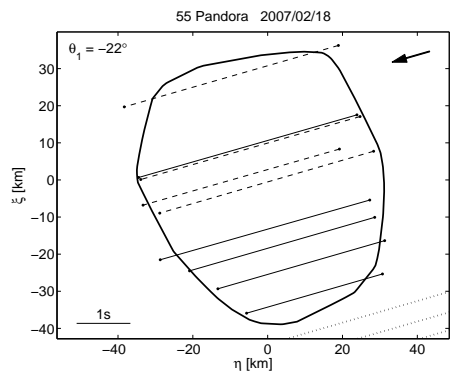}
\caption{(55) Pandora.}
\label{Pandora_fig}
\end{center}
\end{figure}

\begin{figure}
\begin{center}
\includegraphics[width=0.32\columnwidth]{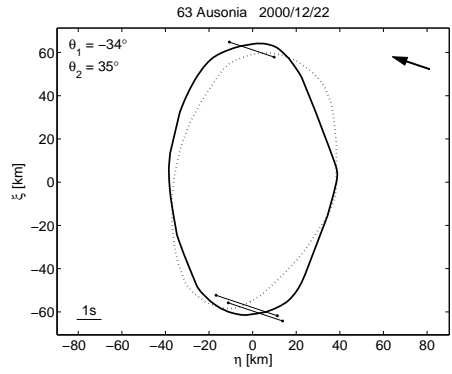}
\caption{(63) Ausonia. The solid profile corresponds to the pole $(120^\circ, -15^\circ)$, the dotted one to $(305^\circ, -21^\circ)$.}
\label{Ausonia_fig}
\end{center}
\end{figure}

\begin{figure}
\begin{center}
\includegraphics[width=0.32\columnwidth]{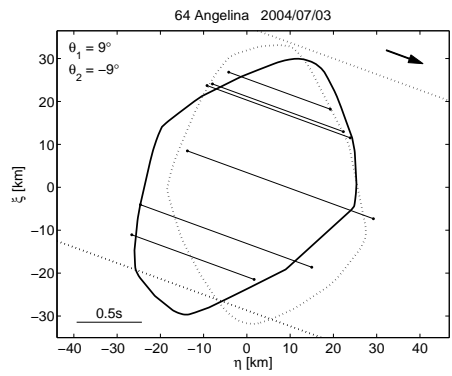}
\caption{(64) Angelina. The solid profile corresponds to the pole $(137^\circ, 14^\circ)$, the dotted one to $(317^\circ, 17^\circ)$.}
\label{Angelina_fig}
\end{center}
\end{figure}

\begin{figure}
\begin{center}
\includegraphics[width=0.32\columnwidth]{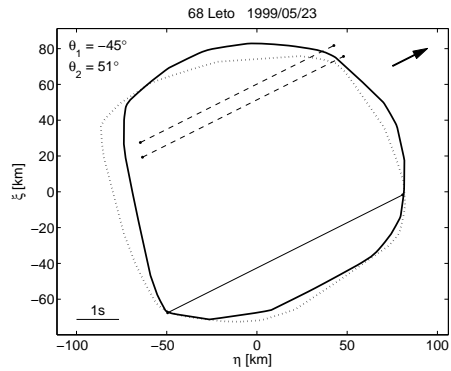}
\caption{(68) Leto. The solid profile corresponds to the pole $(103^\circ, 43^\circ)$, the dotted one to $(209^\circ, 23^\circ)$.}
\label{Leto_fig}
\end{center}
\end{figure}

\begin{figure}
\begin{center}
\includegraphics[width=0.32\columnwidth]{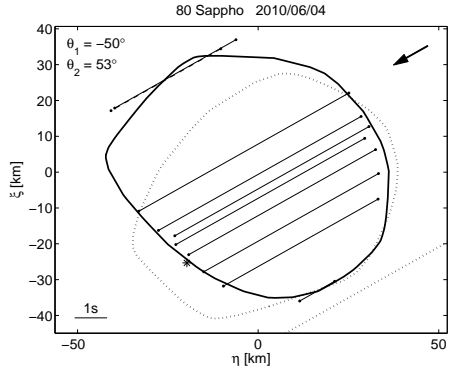}
\caption{(80) Sappho. The solid profile corresponds to the pole $(194^\circ, -26^\circ)$, the dotted one to $(6^\circ, -16^\circ)$.}
\label{Sappho_fig}
\end{center}
\end{figure}

\begin{figure}
\begin{center}
\includegraphics[width=0.32\columnwidth]{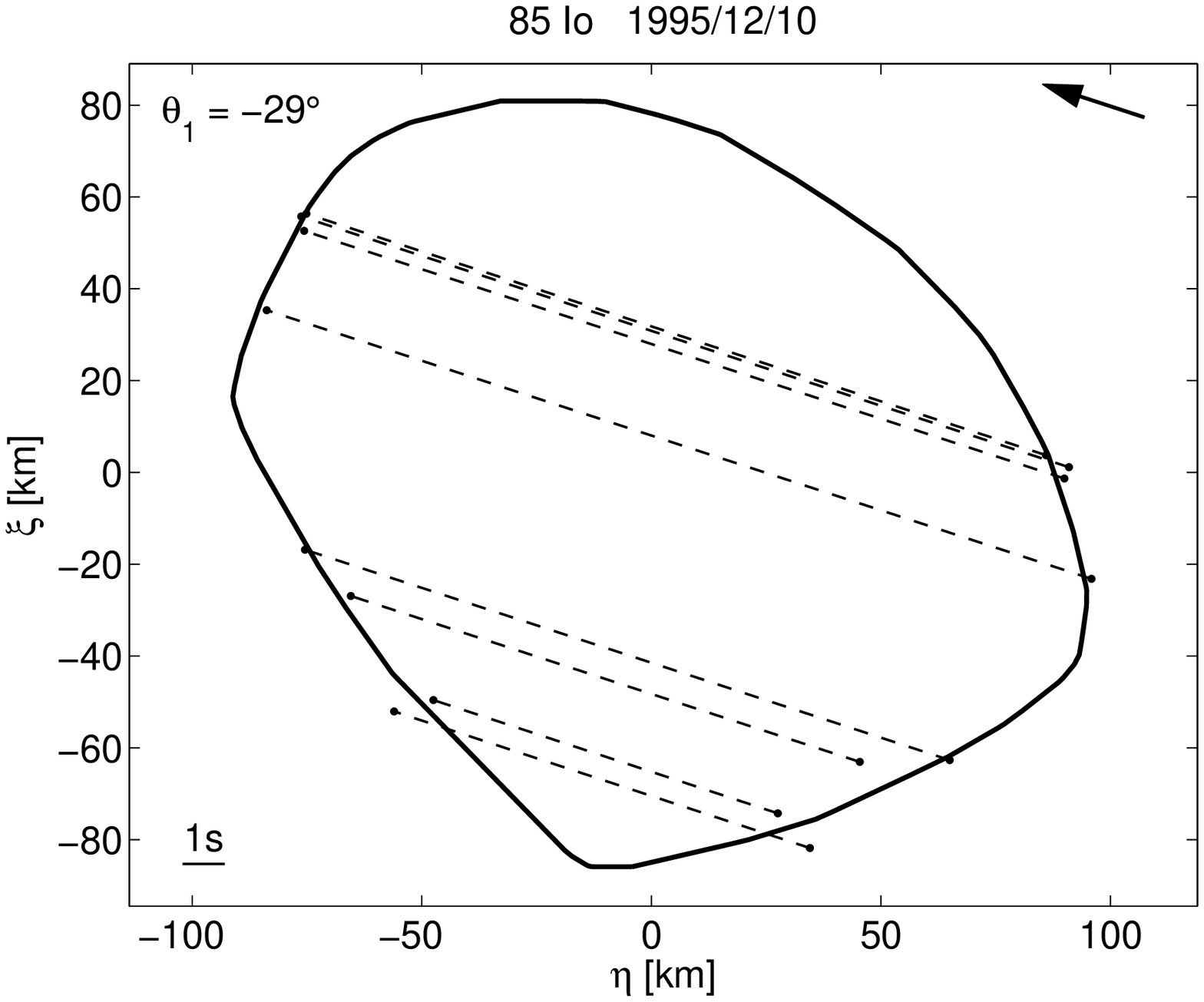}
\includegraphics[width=0.32\columnwidth]{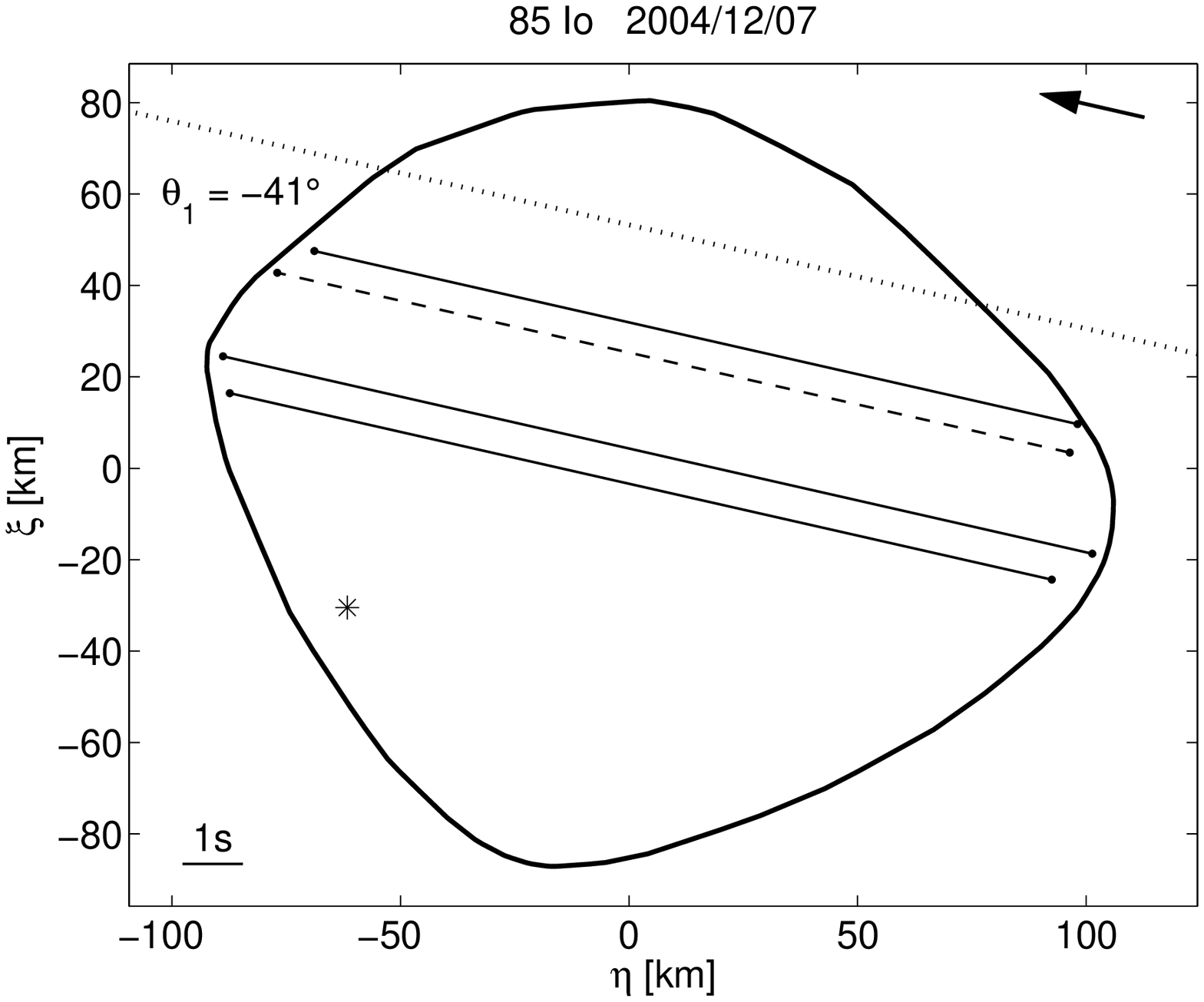}
\includegraphics[width=0.32\columnwidth]{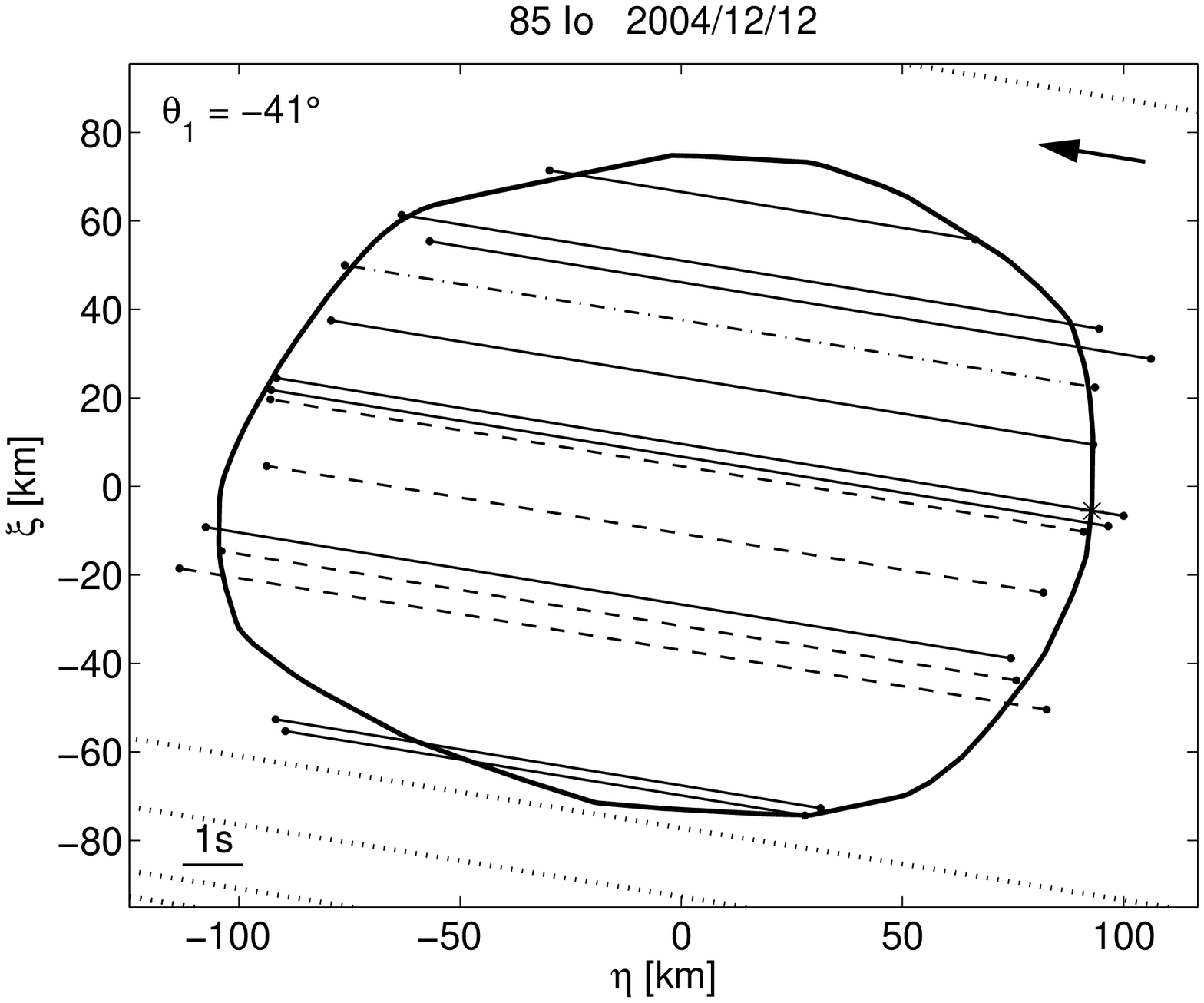}
\includegraphics[width=0.32\columnwidth]{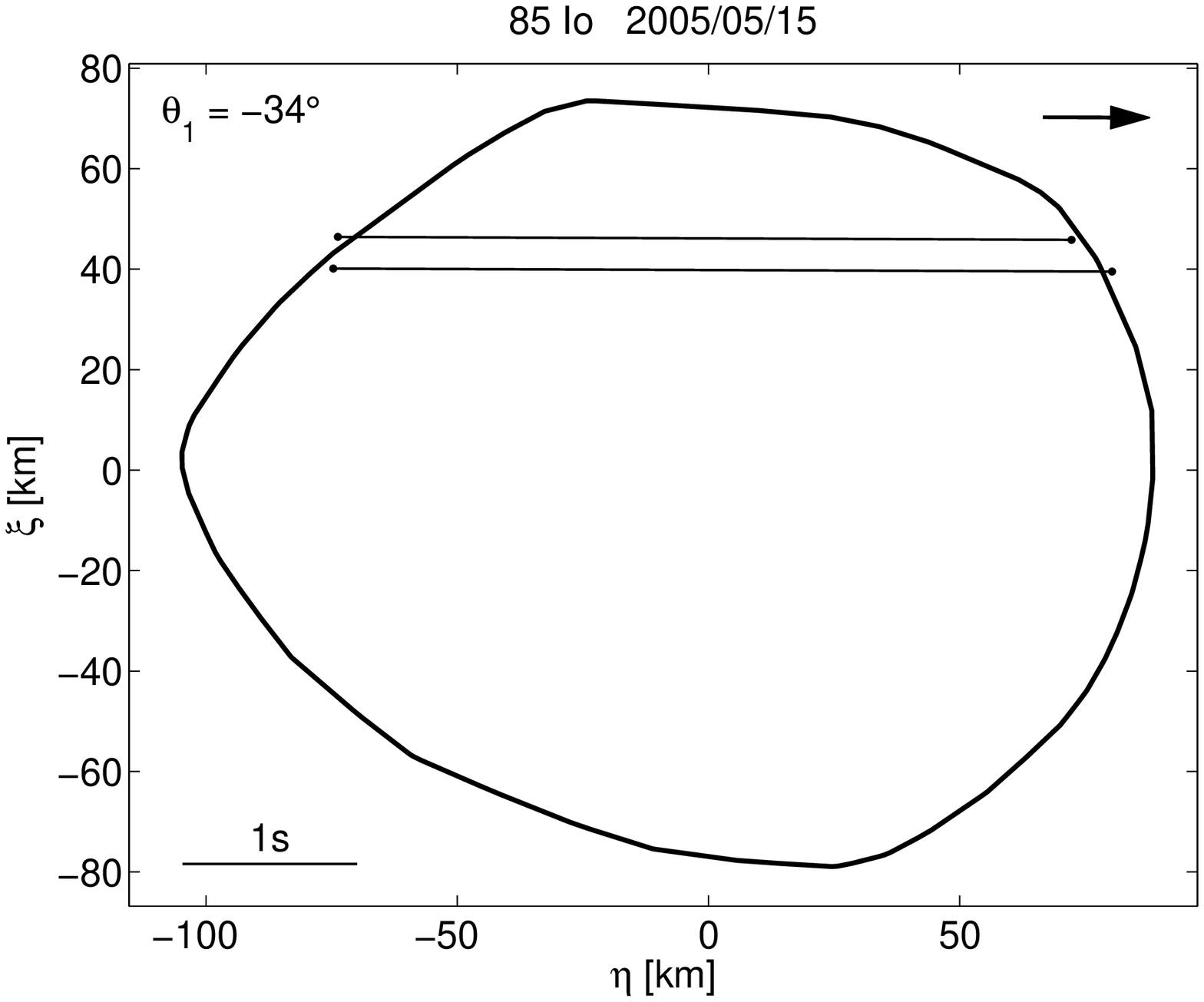}
\caption{(85) Io.}
\label{Io_fig}
\end{center}
\end{figure}

\begin{figure}
\begin{center}
\includegraphics[width=0.32\columnwidth]{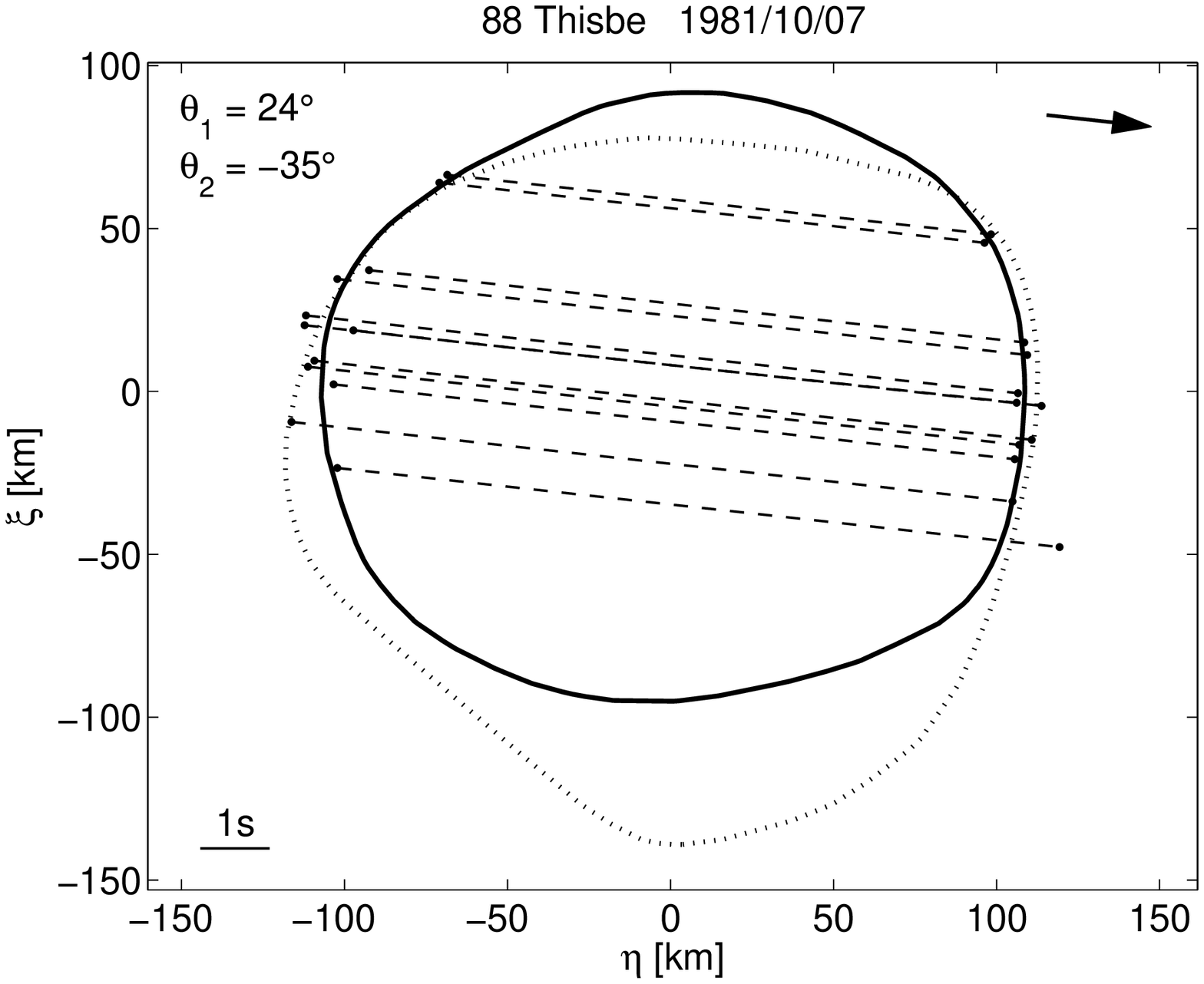}
\includegraphics[width=0.32\columnwidth]{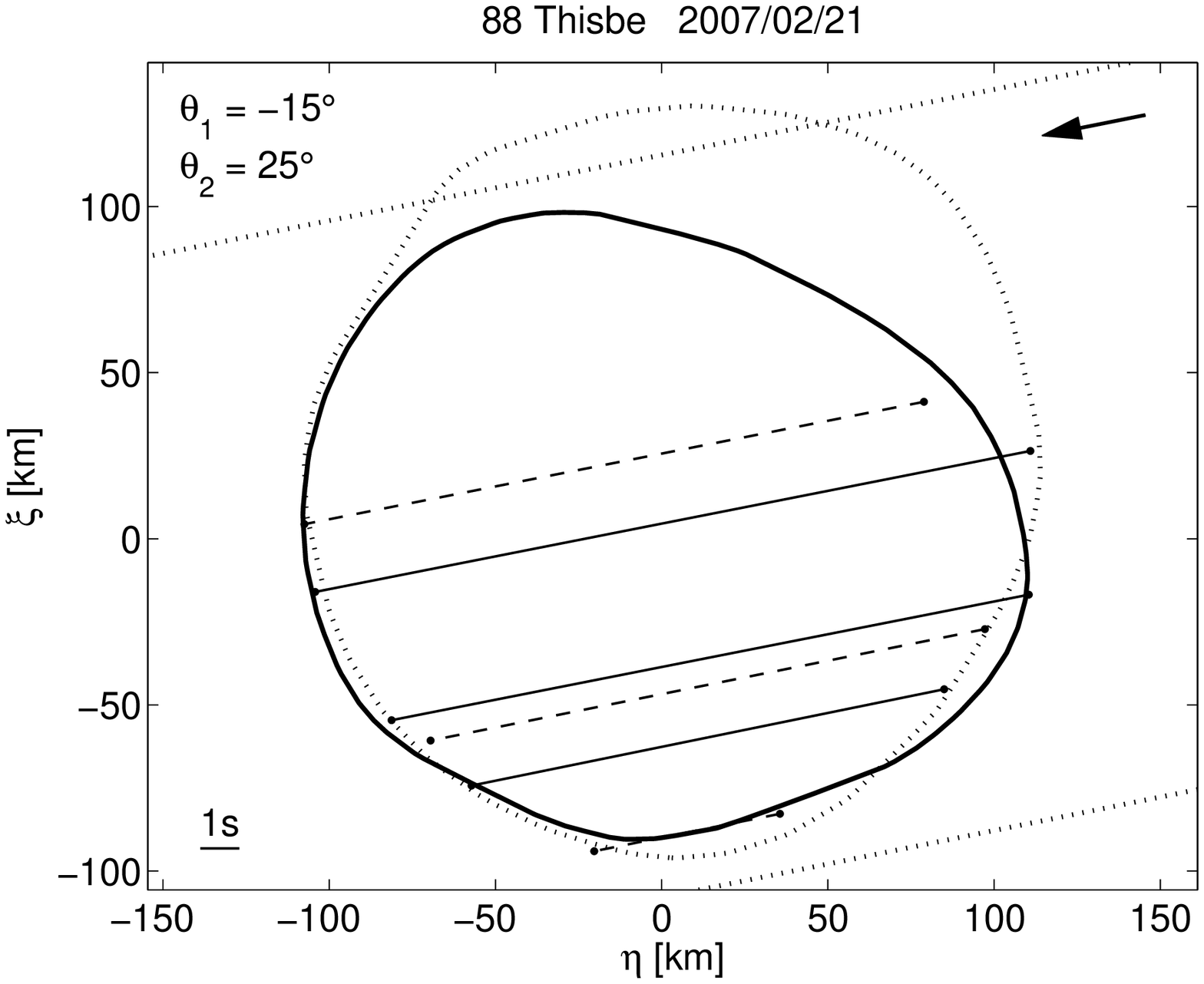}
\caption{(88) Thisbe. The solid profile corresponds to the pole $(72^\circ, 60^\circ)$, the dotted one to $(247^\circ, 50^\circ)$.}
\label{Thisbe_fig}
\end{center}
\end{figure}

\begin{figure}
\begin{center}
\includegraphics[width=0.32\columnwidth]{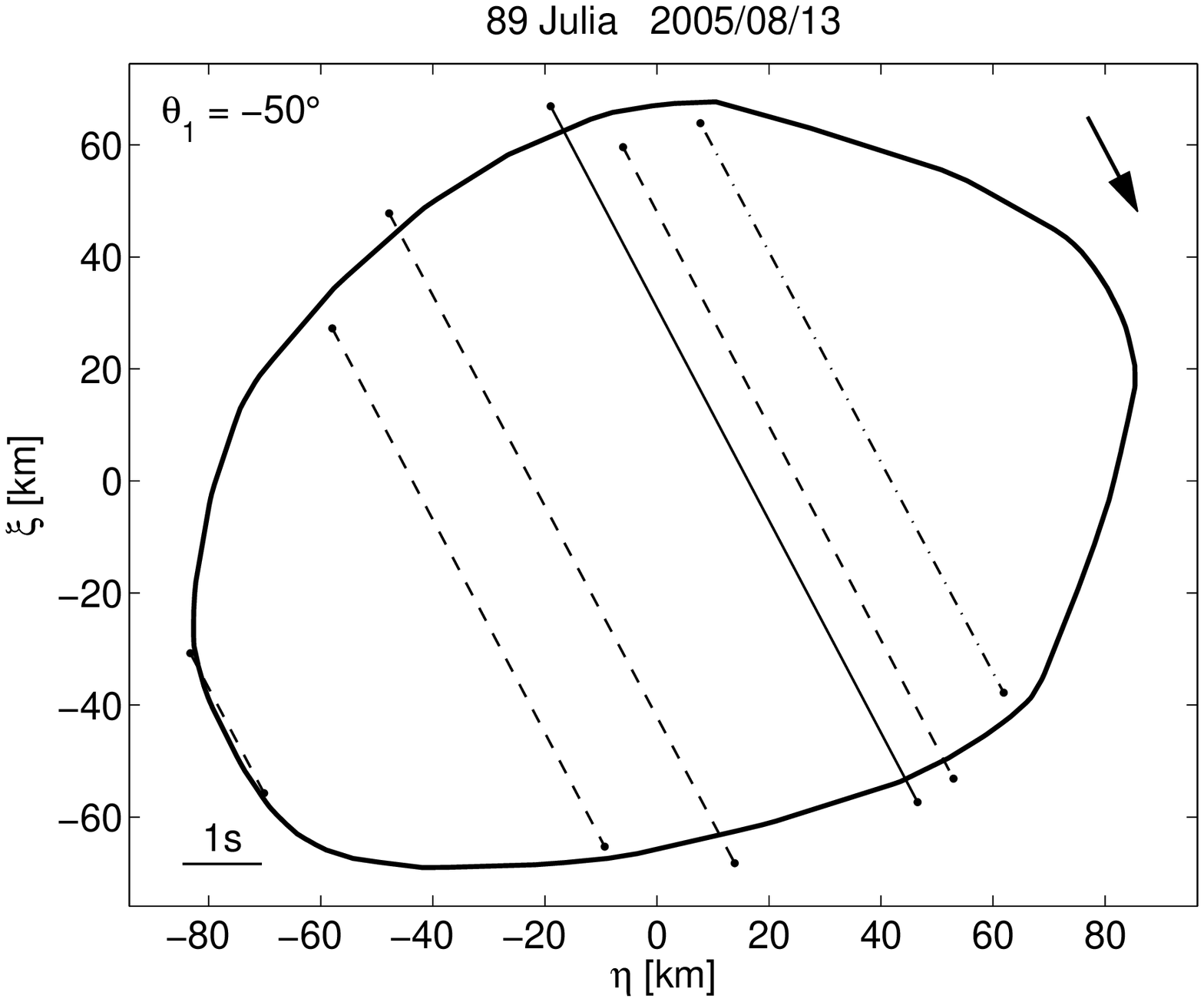}
\includegraphics[width=0.32\columnwidth]{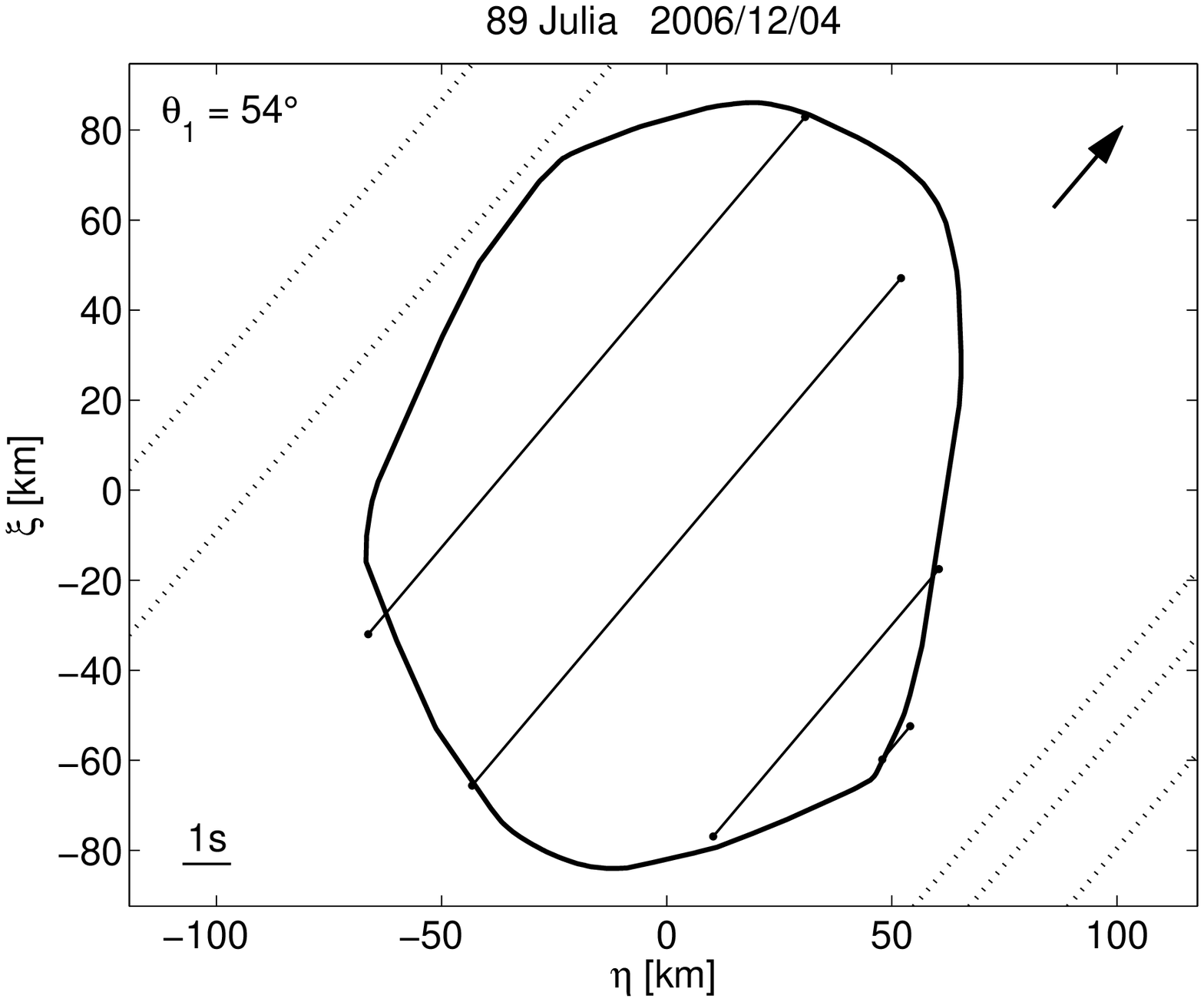}
\caption{(89) Julia}
\label{Julia_fig}
\end{center}
\end{figure}

\begin{figure}
\begin{center}
\includegraphics[width=0.32\columnwidth]{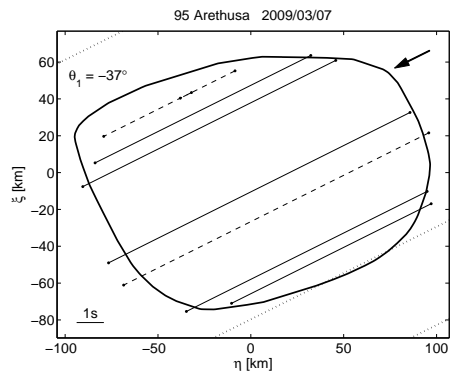}
\caption{(95) Arethusa.}
\label{Arethusa_fig}
\end{center}
\end{figure}

\begin{figure}
\begin{center}
\includegraphics[width=0.32\columnwidth]{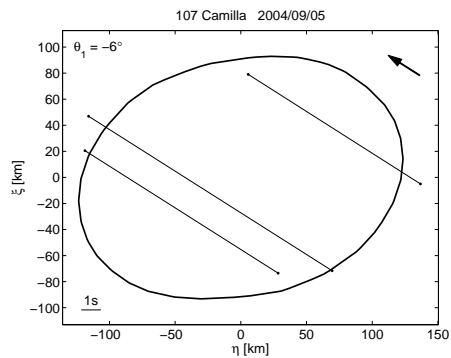}
\caption{(107) Camilla.}
\label{Camilla_fig}
\end{center}
\end{figure}

\begin{figure}
\begin{center}
\includegraphics[width=0.32\columnwidth]{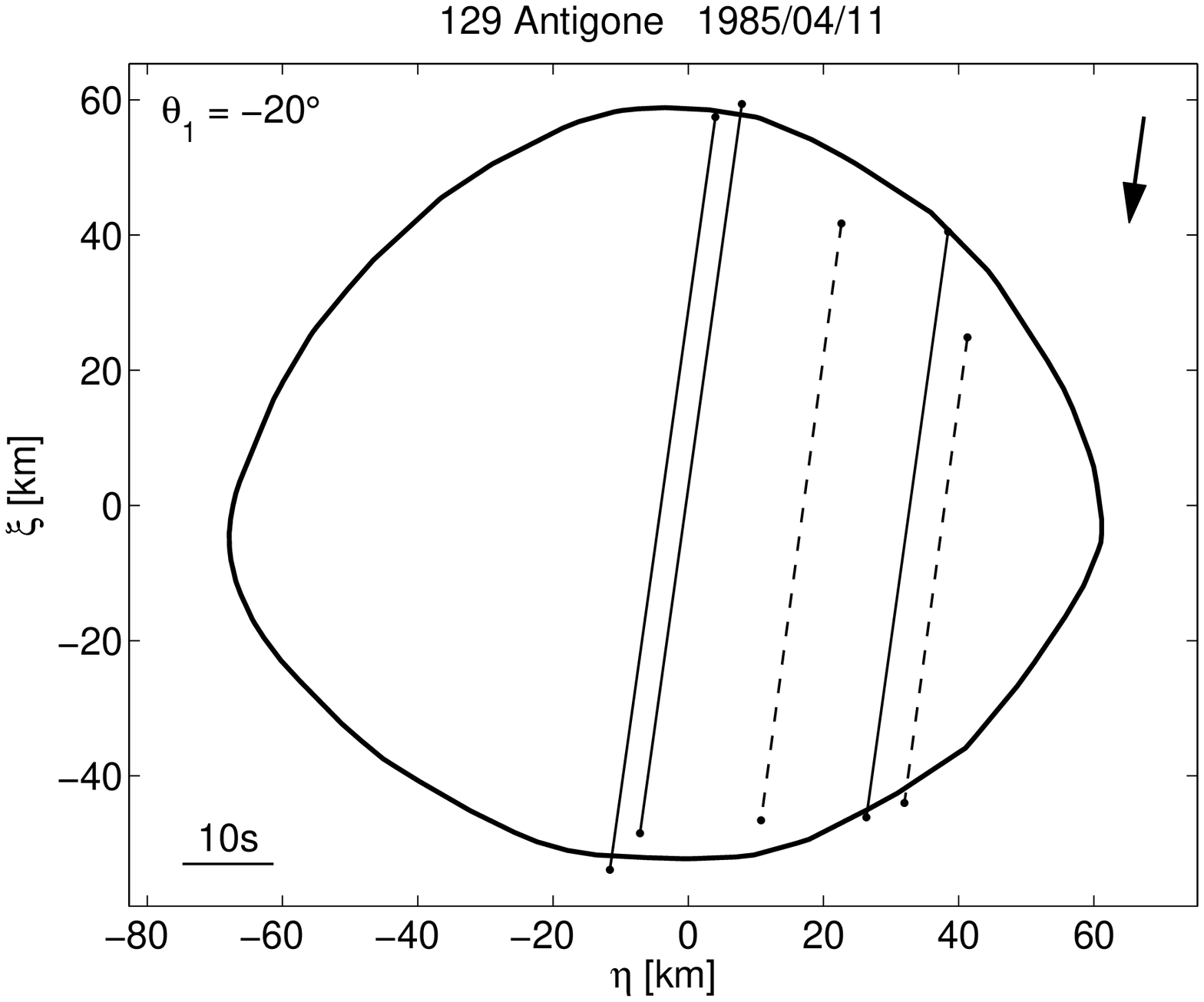}
\includegraphics[width=0.32\columnwidth]{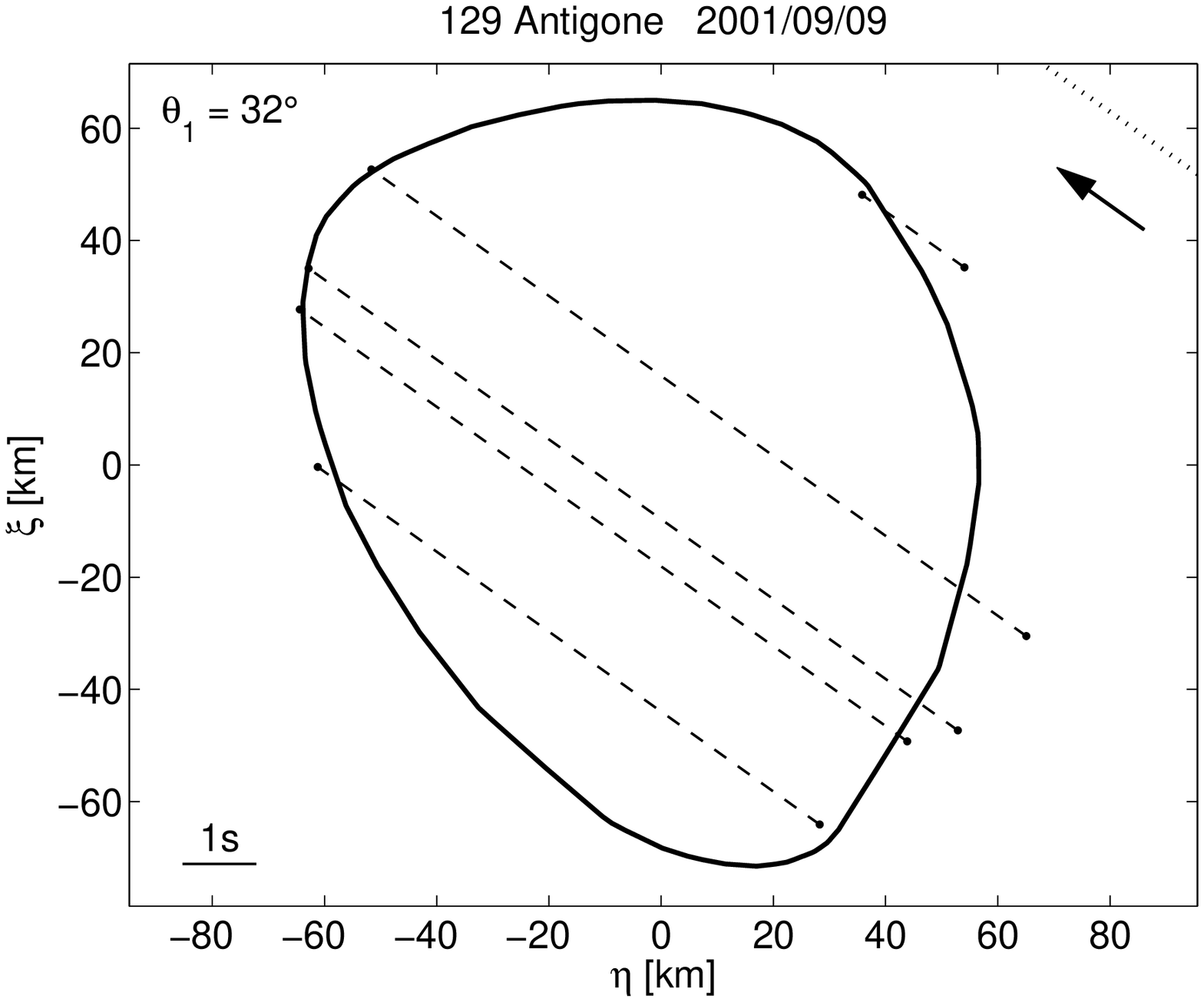}
\includegraphics[width=0.32\columnwidth]{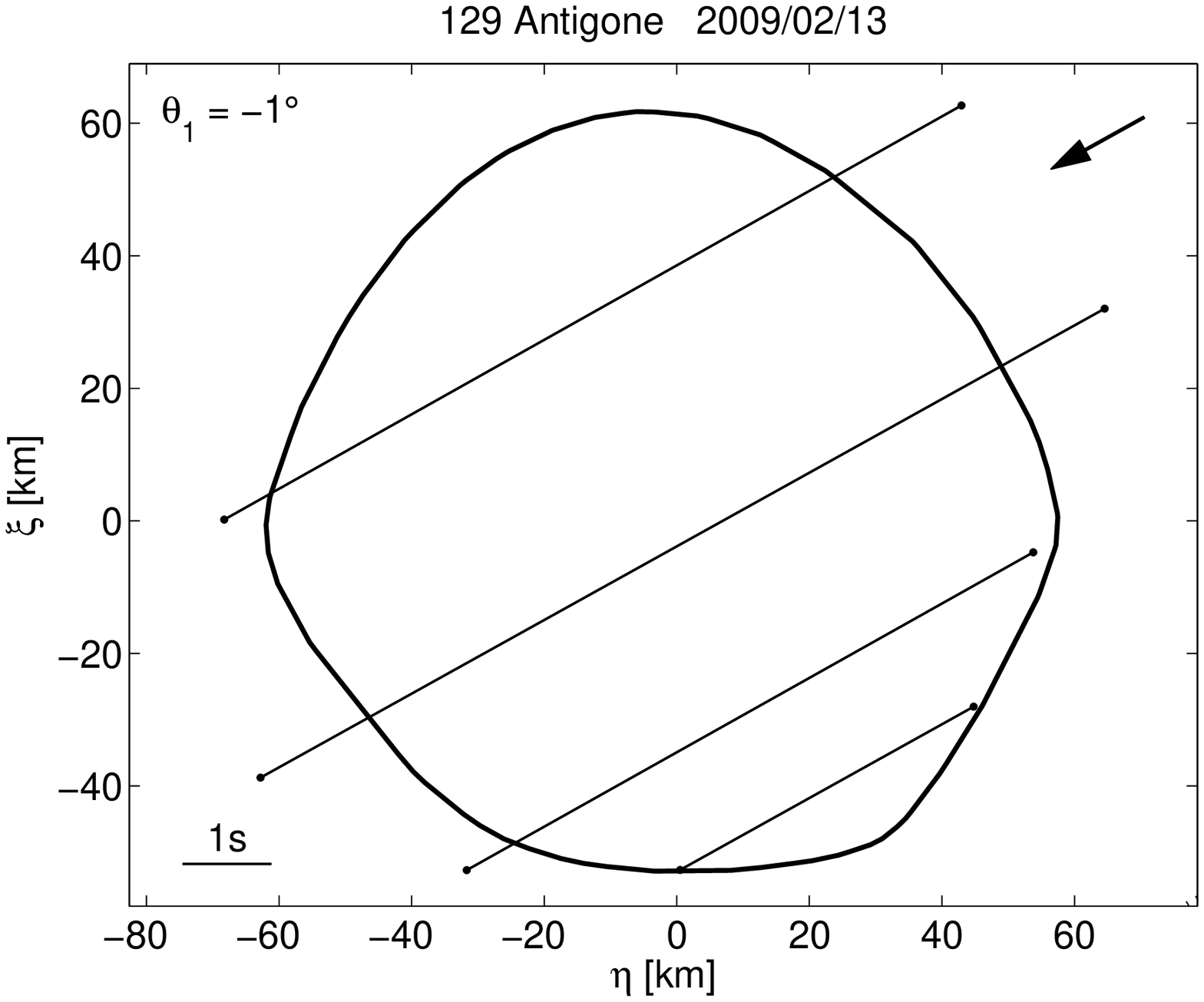}
\caption{(129) Antigone.}
\label{Antigone_fig}
\end{center}
\end{figure}

\begin{figure}
\begin{center}
\includegraphics[width=0.32\columnwidth]{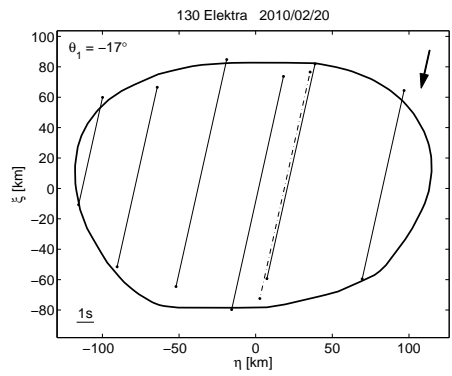}
\caption{(130) Elektra.}
\label{Elektra_fig}
\end{center}
\end{figure}

\begin{figure}
\begin{center}
\includegraphics[width=0.32\columnwidth]{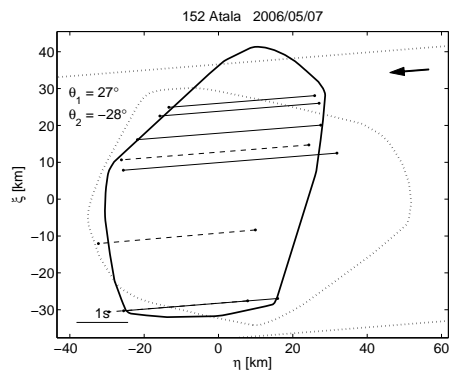}
\caption{(152) Atala. The solid profile corresponds to the pole $(347^\circ, 47^\circ)$, the dotted one to $(199^\circ, 62^\circ)$.}
\label{Atala_fig}
\end{center}
\end{figure}

\begin{figure}
\begin{center}
\includegraphics[width=0.32\columnwidth]{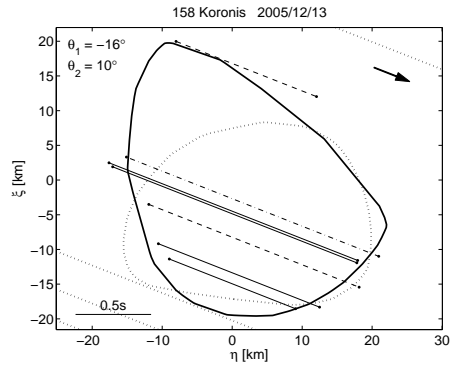}
\caption{(158) Koronis. The solid profile corresponds to the pole $(30^\circ, -64^\circ)$, the dotted one to $(225^\circ, -70^\circ)$.}
\label{Koronis_fig}
\end{center}
\end{figure}

\begin{figure}
\begin{center}
\includegraphics[width=0.32\columnwidth]{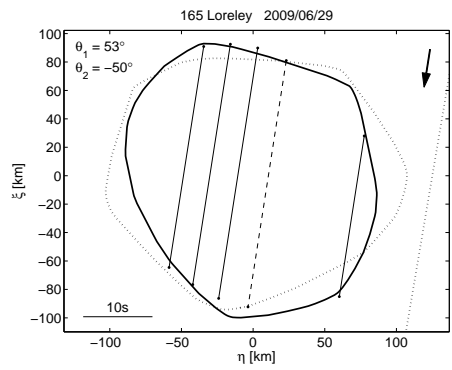}
\caption{(165) Loreley. The solid profile corresponds to the pole $(174^\circ, 29^\circ)$, the dotted one to $(348^\circ, 42^\circ)$.}
\label{Loreley_fig}
\end{center}
\end{figure}

\begin{figure}
\begin{center}
\includegraphics[width=0.32\columnwidth]{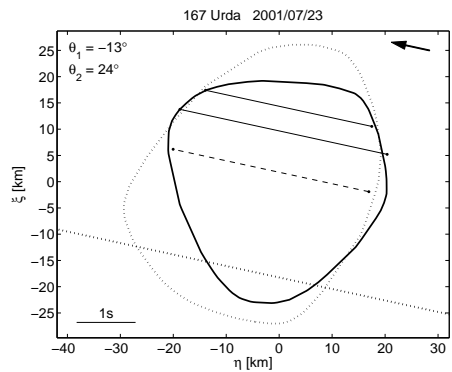}
\caption{(167) Urda. The solid profile corresponds to the pole $(249^\circ, -68^\circ)$, the dotted one to $(107^\circ, -69^\circ)$.}
\label{Urda_fig}
\end{center}
\end{figure}

\begin{figure}
\begin{center}
\includegraphics[width=0.32\columnwidth]{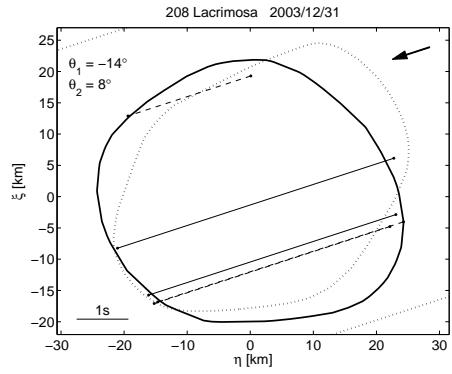}
\caption{(208) Lacrimosa. The solid profile corresponds to the pole $(176^\circ, -68^\circ)$, the dotted one to $(20^\circ, -75^\circ)$.}
\label{Lacrimosa_fig}
\end{center}
\end{figure}

\begin{figure}
\begin{center}
\includegraphics[width=0.32\columnwidth]{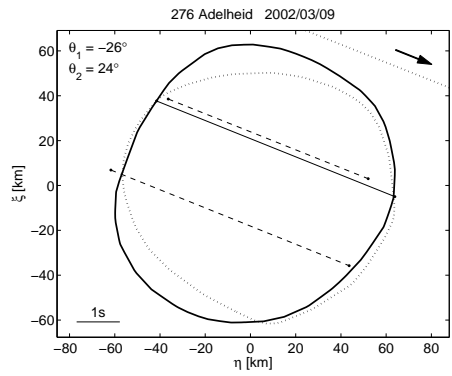}
\caption{(278) Adelheid. The solid profile corresponds to the pole $(9^\circ, -4^\circ)$, the dotted one to $(199^\circ, -20^\circ)$.}
\label{Adelheid_fig}
\end{center}
\end{figure}

\begin{figure}
\begin{center}
\includegraphics[width=0.32\columnwidth]{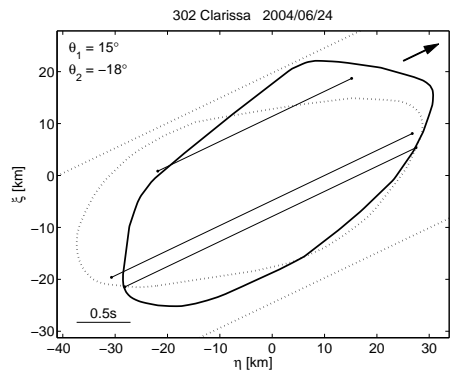}
\caption{(302) Clarissa. The solid profile corresponds to the pole $(28^\circ, -72^\circ)$, the dotted one to $(190^\circ, -72^\circ)$.}
\label{Clarissa_fig}
\end{center}
\end{figure}

\begin{figure}
\begin{center}
\includegraphics[width=0.32\columnwidth]{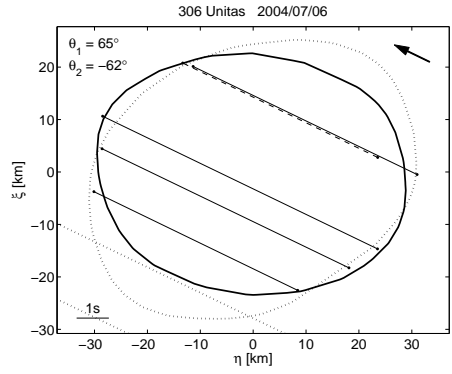}
\caption{(306) Unitas. The solid profile corresponds to the pole $(79^\circ, -35^\circ)$, the dotted one to $(253^\circ, -17^\circ)$.}
\label{Unitas_fig}
\end{center}
\end{figure}

\begin{figure}
\begin{center}
\includegraphics[width=0.32\columnwidth]{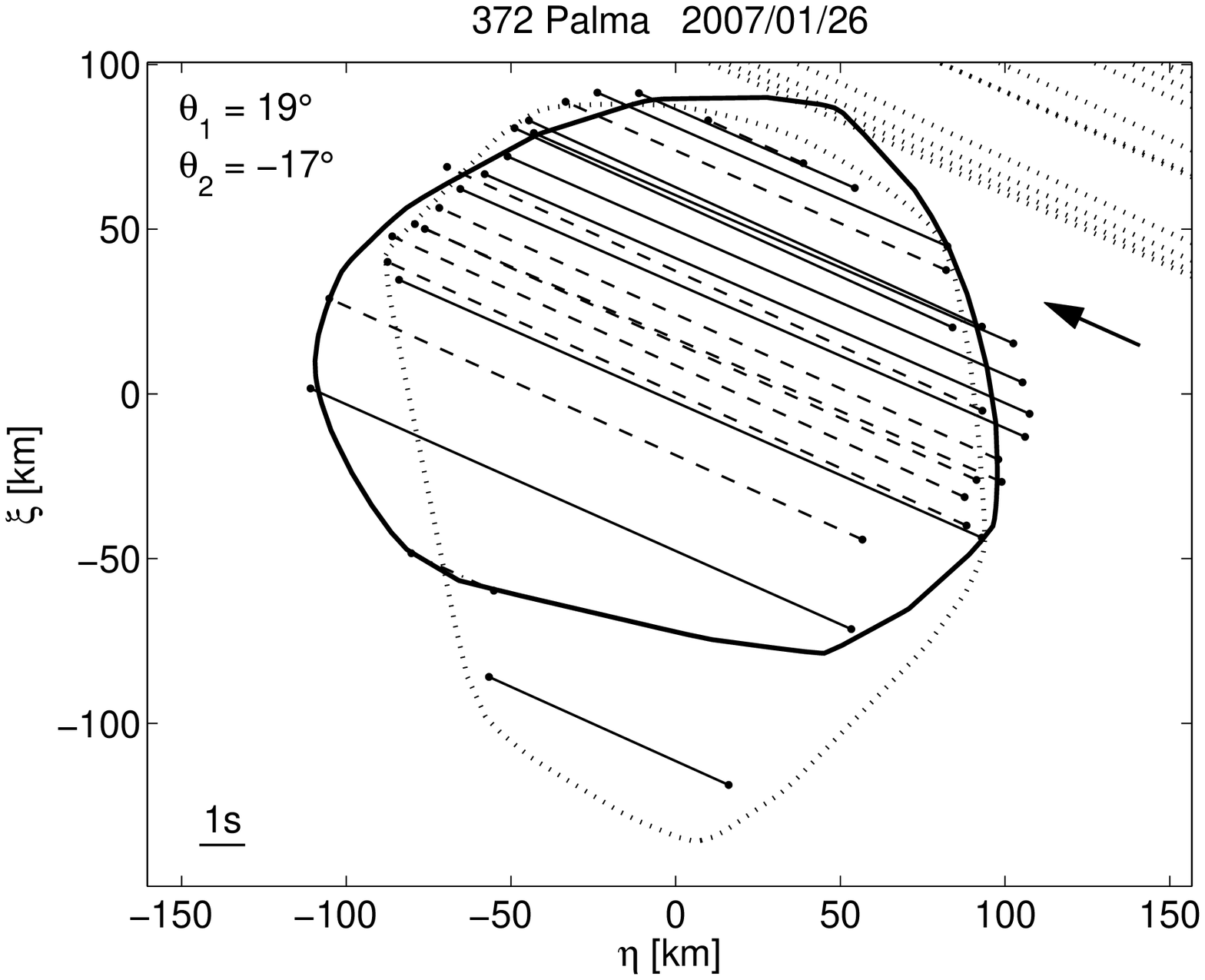}
\includegraphics[width=0.32\columnwidth]{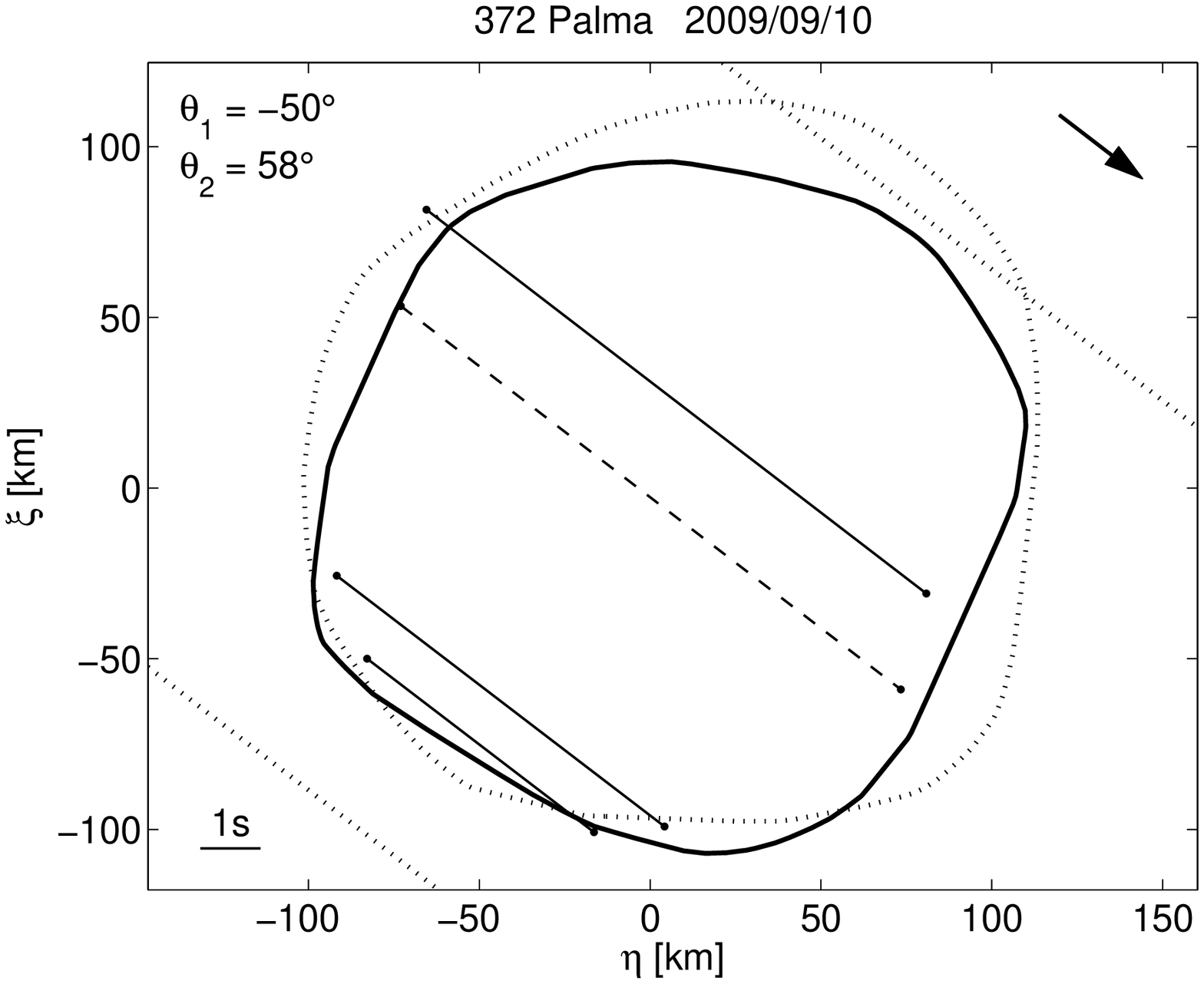}
\caption{(372) Palma. The solid profile corresponds to the pole $(221^\circ, -47^\circ)$, the dotted one to $(44^\circ, 17^\circ)$.}
\label{Palma_fig}
\end{center}
\end{figure}

\begin{figure}
\begin{center}
\includegraphics[width=0.32\columnwidth]{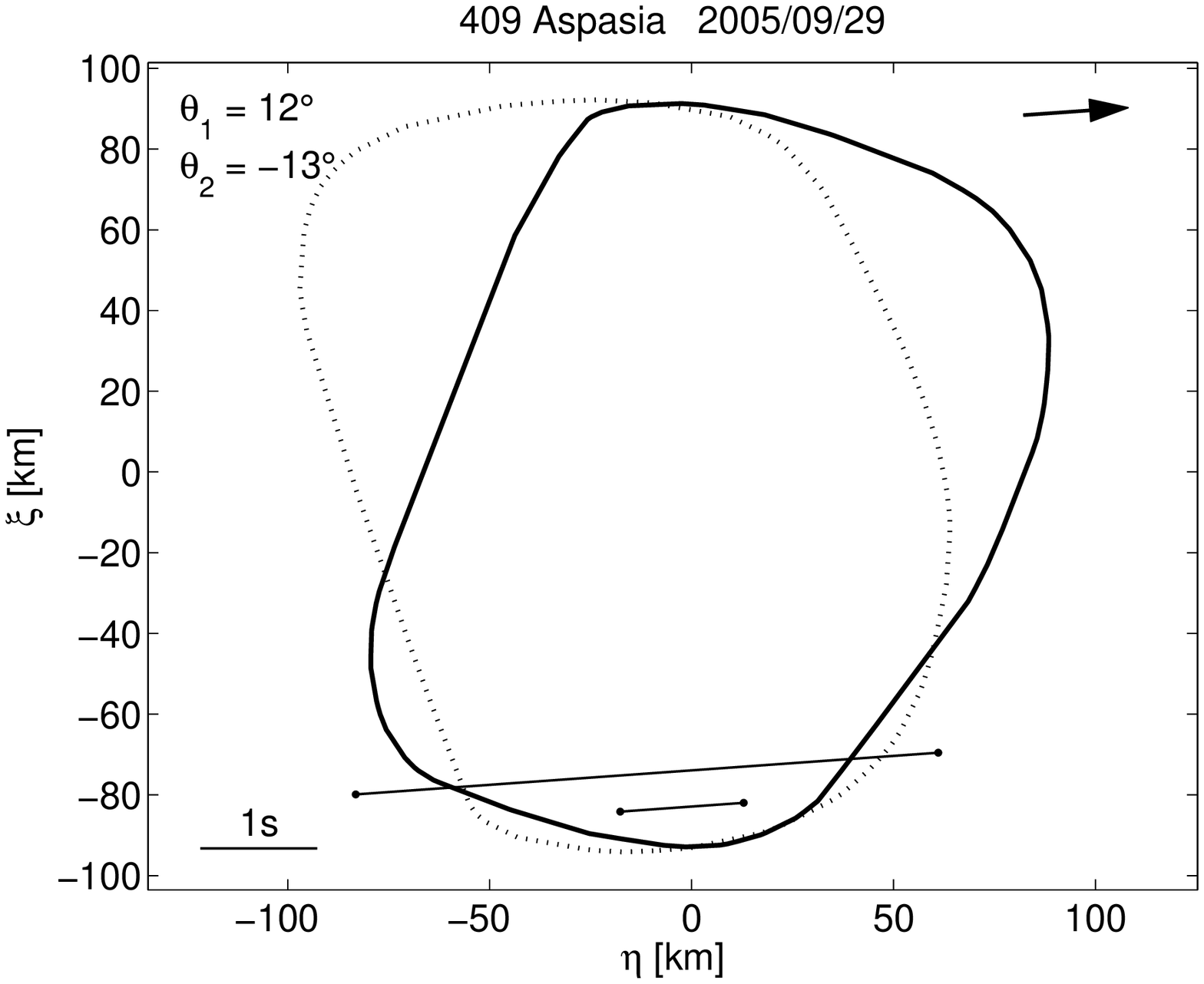}
\includegraphics[width=0.32\columnwidth]{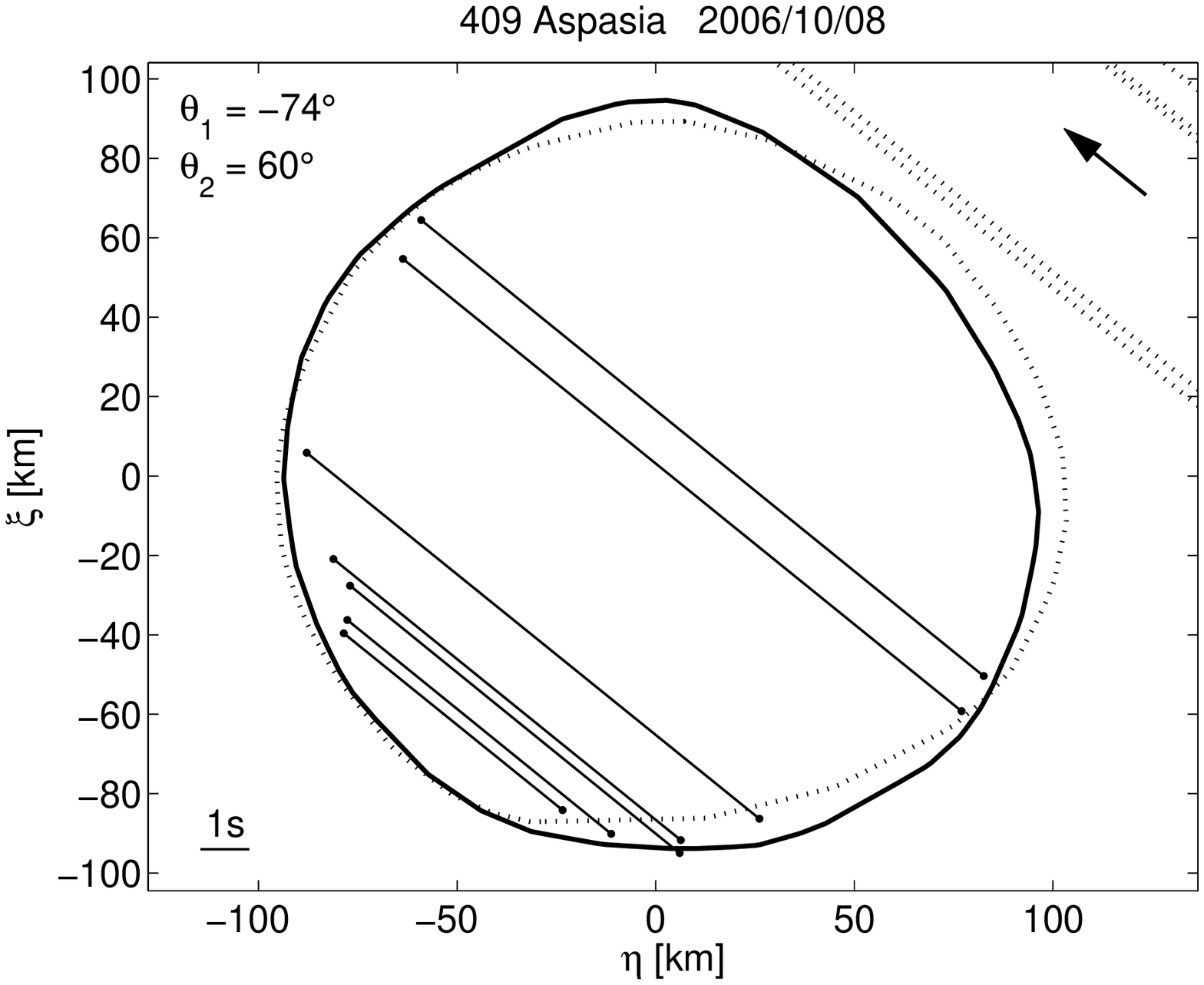}
\includegraphics[width=0.32\columnwidth]{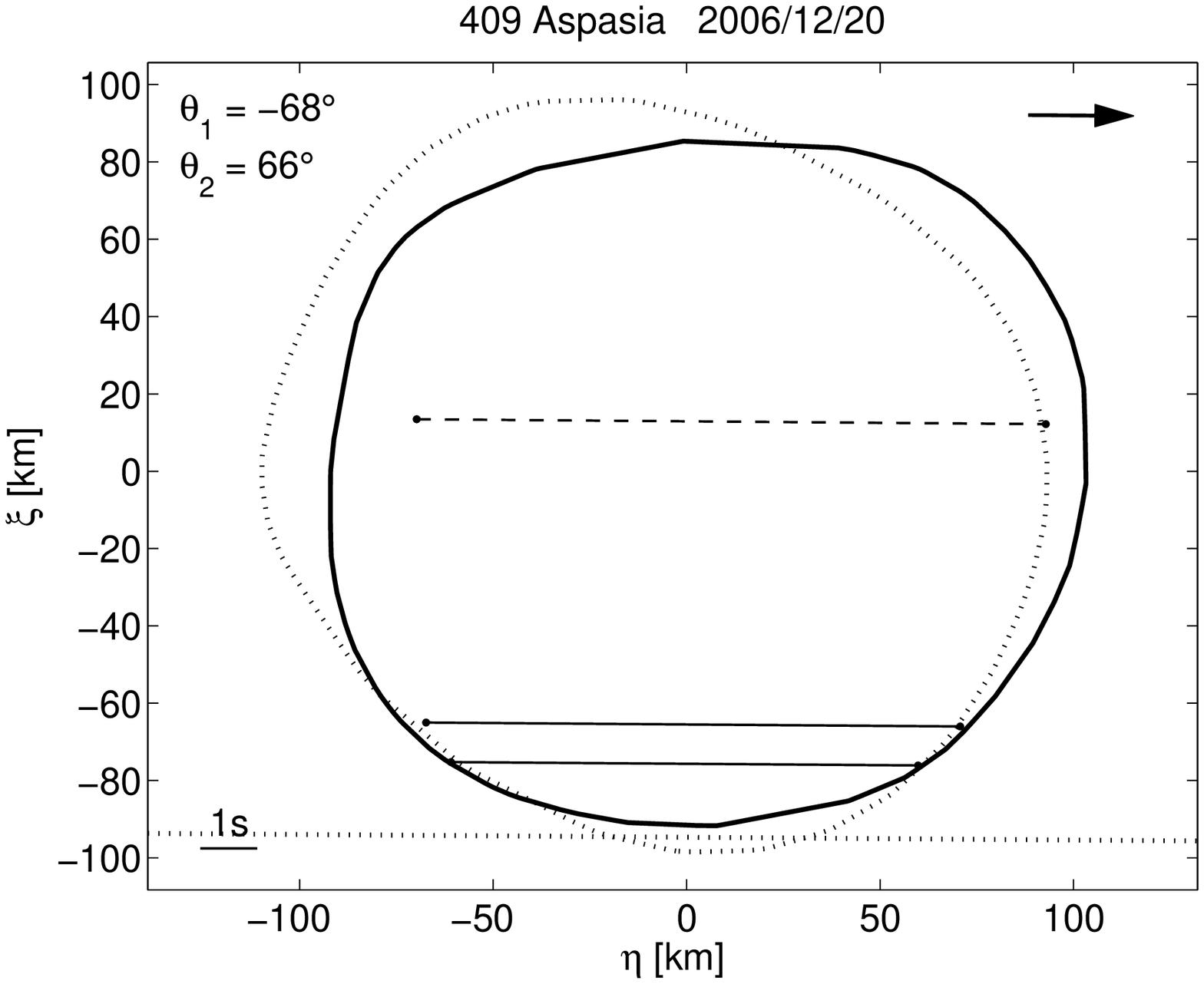}
\includegraphics[width=0.32\columnwidth]{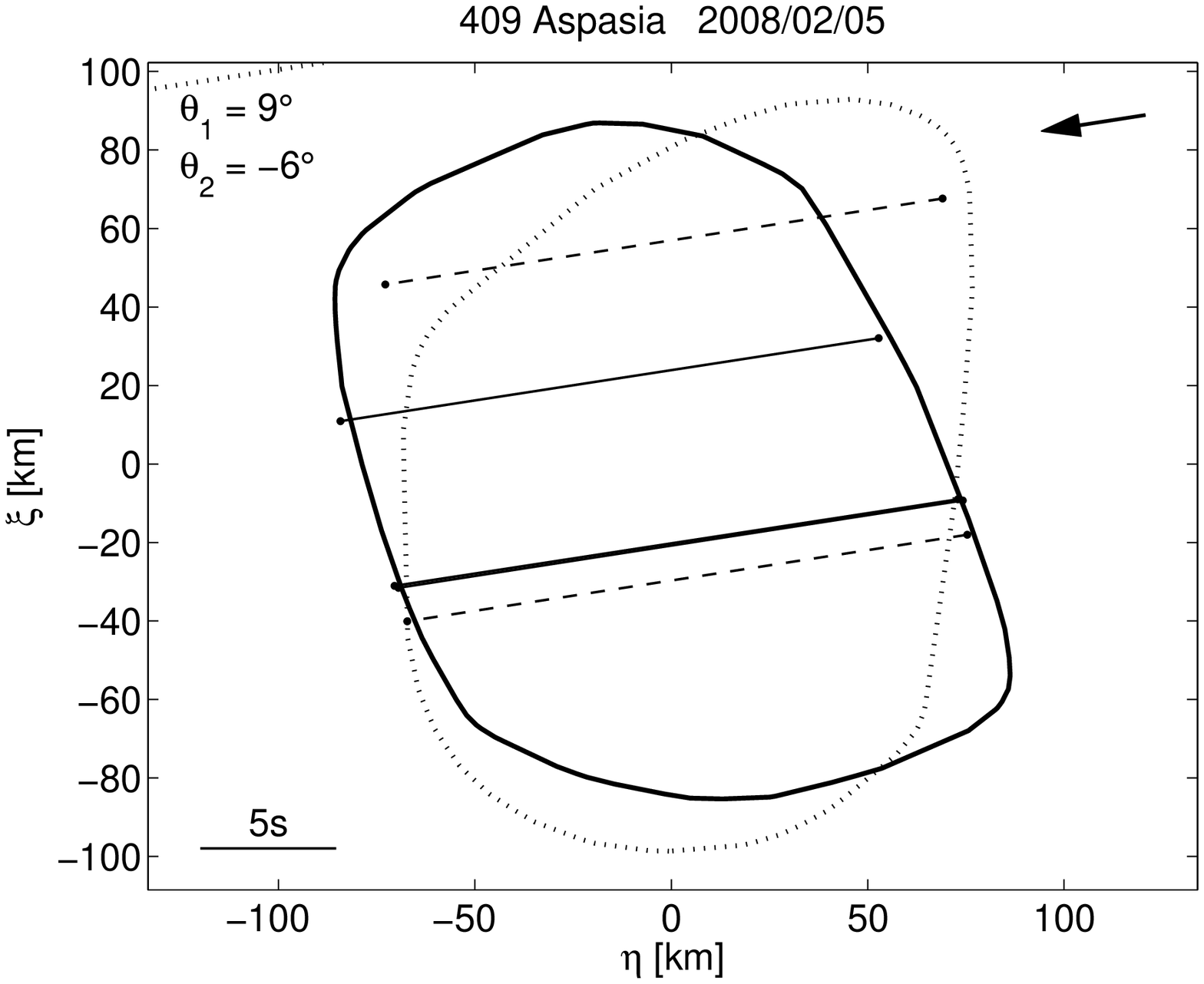}
\includegraphics[width=0.32\columnwidth]{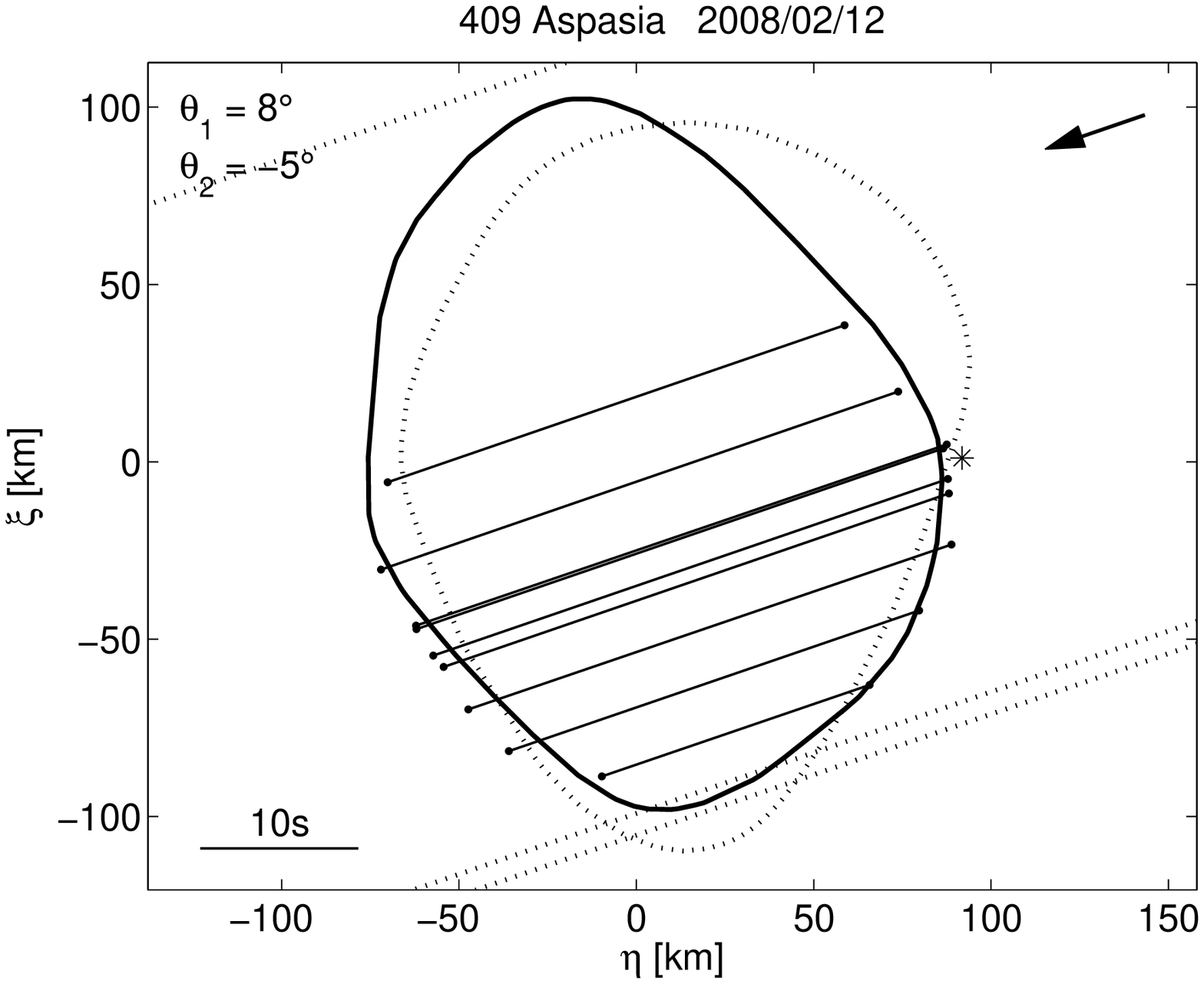}
\includegraphics[width=0.32\columnwidth]{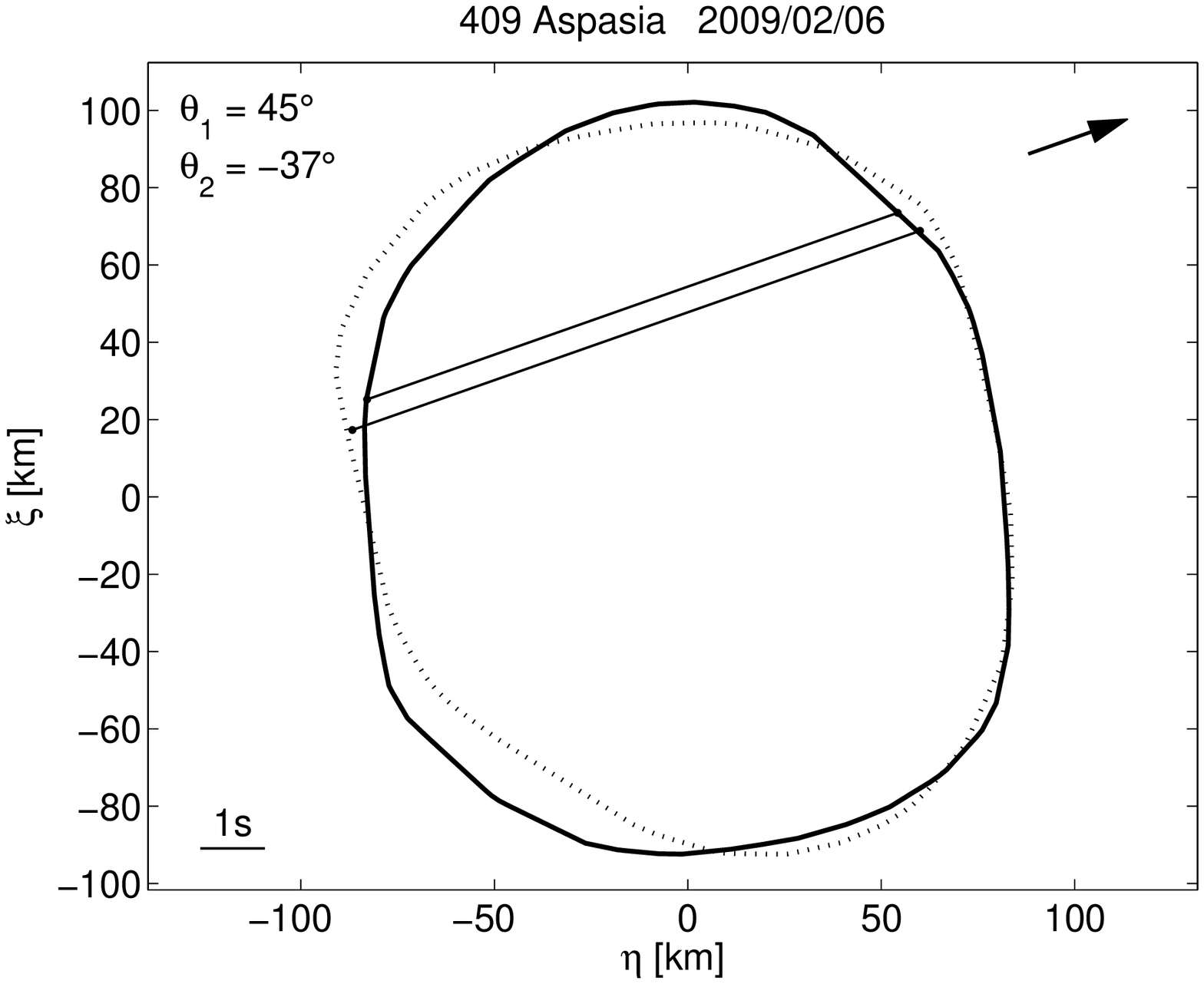}
\caption{(409) Aspasia. The solid profile corresponds to the pole $(3^\circ, 30^\circ)$, the dotted one to $(177^\circ, 15^\circ)$.}
\label{Aspasia_fig}
\end{center}
\end{figure}

\begin{figure}
\begin{center}
\includegraphics[width=0.32\columnwidth]{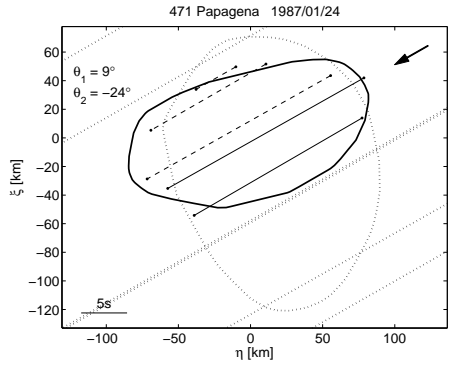}
\caption{(471) Papagena. The solid profile corresponds to the pole $(223^\circ, 67^\circ)$, the dotted one to $(22^\circ, 18^\circ)$.}
\label{Papagena_fig}
\end{center}
\end{figure}

\begin{figure}
\begin{center}
\includegraphics[width=0.32\columnwidth]{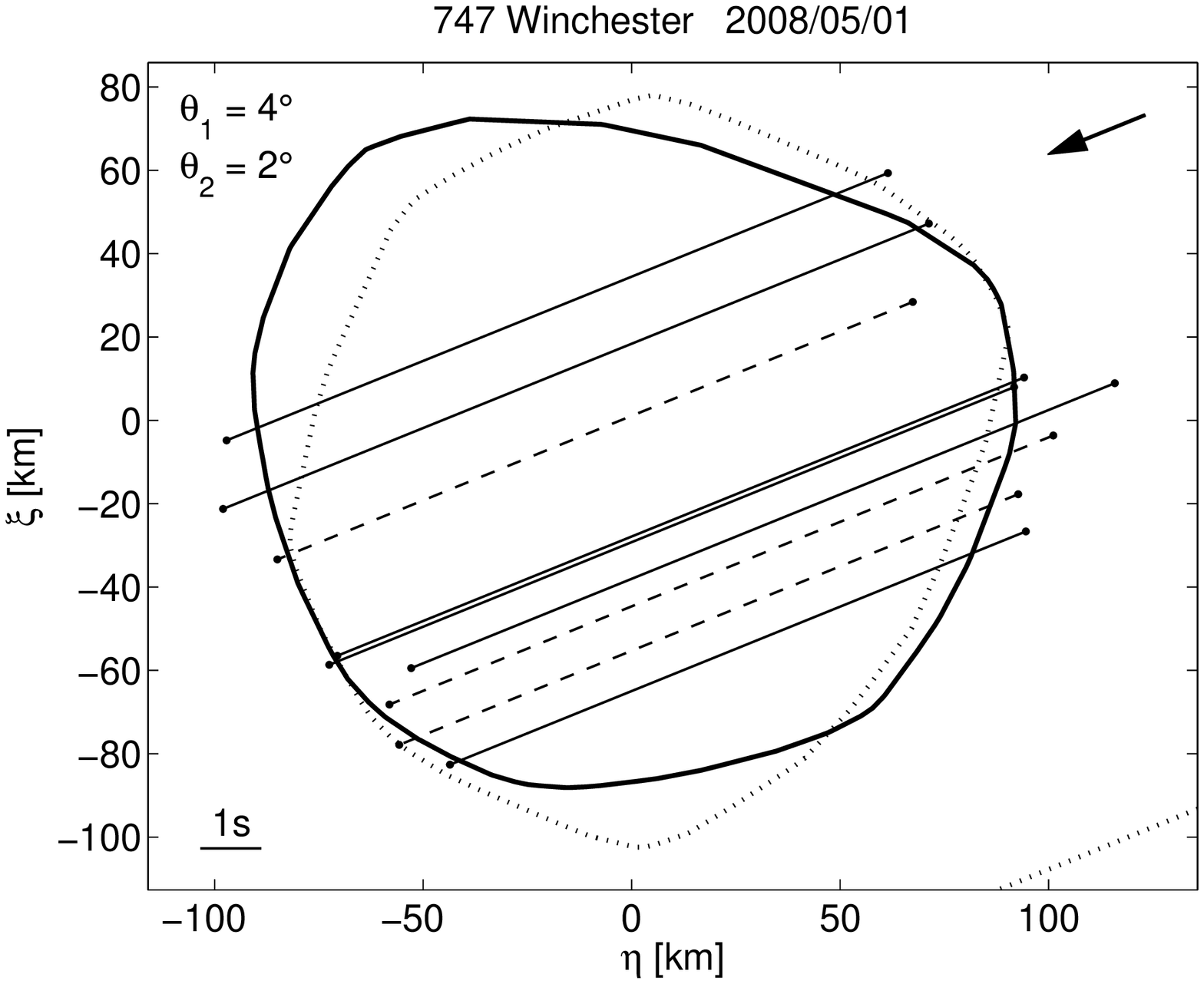}
\includegraphics[width=0.32\columnwidth]{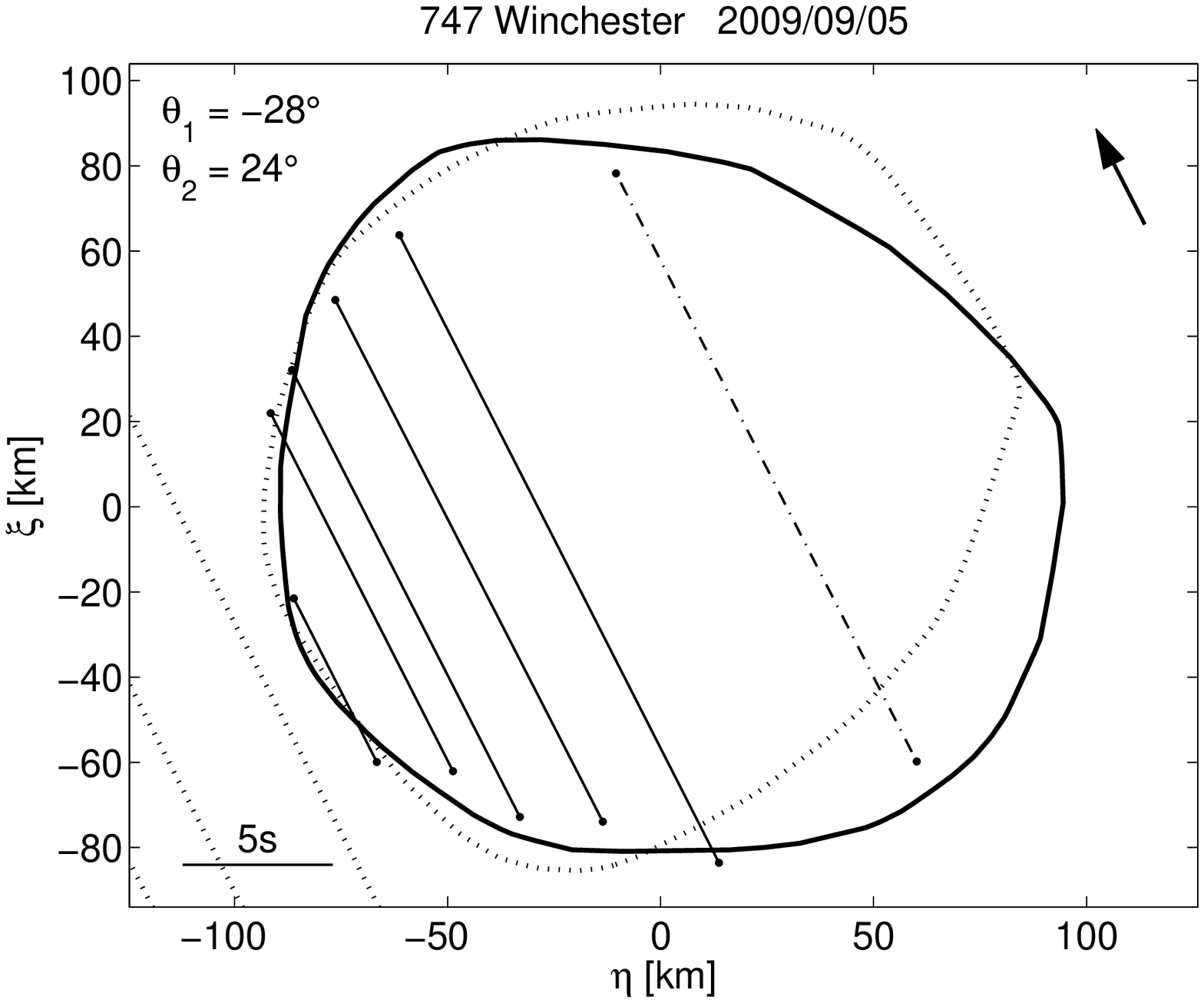}
\caption{(747) Winchester. The solid profile corresponds to the pole $(304^\circ, -60^\circ)$, the dotted one to $(172^\circ, -36^\circ)$.}
\label{Winchester_fig}
\end{center}
\end{figure}

\begin{figure}
\begin{center}
\includegraphics[width=0.32\columnwidth]{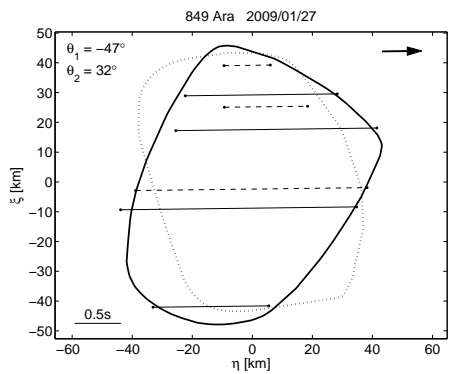}
\caption{(849) Ara. The solid profile corresponds to the pole $(223^\circ, -40^\circ)$, the dotted one to $(10^\circ, -25^\circ)$.}
\label{Ara_fig}
\end{center}
\end{figure}

\begin{figure}
\begin{center}
\includegraphics[width=0.32\columnwidth]{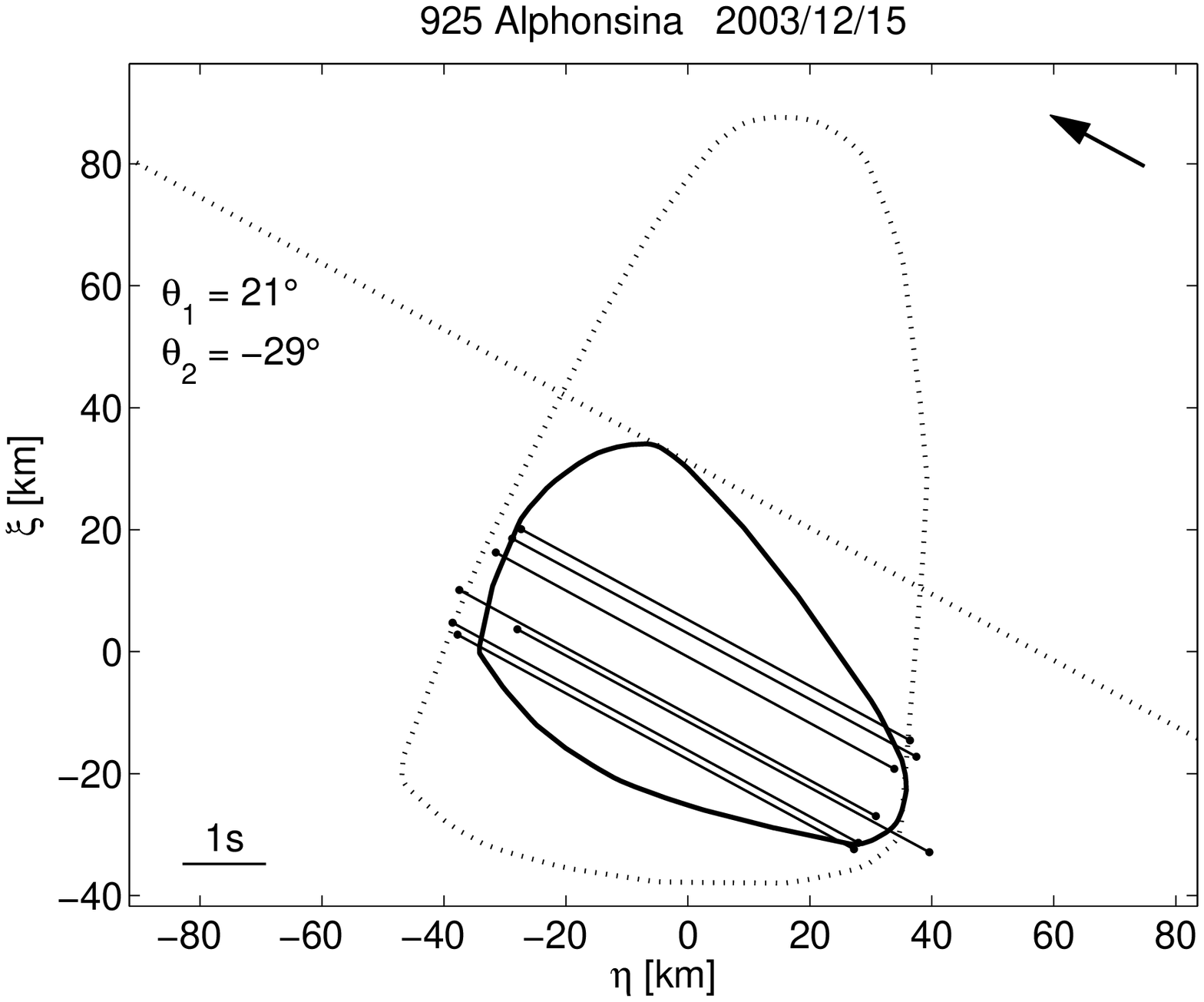}
\includegraphics[width=0.32\columnwidth]{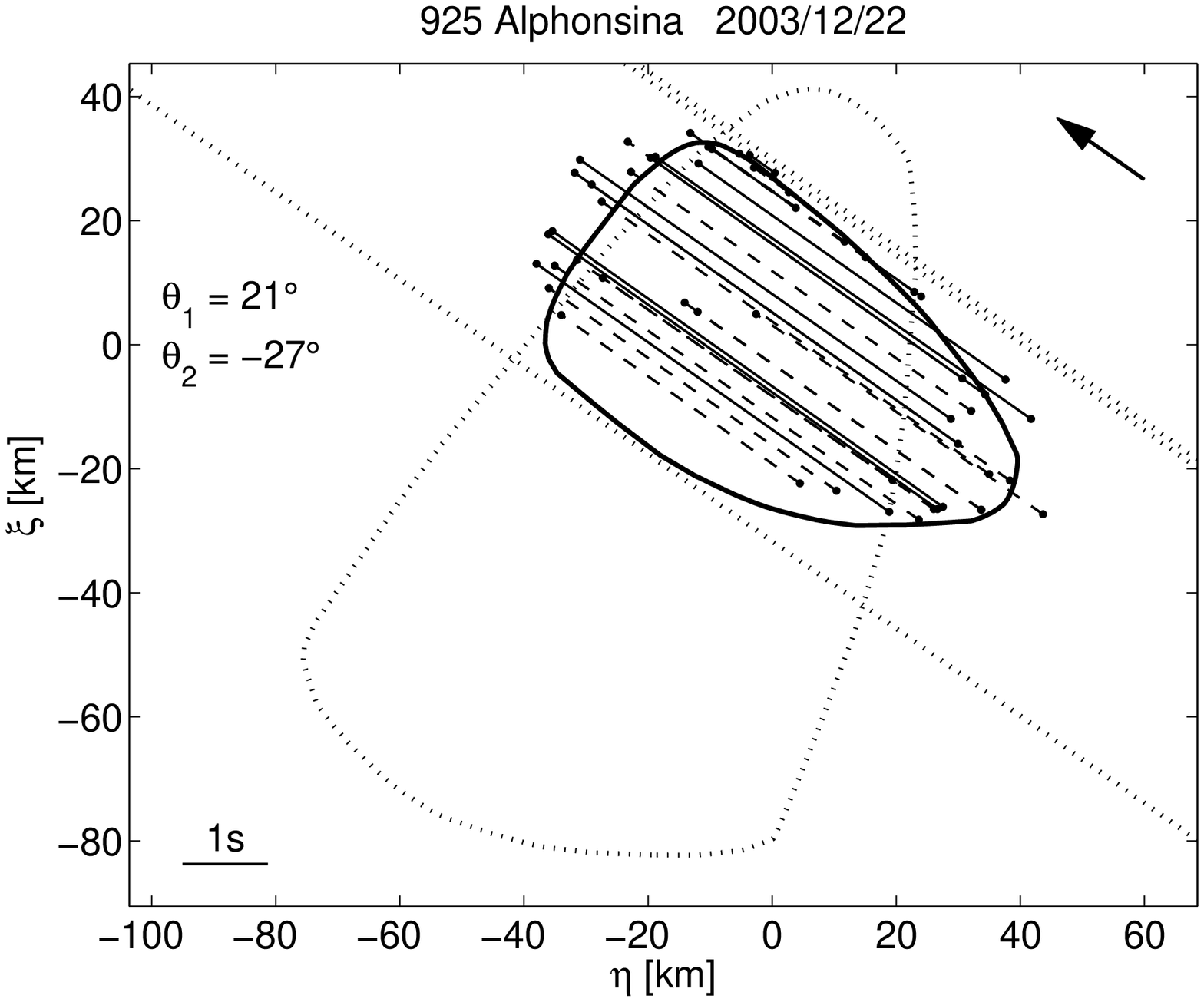}
\caption{(925) Alphonsina. The solid profile corresponds to the pole $(294^\circ, 41^\circ)$, the dotted one to $(148^\circ, 25^\circ)$.}
\label{Alphonsina_fig}
\end{center}
\end{figure}

\begin{figure}
\begin{center}
\includegraphics[width=0.32\columnwidth]{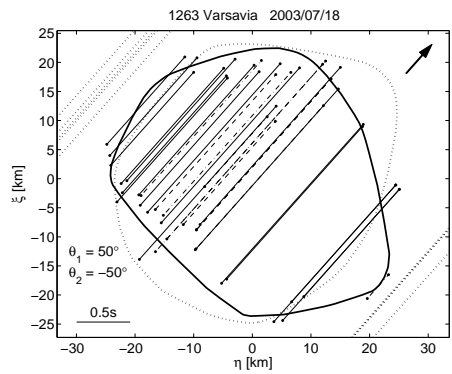}
\caption{(1263) Varsavia. The solid profile corresponds to the pole $(341^\circ, -14^\circ)$, the dotted one to $(172^\circ, -1^\circ)$.}
\label{Varsavia_fig}
\end{center}
\end{figure}

In Figs.~\ref{Pallas_fig}--\ref{Varsavia_fig}, we plot the projected asteroid silhouettes and the chords used for scaling the model. 
The meaning of different types of lines and curves is explained in Fig.~\ref{fig_example}. 
If there are two possible pole solutions, one of the corresponding profiles (usually the one that gives a worse fit) is plotted with
a dotted curve. Chords observed visually are plotted as dashed lines, others (observed by means of CCD or other photodetector) are plotted 
as solid lines. Dotted lines represent negative observations and dot-dashed lines represent observations that report only the duration 
of the event, not absolute timings. Such chords can be arbitrarily shifted in time along the relative velocity vector. If an observer
reported only the beginning or the end of the occultation, we plot it as an asterisk. Each plot contains also the time scale and the latitude of 
the sub-Earth point $\theta$ for the time of occultation and corresponding model. If there are two models in one figure, the visual chords 
are plotted with shifts (Eq.~\ref{eq_shift}) corresponding to the solid-contour model.

All shape models, spin parameters, and lightcurve data we present are available online in DAMIT.

\paragraph*{(2) Pallas}
The occultation of SAO~85009 by asteroid Pallas in 29 May 1978 was one of the first successfully observed occultations with 
photoelectrically measured timings and a very good
coverage of the whole projected silhouette. The next event in 29 May 1983 is the most densely covered occultation ever observed. 
More than one hundred chords were measured and the size of Pallas was accurately determined \citep{Dun.ea:90}. 
The other three events from 1985, 2001, and 2006 provide additional projection geometry.

Two models of Pallas were derived by \cite{Tor.ea:03},  with the preference of the pole $(35^\circ, -12^\circ)$. 
Because the shape of Pallas is rather spherical and the occultation did not 
reveal any distinct features, both models fit the chords almost equally well. However, only one pole solution
is consistent with the adaptive optics images by \cite{Car.ea:10}. They derived a nonconvex model of Pallas with the pole direction 
$(30^\circ, -16^\circ) \pm 5^\circ$ and the equivalent diameter $512 \pm 6$\,km, which is close to the diameter  $539 \pm 28$\,km 
derived from occultations (Fig.~\ref{Pallas_fig}). \cite{Sch.ea:09} derived an ellipsoidal model of Pallas from HST observations. 
The model has the equivalent diameter of $545 \pm 18\,$km, which is also consistent with our value. 

\paragraph*{(3) Juno}
The 11 December 1979 occultation event provided excellent photoelectric measurements
of 15 chords distributed over the whole projection. 
The second observed occultation on 24 May 2000 provided only four visual chords and
three of them were placed close together. 
The shape model of Juno derived by \cite{Kaa.ea:02} fits both occultations well
(Fig.~\ref{Juno_fig}).

\paragraph*{(5) Astraea}
A recent occultation by Astraea was observed in 6 June 2008 with the timings determined mainly visually. 
From the two shape models that were derived by \cite{Dur.ea:09}, we can reject the pole $(310^\circ, 44^\circ)$, 
because the shape model with the pole direction $(126^\circ, 40^\circ)$ clearly gives a better fit (Fig.~\ref{Astraea_fig}). 

\paragraph*{(6) Hebe}
The shape model derived by \cite{Tor.ea:03} agrees with the occultation data. However, scaling of the model is not very accurate, because
the occultation duration was only a few seconds and all observations were visual (Fig.~\ref{Hebe_fig}).

\paragraph*{(7) Iris}
The shape model derived by \cite{Kaa.ea:02} as well as its mirror solution can be scaled to fit the two 
occultations. The second
occultation was short and there were large errors in the visually determined timings (Fig.~\ref{Iris_fig}). \cite{Ost.ea:10} derived from radar observations
a shape models of Iris with the equivalent diameter of $208 \pm 35\,$km and the pole direction $\lambda = 15 \pm 5^\circ$ and $\beta = 25 \pm 15^\circ$ that is
consistent with the first model of Iris in Table~\ref{table2}.

\paragraph*{(8) Flora}
Although \cite{Tor.ea:03} give only one pole solution $(160^\circ, 16^\circ)$, the mirror solution at $(335^\circ, -5^\circ)$ 
fits the lightcurves equally well and
gives a slightly better fit to the occultation chords (Fig.~\ref{Flora_fig}). 

\paragraph*{(9) Metis}
The lightcurve inversion led to two possible pole solutions \citep{Tor.ea:03}, but only one agreed with the adaptive optics images \citep{Mar.ea:06}.
The consistency of this shape model with the occultation chords from 12 September 2008 was demonstrated by \cite{Tim.ea:09}. In Fig.~\ref{Metis_fig}, we present
the fit to four occultation events. 

\paragraph*{(10) Hygiea}
\cite{Han.ea:10} derived two possible models of Hygiea by lightcurve inversion of archived data and sparse photometry.
Both models fit the occultation data similarly well, but with very different sizes of 351\,km and 443\,km 
(Fig.~\ref{Hygiea_fig}). The diameter derived from IRAS measurements was 407\,km \citep{Ted.ea:04}, which is closer to the second (less preferred) model in
Table~\ref{table2}.

\paragraph*{(16) Psyche}
Only one of the two models reported by \cite{Kaa.ea:02} is consistent with the two occultations observed in 2004 and 2010 (Fig.~\ref{Psyche_fig}). 
For the occultation from 2004, one chord does not match with the model's
profile. This may be caused either by an error in timing or by an error of the shape model.

\paragraph*{(17) Thetis}
\cite{Dur.ea:09} derived two models, but only one, with the pole direction $(236^\circ, 20^\circ)$,
gives a good match with the occultation chords (Fig.~\ref{Thetis_fig}).

\paragraph*{(22) Kalliope}
One of the models derived by \cite{Kaa.ea:02} with the pole direction $(196^\circ, 3^\circ)$ is consistent with the satellite orbit analysis \citep{Des.ea:08} 
and also with the occultation observed in 2006 (Fig.~\ref{Kalliope_fig}). However, the equivalent diameter $166.2 \pm 2.8\,$km derived by \cite{Des.ea:08}
is significantly larger than our value $143 \pm 10\,$km.

\paragraph*{(28) Bellona}
We derived two possible models from the Asteroid Photometric Catalogue (APC) data \citep{Lag.ea:01b}, 
new observations (Table~\ref{table_aspects}), and US Naval Observatory (USNO) sparse photometry downloaded from 
the Asteroids Dynamic Site\footnote{\tt http://hamilton.dm.unipi.it/astdys/}.
Both models give similar fits to the data (Fig.~\ref{Bellona_fig}).

\paragraph*{(34) Circe}
Both shape models derived by \cite{Dur.ea:09} are possible, because the four chords observed in 2004 are grouped 
into two close pairs and do not allow us to select the correct pole. 
The two models differ significantly in the size (Fig.~\ref{Circe_fig}). The formal error of the fit (and thus of the determined size)
is only 4\,km, but the real uncertainty in the size is at least $\sim 10$\,km.

\paragraph*{(39) Laetitia}
The shape model derived by \cite{Kaa.ea:02} agrees very well with the observed silhouette (Fig.~\ref{Laetitia_fig}).
 
\paragraph*{(41) Daphne} 
The model derived by \cite{Kaa.ea:02b} is consistent with the occultation silhouette (Fig.~\ref{Daphne_fig}).

\paragraph*{(52) Europa} 
The pole direction $(252^\circ, 38^\circ)$ and the corresponding shape model 
derived by \cite{Mic.ea:04} is consistent with the AO image \citep{Mar.ea:06} and also with the
two occultation events (Fig.~\ref{Europa_fig}).

\paragraph*{(54) Alexandra} 
\cite{War.ea:08} derived two shape models. The model with the pole direction $(318^\circ, 23^\circ)$ fits the chords better but the other pole solution $(156^\circ, 13^\circ)$ cannot be rejected (Fig.~\ref{Alexandra_fig}).

\paragraph*{(55) Pandora} 
The shape model derived by \cite{Tor.ea:03} is consistent with the occultation chords (Fig.~\ref{Pandora_fig}).

\paragraph*{(63) Ausonia} 
The fit is not very good and there are only three chords observed in 2000, 
but the model with the pole $(120^\circ, -15^\circ)$ fits much better than the second possible 
pole $(305^\circ, -21^\circ)$ (Fig.~\ref{Ausonia_fig}). The size determination is not very accurate, the formal error 
of 18\,km given in Table~\ref{table2} is only the lowest limit. Our result is in agreement with results of \cite{Tan.ea:03} based on HST observations, 
who derived an ellipsoidal model of Ausonia with the pole direction $(119^\circ, -29^\circ)$ and the equivalent diameter 87\,km.

\paragraph*{(64) Angelina} 
We derived a new model of asteroid Angelina from the lightcurves in the APC and sparse data from USNO. From the two pole solutions
$(137^\circ, 14^\circ)$ and $(317^\circ, 17^\circ)$, the former one is
preferred, because it fits the chords significantly better that the latter pole (Fig.~\ref{Angelina_fig}). The
negative chord that intersects the silhouette was observed by two independent observers. The model has to be refined by further observations 
to match better with the occultation. 
The uncertainty of the size determination is likely to be larger than the formal value determined from the rms residuum.

\paragraph*{(68) Leto}
The three chords observed during the occultation in 1999 are not enough to distinguish between two pole solution derived
by \cite{Han.ea:10} (Fig.~\ref{Leto_fig}). The size is not determined very accurately.

\paragraph*{(80) Sappho}
The recently observed occultation clearly shows that only one of the two shape solutions derived by \cite{Dur.ea:09} is acceptable
(Fig.~\ref{Sappho_fig}).

\paragraph*{(85) Io} 
We derived a new model that slightly differs from that derived by \cite{Tor.ea:03} and that is in agreement 
with all four occultation events (Fig.~\ref{Io_fig}). The negative observation that intersect the model's 
projection in 7 December 2004 was made visually and is probably wrong. 

\paragraph*{(88) Thisbe} 
The mirror pole solution with the pole direction $(72^\circ, 60^\circ)$ is preferred over that reported by \cite{Tor.ea:03}, 
because it not only
fits the chords well, but also does not overlap with the negative observations (Fig.~\ref{Thisbe_fig}).

\paragraph*{(89) Julia}
We derived a new model of this asteroid from archived APC data and sparse photometry from USNO. From three possible pole solutions derived from 
photometry, only one is fully consistent with the occultation data (Fig.~\ref{Julia_fig}).

\paragraph*{(95) Arethusa}
From lightcurves only, we were not able to derive a unique model. Although the rotation period was determined uniquely, there were four
different pole directions and corresponding shapes that all fitted the photometric data equally well. However, when comparing
the predicted profiles with the occultation from 2009, only one model with the pole direction $(149^\circ, 33^\circ)$ was acceptable
(Fig.~\ref{Arethusa_fig}).

\paragraph*{(107) Camilla} 
The three chords observed during the occultation in 2004 allowed only a rough scaling of an updated version of the shape model derived 
originally by \cite{Tor.ea:03} (Fig.~\ref{Camilla_fig}).

\paragraph*{(129) Antigone} 
The projections of the shape model derived by \cite{Tor.ea:03} are consistent with the three occultations, but the size cannot be
fitted well. The size that fits the two occultations from 1985 and 2001 well gives a poor fit to the third occultation form 2009
(Fig.~\ref{Antigone_fig}). When fitting only the first two occultations, the size is $(118 \pm 14)$\,km, while the third event is 
fitted at best with the equivalent size $(148 \pm 11)$\,km. This inconsistency might be caused by albedo variegation (reported by 
\cite{Tor.ea:03}) -- the real shape is different from the model that assumes that albedo distribution on the surface is uniform.

\paragraph*{(130) Elektra} 
The shape model derived by \cite{Dur.ea:07} agrees well with the six chords observed in 2010 
(Fig.~\ref{Elektra_fig}). The model is also consistent with the adaptive optics images obtained by 
\cite{Mar.ea:06}.

\paragraph*{(152) Atala} 
A preliminary model was derived by \cite{Dur.ea:09} with two possible spin/shape solutions. Only one model is consistent with the occultation 
data (Fig.~\ref{Atala_fig}). One negative report intersects the model projection at one end. Modifying the model slightly should solve this
inconsistency.

\paragraph*{(158) Koronis} 
From two possible spin vector solutions derived by \cite{Sli.ea:03}, only one is consistent with 
the occultation chords (Fig.~\ref{Koronis_fig}). However, even this model fails to fit the northernmost chord that is longer than the shape model.

\paragraph*{(165) Loreley}
We derived a new model of Loreley with two possible pole directions $(174^\circ, 29^\circ)$ and $(348^\circ, 42^\circ)$. Only the first
pole solution and the corresponding model is consistent with the occultation data (Fig.~\ref{Loreley_fig}). This new model is different from
that published by \cite{Dur.ea:07}.

\paragraph*{(167) Urda} 
One of the pole solutions gives a better fit, but the rival pole cannot be rejected (Fig.~\ref{Urda_fig}). The negative chord that 
intersects the model profile at one end was observed visually. The size determination is not very accurate.

\paragraph*{(208) Lacrimosa} 
The four chords do not enable us to clearly reject one of the two possible pole solutions, although the pole $(176^\circ, -68^\circ)$ is 
preferred (Fig.~\ref{Lacrimosa_fig}).

\paragraph*{(276) Adelheid} 
Only three chords, two of them very close to each other, cannot distinguish between two shape models (Fig.~\ref{Adelheid_fig}). Both models
fit the occultation equally well.

\paragraph*{(302) Clarissa}
\cite{Han.ea:10} derived a very elongated shape model of Clarissa. Both pole solutions are consistent with the occultation observed in 2004,
one of them fitting very well (Fig.~\ref{Clarissa_fig}). Although the chords were observed using video, the reported timing errors are
of tenths of second.

\paragraph*{(306) Unitas} 
\cite{Dur.ea:09} derived two possible shape models. \cite{Del.Tan:09} showed that one of them -- with the pole direction 
$(79^\circ, -35^\circ)$ -- was much more consistent 
with the thermal infrared measurements made by IRAS. They also derived the effective diameter $55-57$\,km. This is also confirmed by the occultation data (Fig.~\ref{Unitas_fig}), because the profile corresponding to this pole does
not intersect with the negative chord.

\paragraph*{(372) Palma}
\cite{Han.ea:10} derived two possible shape models of Palma with the pole directions $(221^\circ, -47^\circ)$ and $(44^\circ, 17^\circ)$. The first shape model
seems to be consistent with most of the chords observed during two occultations in 2007 and 2009 (Fig.~\ref{Palma_fig}), but the southernmost chord 
from the 2007 event lies far outside the projected best-fit contour of the model. The second model fits well the southernmost chord from 2007, but its fit to
the rest of the chords is slightly worse than for the first model.

\paragraph*{(409) Aspasia} 
With six successfully observed occultations, Aspasia has the highest number of occultation events in our set.
From the two models derived by \cite{War.ea:09}, the one with the pole direction $(3^\circ, 30^\circ)$ fits
the chords observed during two events in 2008 clearly better than the second pole (Fig.~\ref{Aspasia_fig}).

\paragraph*{(471) Papagena}
From the two possible models derived by \cite{Han.ea:10}, only one is consistent with the occultation observed in 1987 
(Fig.~\ref{Papagena_fig}).

\paragraph*{(747) Winchester} 
We updated the model by \cite{Mar.ea:09} with new observations by F.~Pilcher and A.~Gal\'ad (Table~\ref{table_aspects}). Lightcurve inversion 
yields two pole directions. The second
pole solution $(172^\circ, -36^\circ)$ can be rejected, because the shape projection for this pole fits the occultation data
significantly worse than the pole $(304^\circ, -60^\circ)$ (Fig.~\ref{Winchester_fig}).

\paragraph*{(849) Ara}
Only one model, out of the two derived by \cite{Mar.ea:09}, is consistent with the occultation (Fig.~\ref{Ara_fig}).

\paragraph*{(925) Alphonsina}
\cite{Han.ea:10} derived a new model from combined sparse and dense photometry. 
Although both occultations suffer from large errors in timings, only one of the pole
solutions derived from photometry is consistent with the occultations (Fig.~\ref{Alphonsina_fig}).

\paragraph*{(1263) Varsavia}
We derived two new models from USNO sparse photometry but only one was consistent with the occultation 
data (Fig.~\ref{Varsavia_fig}).

\section{Conclusions}
By combining asteroid shape models derived from photometry with the occultation data, we can obtain 
unique physical models of asteroids. Lightcurve data enable us to derive asteroid shape and spin state and the occultations
are used for scaling the shape and for solving the pole ambiguity. 
As can be seen from Table~\ref{table2}, the uncertainty of the effective diameter determination is
$\sim 5\,\%$ for the best cases with many accurate chords, $\sim 10\,\%$ for a typical occultation, 
and $\sim 20\,\%$ for occultations with only a few chords. Contrary to other indirect methods for asteroid size determination, 
occultation timings are direct `measurements' of asteroid dimensions.  
 
 The number of successfully observed occultations with many chords steadily grows as does the accuracy of timings -- observers routinely 
 use video or CCD techniques and the timings are more accurate and free of reaction times. The use of automatic observing stations equipped with a small
 telescope will revolutionize the way asteroidal occultations are observed \citep{Deg:09}.
 Asteroidal occultations are no longer `sporadic'
 events, rather they are systematic and reliable astronomical measurements. 
 With the increasing number of asteroid models and successfully observed occultations, we expect that the scaling of models derived by 
 lightcurve inversion using occultations will become routine. 

\section*{Acknowledgements}
We thank all the observers for their efforts and for providing their observations.
We appreciate the help of Jan M\'anek, who provided us with a valuable insight into observation and reduction of occultations. 
We thank Benoit Carry, one of the referees, whose critical comments helped us to improve the manuscript.
The work of J.\v{D}. was supported by the grant GACR P209/10/0537 of the Czech Science Foundation and by the Research Program MSM0021620860
of the Ministry of education. The work of M.K. was supported by the Academy of Finland. 
Funding for B.W. observations at the Palmer Divide Observatory
was provided by NASA grant NNX 09AB48G, by National Science
Foundation grant AST-1032896, and by a 2007 Gene Shoemaker
NEO Grant from the Planetary Society.
The work of A.G. was supported by the grant 2/0016/09
of the Slovak Grant Agency for Science VEGA and by the grant 205/09/1107 of the Czech Science Foundation.  
J.H. was supported by the grant GACR 205/08/H005 and by the grant GAUK 134710 of the Grant agency of the Charles University.

\newcommand{\SortNoop}[1]{}

\clearpage	

\onecolumn
\begin{longtable}{llrrrrrrl}
\multicolumn{9}{c}{\textbf{List of occultations}}\\
\hline
\hline
Asteroid & \multicolumn{1}{c}{Date} 	& $N_\mathrm{total}$ 	& $N_\mathrm{phot}$ 	& \multicolumn{1}{c}{$\Delta\phi$}	& \multicolumn{1}{c}{$t_\mathrm{mean}$}	& \multicolumn{2}{c}{$|\Delta \tau|$ [s]} 	& \multicolumn{1}{c}{Reference} 	\\
\cline{7-8}
	 & 				& 			& 			& \multicolumn{1}{c}{[deg]}		& \multicolumn{1}{c}{[s]}		& model 1	& model 2		& 	 			\\
\hline
\endhead
\hline 
\endfoot
2 Pallas      	& 1978/05/29 	&   8 	&  8 	& 2.6	 & 30.1	& 	&	& \citet{Was.ea:79} \\
              	& 1983/05/29 	& 121 	& 18 	& 2.5	 & 34.4	& 0.24	&	& \citet{Dun.ea:90} \\
              	& 1985/10/24 	&   3 	&  0 	& 0.4	 & 27.9	& 0.02	&	& \citet{Sta:85} \\
              	& 2001/06/09 	&   3 	&  3 	& 0.8	 & 25.1	&  	&	& \\
              	& 2006/06/12	&   4	&  4	& 0.5	 & 35.4	& 	&	& \\
3 Juno        	& 1979/12/11 	&  17 	& 14 	& 13.6	 & 57.6	& 1.42	& 	& \citet{Mil.ea:81} \\
            	& 2000/05/24 	&   5 	&  0 	& 0.6	 & 17.0	& 0.19	&	& \\
5 Astraea	& 2008/06/06	& 13	& 3	& 0.3	 & 13.8	& 0.11	& 0.23	& \\
6 Hebe		& 1977/03/05	& 5 	& 0	& 0.1	 & 3.3	& 0.12	& 	& \citet{Tay.Dun:78} \\
7 Iris		& 2005/02/17	& 4	& 0	& 0.1	 & 7.2	& 0.20	& 0.12	& \\
		& 2006/05/05	& 4	& 1	& 0.1	 & 4.2	& 0.59 	& 0.54	& \citet{Tho.Yee:06} \\
8 Flora		& 2004/10/29	& 7	& 6	& 0.1	 & 6.2	& 0.03	& 0.27	& \citet{Dun:05} \\
9 Metis		& 1984/02/19	& 8	& 1	& 1.2	 & 16.6	& 0.98	&	& \citet{Kri:84}	\\
		& 1989/08/06	& 5	& 1	& 3.2	 & 12.7	& 0.31	&	& \citet{Sta:89}	\\
		& 2001/09/07	& 7	& 5	& 0.2	 & 3.1	& 0.07	&	& \citet{Dun:02} \\
		& 2008/09/12	& 20	& 15	& 1.8	 & 35.5	& 0.67	&	& \citet{Tim.ea:09} \\		
10 Hygiea	& 2002/09/07	& 8	& 6	& 1.9	 & 110.1& 0.66	& 2.17	& \\
16 Psyche	& 2004/05/16	& 4	& 1	& 0.7	 & 12.9	& 0.15	& 0.30	& \\
		& 2010/08/21	& 14	& 14	& 1.9	 & 7.1	&	&	& \\		   
17 Thetis	& 2007/04/21	& 16	& 9	& 0.6	 & 6.0	& 0.59	& 0.46	& \citet{Kos.ea:09} \\
22 Kalliope	& 2006/11/07	& 8	& 5	& 1.4	 & 24.2	& 0.53	&	& \citet{Som.ea:07} \\
28 Bellona	& 2002/05/05	& 8	& 4	& 0.7	 & 8.1	& 0.47	& 0.54	& \\ 
34 Circe	& 2004/02/10	& 4	& 4	& 0.4	 & 29.1	& 	& 	& \\
39 Laetitia   	& 1998/03/21 	&  19 	&  2 	& 1.4	 & 9.7	& 0.37	&	& \citet{Dun:99} 	\\
41 Daphne     	& 1999/07/02 	&  16 	&  3 	& 3.6	 & 32.7	& 0.50	& 	& \citet{Dun.ea:02} \\
52 Europa     	& 1983/04/26 	&  10 	&  2 	& 0.8	 & 9.5	& 0.47	&	& \\ %\citet{Dun:83}	\\
		& 2005/12/03	& 12	& 10	& 0.4	 & 6.7	& 0.08	&	& \\
54 Alexandra	& 2005/05/17	& 11	& 6	& 10.3	 & 41.7	& 0.22	& 0.04	& \citet{Dun:06} \\
55 Pandora	& 2007/02/18	& 9 	& 5	& 3.0	 & 3.8	& 0.63 &	& \\
63 Ausonia	& 2000/12/22	& 3	& 3	& 0.2	 & 2.3	& 	& 	& \\
64 Angelina	& 2004/07/03	& 6 	& 6	& 0.0	 & 1.2	& 	& 	& \\
68 Leto		& 1999/05/23	& 3	& 1	& 0.1	 & 5.5	& 0.10	& 0.19	& \\
80 Sappho	& 2010/06/04	& 11	& 9	& 1.0	 & 5.6	& 0.94	& 0.24	& \\
85 Io         	& 1995/12/10 	&   8 	&  0 	& 3.7	 & 15.6	& 0.30	&	& \citet{Dun:98} \\
		& 2004/12/07	& 5	& 4	& 0.7	 & 13.8	& 0.36 &	& \\
		& 2004/12/12	& 15	& 9	& 1.4	 & 12.2	& 0.10	&	& \\
		& 2005/05/15	& 2	& 2	& 0.1	 & 4.3	& 	&	& \\
88 Thisbe	& 1981/10/07	& 12 	& 0	& 1.2	 & 9.8	& 0.32	& 0.35	& \citet{Mil.ea:83}	\\
		& 2007/02/21	& 6	& 3	& 3.7	 & 14.0	& 0.52	& 0.85	& \\		
89 Julia	& 2005/08/13	& 6	& 1	& 1.9	 & 7.6	& 0.06	&	& \\
		& 2006/12/04	& 4	& 4	& 0.2	 & 9.0 	& 	&	& \\
95 Arethusa	& 2009/03/07	& 8	& 5	& 0.6	 & 8.5	& 0.29	&	& \\		
107 Camilla	& 2004/09/05	& 3	& 3	& 1.3	 & 12.3	& 	&	& \\	
129 Antigone  	& 1985/04/11 	&   5 	&  3 	& 4.0	 & 69.5	& 6.75	&	& \citet{Was.ea:86}\\
              	& 2001/09/09 	&   5 	&  0 	& 2.5	 & 8.4	& 0.23	&	& \\
              	& 2009/02/13	& 4	& 4	& 1.8	 & 7.9	& 	&	& \\
130 Elektra	& 2010/02/20	& 7	& 6	& 0.7	 & 11.7	& 	&	& \\
152 Atala	& 2006/05/07	& 8	& 5 	& 7.5	 & 3.3	& 0.69	& 0.75	& \\
158 Koronis	& 2005/12/13	& 7	& 4	& 0.1	 & 1.4	& 0.10	& 0.12 	& \\
165 Loreley	& 2009/06/29	& 5	& 4	& 0.6	 & 33.0	& 0.29	& 0.08	& \\
167 Urda	& 2001/07/23	& 3	& 2	& 0.1	 & 3.3	& 0.08	& 0.01	& \\
208 Lacrimosa 	& 2003/12/31 	&   5 	&  2 	& 0.2	 & 4.6	& 0.14	& 0.30	& \\
276 Adelheid	& 2002/03/09	& 3 	& 1 	& 1.3	 & 5.7	& 0.41	& 0.45	& \\
302 Clarissa	& 2004/06/24	& 3	& 3	& 0.1	 & 2.7	& 	& 	& \\
306 Unitas	& 2004/07/06	& 5	& 4	& 0.9	 & 8.0	& 0.94	& 1.10	& \\
372 Palma	& 2007/01/26	& 21	& 11	& 3.0	 & 10.8	& 0.14	& 0.23	& \\
		& 2009/09/10	& 4 	& 3	& 0.6	 & 8.2	& 0.07	& 0.13	& \\
409 Aspasia	& 2005/09/29	& 2 	& 2	& 0.1	 & 3.0	& 	&	& \\	
		& 2006/10/08	& 7 	& 7	& 0.8	 & 10.3	& 	&	& \\
		& 2006/12/20	& 3	& 2	& 0.4	 & 9.6	& 0.48	& 0.38	& \\
		& 2008/02/05	& 5	& 3	& 0.9	 & 20.8	& 0.50	& 0.22	& \\
		& 2008/02/12	& 10	& 10	& 2.8	 & 31.4	& 	& 	& \\
		& 2009/02/06	& 2	& 2	& 0.7    & 9.0	& 	& 	& \\	
471 Papagena	& 1987/01/24	& 5	& 2	& 3.3	 & 17.7	& 0.84	& 0.37	& \\		
747 Winchester	& 2008/05/01	& 9	& 6	& 0.8	 & 11.7	& 0.25	& 0.37	& \\		
		& 2009/09/05	& 6	& 5	& 1.3	 & 16.9	& 	&	& \\
849 Ara		& 2009/01/27	& 7	& 4	& 0.6	 & 1.6	& 0.75	& 0.34	& \\
925 Alphonsina	& 2003/12/15	& 7	& 7	& 0.6	 & 5.5	& 	& 	& \\
		& 2003/12/22	& 27	& 12	& 1.8	 & 3.9	& 0.46	& 0.69	& \\
1263 Varsavia	& 2003/07/18	& 31	& 20	& 1.4	 & 1.6	& 0.04	& 0.04	& \\
\hline
\caption{List of occultations used for deriving asteroid sizes.
Here $N_\mathrm{total}$ is the total number of chords used for fitting, $N_\mathrm{phot}$ is 
the number of chords observed by means of a photometric detector, video, or CCD,  $\Delta\phi$ is the angle 
of asteroid's rotation between the first and the last reported timings, and $t_\mathrm{mean}$ is the the mean duration of occultation. 
For each occultation observed visually, the average absolute time shift $|\Delta \tau|$
of visual chords is listed.}
\label{table1}
\end{longtable}
\twocolumn

\clearpage	

\onecolumn
\begin{longtable}{l}
\multicolumn{1}{c}{\textbf{List of observers}}\\
\hline
\hline
\endhead
\hline
\endfoot
\multicolumn{1}{c}{\textbf{(5) Astraea } 2008-06-06 }\\
T. Jan\'ik, CZ \\
Z. Moravec, CZ \\
H. Raab, AT \\
V. \v{C}ejka, CZ \\
J. M\'anek, CZ \\
F. Lomoz, CZ \\
J. Jindra, CZ \\
P. Ku\v{s}nir\'ak, CZ \\
J. Urban, CZ \\
R. Piffl \& T. Maru\v{s}ka \& I. Majchrovi\v{c}, AT \\
M. Anto\v{s}, CZ \\
G. Dangl, AT \\
J. \v{D}urech, CZ \\
P. Zelen\'y, CZ \\
J. Mocek, CZ \\
M. Kro\v{c}il, CZ \\
M. Kapka, SK \\
D. Kapetanakis, GR \\
H. Denzau, DE \\
J. Kopplin, DE \\

\\
\multicolumn{1}{c}{\textbf{(16) Psyche } 2010-08-21 }\\
J. Brooks, Winchester, VA, USA \\
S. Conard, Gamber, MD, USA \\
D. Dunham, Seymour, TX, USA (5) \\
A. Scheck, Scaggsville, MD, USA \\
C. Ellington, Highland Village, TX, USA \\
P. Maley, Annetta South, TX, USA \\
R. Tatum, Richmond, VA, USA \\
P. Maley, Godley, TX, USA \\
H. \& K. Abramson, Mechanicsville, VA, USA \\
D. Caton, Boone, NC, USA \\
E. Iverson, Athens, TX, USA \\
R. Suggs \& B. Cooke, Huntsville, AL, USA \\
J. Faircloth, Kinston, NC, USA \\

\\
\multicolumn{1}{c}{\textbf{(80) Sappho } 2010-06-04 }\\
P. Birtwhistle, UK \\
C. Ratinaud, FR \\
T. Haymes, UK \\
J. Lecacheux, FR \\
G. Regheere, FR \\
O. Dechambre FR \\
F. Vachier \& S. Vaillant \& J. Berthier, FR \\
T. Midavaine, FR \\
E. Bredner \& F. Colas, FR \\
A. Leroy \& S. Bouley \& R. Palmade \& G. Canaud, FR \\
E. Frappa, FR \\
G. Bonatti \& D. Del Vecchio, IT \\
P. Baruffetti \& A. Bugliani \& G. Tonlorenzi, IT \\
 
\\
\multicolumn{1}{c}{\textbf{(95) Arethusa } 2009-03-07 }\\
A. Carcich, Wantage, NJ, USA \\
W. Rauscher, Doylestown, PA, USA \\
G. Nason, West Lorne, ON, Canada \\
R. Sauder, Narvon, PA, USA \\
S. Conard, Gamber, MD, USA \\
B. Huxtable, Gambrills, MD, USA \\
D. Dunham, MD \& VA, USA (3) \\
H. Abramson, Smithfield, VA, USA \\
 
\\
\multicolumn{1}{c}{\textbf{(129) Antigone } 2009-02-13 }\\
R. Cadmus, Grinnell, IA, USA \\
A. Carcich, Lacey, NJ, USA \\
S. Messner, Morning Sun, IA, USA \\
D. Dunham, Glen Rock, PA, USA \\
S. Conard, Gamber, MD, USA \\
B. Huxtable,Gambrills, MD, USA \\
A. Olsen, Urbana, IL, USA \\
 
\\
\multicolumn{1}{c}{\textbf{(130) Elektra } 2010-02-20 }\\
A. Elliott, UK \\
C. Ratinaud, FR \\
R. Miles, UK \\
M. Cole, UK \\
A. Pratt, UK \\
P. Birtwhistle, UK \\
T. Haymes, UK \\
J.-M. Winkel, NL \\
R. \& E. Simonson, UK \\
G. Rousseau, FR \\
 
\\
\multicolumn{1}{c}{\textbf{(165) Loreley } 2009-06-29 }\\
R. Peterson, Scottsdale, AZ, USA \\
W. Morgan, Wilton, CA, USA \\
D. Machholz, Colfax, CA, USA \\
D. Dunham, Blue Canyon, CA \& NV, USA (3) \\
 
\\
\multicolumn{1}{c}{\textbf{(372) Palma } 2009-09-10 }\\
D. Herald, Kambah, ACT, AU \\
H. Pavlov, Marsfield, AU \\
D. Gault, Hawkesbury Heights, AU \\
J. Broughton, Reedy Creek, QLD, AU \\
C. Wyatt, Walcha, NSW, AU \\
J. Bradshaw, Samford, QLD, AU \\
 
\\
\multicolumn{1}{c}{\textbf{(409) Aspasia } 2008-02-05 }\\
C. Ninet, FR \\
M. Boutet, FR \\
J. Sanchez, FR \\
G. Faure, FR \\
E. Frappa, FR \\
S. Bolzoni, IT \\
S. Sposetti, CH \\
G. Sbarufatti, IT \\
R. Di Luca, IT \\
 
\\
\multicolumn{1}{c}{\textbf{(409) Aspasia } 2008-02-12 }\\
J. Denis, FR \\
T. Flatr\`es \& J.-J. Sacr\'e, FR \\
M. Jennings, UK \\
P. Baudouin, FR \\
M. Audejean, FR \\
O. Dechambre, FR \\
B. Christophe, FR \\
T. Midavaine, J. Langlois, FR \\
D. Fiel, FR \\
A. Leroy \& G. Canaud, FR \\
J. Lecacheux, FR \\
E. Frappa, FR \\
F. Van Den Abbeel, BE \\
C. Gros, FR \\
E. Bredner, FR \\
C. Demeautis \& D. Matter, FR \\
S. Bolzoni, IT \\
A. Manna, CH \\
S. Klett, CH \\
S. Sposetti, CH \\
M. Parl, DE \\
P. Corelli, IT \\
R. Di Luca, IT \\
 
\\
\multicolumn{1}{c}{\textbf{(409) Aspasia } 2009-02-06 }\\
D. Snyder, Bisbee, AZ, USA \\
P. Sada, Observatorio UDEM, Mexico \\
 
\\
\multicolumn{1}{c}{\textbf{(747) Winchester } 2008-05-01 }\\
D. Koschny, NL \\
H. De Groot, NL \\
A. Scholten, NL \\
H. Rutten, NL \\
M. Rain, DE \\
E. Frappa \& A. Klotz, FR \\
H. Bill \& M. Jung, DE \\
J. M\"uller \& U. Appel, DE \\
W. Rothe, DE \\
I. Majchrovi\v{c} \& R. Piffl, SK \\
A. Gal\'ad, SK \\
T. Pauwels \& P. De Cat, BE \\
 
\\
\multicolumn{1}{c}{\textbf{(747) Winchester } 2009-09-05 }\\
D. Gault, Macquarie Woods, NSW, AU \\
J. Broughton, Reedy Creek, QLD, AU \\
D. Lowe, Leyburn, QLD, AU \\
S. Quirk, Mudgee, NSW, AU \\
P. Purcell, Weston Creek, ACT, AU \\
D. Herald, ACT, AU (2) \\
C. Wyatt, Narrabri, NSW, AU \\
J. Betts, Hawkesbury Heights, NSW, AU \\
S. Russell, Orange, NSW, AU \\
 
\\
\multicolumn{1}{c}{\textbf{(849) Ara } 2009-01-27 }\\
R. Stanton, Three Rivers, CA, USA \\
D. Breit, Morgan Hill, CA, USA \\
R. Royer, Springville, CA, USA \\
R. Nolthenius, Cabrillo College, CA, USA \\
B. Stine, Weldon, CA, USA \\
R. Peterson, AZ, USA (2) \\

\hline
\caption{List of observers who participated in observations presented in Table~\ref{table1} that were not included in the
database published by \cite{Dun.Her:09}. If some observer provided more than one chord, the number of chords is given in parentheses.}
\label{table_observers}
\end{longtable}
\twocolumn

\clearpage

\onecolumn
\begin{longtable}{lp{2}rd{0}d{6}p{2}d{1}d{1}l}
\multicolumn{9}{c}{\textbf{List of results}}\\
\hline
\hline
Asteroid       	& \multicolumn{1}{c}{$D$} 	& \multicolumn{1}{c}{$\lambda_\mathrm{p}$}	& \multicolumn{1}{c}{$\beta_\mathrm{p}$}	& \multicolumn{1}{c}{$P$}	& \multicolumn{1}{c}{$D_\mathrm{IRAS}$}	& \multicolumn{1}{c}{rms$_1$}	& \multicolumn{1}{c}{rms$_2$}	&	Reference	\\
		& \multicolumn{1}{c}{[km]} 	& \multicolumn{1}{c}{[deg]}			& \multicolumn{1}{c}{[deg]}			& \multicolumn{1}{c}{[hr]}	& \multicolumn{1}{c}{[km]}		& \multicolumn{1}{c}{[km]}	& \multicolumn{1}{c}{[km]}	&			\\
\hline
\endhead
\hline
\endfoot
2 Pallas      	& 539 p 28	& 35		& -12		& 7.81323	& 498 p 19	& 14.2	& 16.6	& \citet{Tor.ea:03}\\
3 Juno        	& 252 p 29 	& 103		& 27 		& 7.209531	& 234 p 11	& 14.3	& 17.4	& \citet{Kaa.ea:02}\\
5 Astraea	& 115 p 6	& 126		& 40		& 16.80061	& 119 p 7	& 3.2	& 3.4	& \citet{Dur.ea:09}\\
6 Hebe		& 180 p 40	& 340		& 42		& 7.274471	& 185 p 3	& 7.7	& 9.5	& \citet{Tor.ea:03}\\
7 Iris		& 198 p 27	& 20		& 14		& 7.138840	& 200 p 10	& 13.7	& 26.2	& \citet{Kaa.ea:02}\\
		& 199 p 26	& 199 		& -2 		& 		&		& 13.2	& 27.7	& \\
8 Flora		& 141 p 10	& 155		& 6		& 12.86667	& 136 p 2	& 4.8	& 5.1	& \citet{Tor.ea:03}\\
		& \bf140 p \bf7	& \bf 335	& \bf-5		&		& 		& 3.7	& 3.8	&  \\			
9 Metis		& 169 p 20	& 180		& 22		& 5.079176	& 		& 10.2	& 12.4	& \citet{Tor.ea:03} \\
		&		&		&		&		&		&	& 	& \citet{Tim.ea:09}\\
10 Hygiea	& \bf351 p \bf27& \bf122	& \bf-44	& 27.65905	& 407 p 7	& 13.8	& 14.0	& \citet{Han.ea:10} \\
		& 443 p 45	& 312		& -42		&		&		& 22.5	& 24.1	& \\	
16 Psyche	& \bf225 p \bf20& 33		& -7		& 4.195948	& 253 p 4	& 10.0	& 11.1	& \citet{Kaa.ea:02}\\
		& 225 p 36	& 213		& 1		&		&		& 17.8	& 18.0	& \\
17 Thetis	& 77 p 8	& 236		& 20		& 12.26603	& 90 p 4	& 4.0	& 6.1	& \citet{Dur.ea:09}\\
22 Kalliope 	& 143 p 10	& 196		& 3		& 4.148200	& 181 p 5	& 5.1	& 5.7	& \citet{Kaa.ea:02}\\
28 Bellona	& 97 p 11	& 282		& 6		& 15.70785	& 121 p 3	& 5.4	& 5.8	& this work\\
		& 100 p 10	& 102 		& -8		& 		&		& 5.1	& 6.1	& \\
34 Circe	& 96 p 10	& 94		& 35		& 12.17458	& 114 p 3	& 1.8	&	& \citet{Dur.ea:09}\\
		& 107 p 10 	& 275		& 51		&		&		& 2.0	&	& \\
39 Laetitia	& 163 p 12	& 323		& 32		& 5.138238	& 150 p 9	& 6.2	& 14.4	& \citet{Kaa.ea:02}\\			
41 Daphne     	& 187 p 20 	& 198		& -32 		& 5.98798	& 174 p 12	& 10.2	& 11.0	& \citet{Kaa.ea:02b}\\
52 Europa     	& 293 p 30 	& 251		& 35 		& 5.629958	& 303 p 5	& 14.9	& 16.5	& \citet{Mic.ea:04}\\
54 Alexandra	& 135 p 20	& 156		& 13		& 7.022641	& 166 p 3	& 10.1	& 10.1	& \citet{War.ea:08}\\
		& \bf142 p \bf9	& \bf318	& \bf23		& 7.022649	& 		& 4.7	& 4.9	& \\
55 Pandora	& 70 p 7	& 223		& 18		& 4.804043	& 67 p 3	& 3.3	& 6.0	& \citet{Tor.ea:03}\\
63 Ausonia	& 90 p 18	& 120		& -15		& 9.29759	& 103 p 2	& 8.9	&	& \citet{Tor.ea:03}\\
64 Angelina	& 51 p 10	& 317		& 17		& 8.75032	& 		& 5.1	&	& this work\\
		& \bf52 p \bf10	& \bf137	& \bf14		&		& 		& 2.7	&	& \\
68 Leto		& 148 p 25	& 103		& 43		& 14.84547	& 123 p 5	& 6.5	& 8.3	& \citet{Han.ea:10} \\
		& 151 p 25	& 290		& 23		&		&		& 12.6	& 15.9	& \\		
80 Sappho	& 67 p 11	& 194  		& -26  		& 14.03087  	& 78 p 2	& 5.6	& 8.1	& \citet{Dur.ea:09} \\
85 Io         	& 163 p 15 	& 95		& -65 		& 6.874783	& 155 p 4	& 7.4	& 7.9	& this work\\
88 Thisbe	& \bf204 p \bf14& \bf72		& \bf60		& 6.04131	& 201 p 5	& 7.1	& 12.5	& this work \\
		& 220 p 16	& 247		& 50		&		&		& 6.5	& 13.9	& \citet{Tor.ea:03}\\
89 Julia	& 140 p 10	& 8		& -13		& 11.38834	& 151 p 3	& 4.8	& 5.0	& this work \\		
95 Arethusa	& 147 p 32	& 149		& 33		& 8.70221	& 136 p 10	& 15.8	& 18.7	& this work \\
107 Camilla	& 214 p 28	& 73		& 54		& 4.843928	& 223 p 17	& 14.2	&	& \citet{Tor.ea:03}\\
129 Antigone	& 118 p 19	& 207		& 58		& 4.957154	&		& 9.4	& 10.7	& \citet{Tor.ea:03}\\
130 Elektra	& 191 p 14	& 64  		& -88  		& 5.224664  	& 182 p 12	& 7.2	&	& \citet{Dur.ea:07}\\
152 Atala	& 65 p 8	& 347		& 47		& 6.24472	&		& 3.9	& 9.3	& \citet{Dur.ea:09}\\
158 Koronis	& 38 p 5	& 30		& -64		& 14.20569	& 35 p 1	& 1.7	& 2.4	& \citet{Sli.ea:03}\\
165 Loreley	& 171 p 8	& 174		& 29		& 7.224387	& 155 p 5	& 3.9	& 4.2	& this work \\
167 Urda	& 51 p 15	& 107		& -69		& 13.06133	& 40 p 2	& 1.2	& 2.0	& \citet{Sli.ea:03} \\
		& \bf44 p \bf15	& \bf249	& \bf-68	&		&		& 1.1	& 1.4	& \citet{War.ea:08}\\
208 Lacrimosa 	& 45 p 10	& 20		& -75 		& 14.0769	& 41 p 2	& 3.7	& 5.6	& \citet{Sli.ea:03}\\
		& \bf45 p \bf10	& 176		& \bf-68	&		&		& 1.7	& 2.4	& \\
276 Adelheid	& 125 p 15	& 9		& -4		& 6.31920	& 122 p 8	& 5.0	& 8.9	& \citet{Mar.ea:07}\\
		& 117 p 15	& 199		& -20		&		&		& 4.8	& 9.5	& \\		
302 Clarissa	& \bf43 p \bf4	& \bf28		& \bf-72	& 14.47670	& 39 p 3	& 2.0	&	& \citet{Han.ea:10} \\
		& 43 p 11	& 190		& -72		& 		&		& 5.5	&	& \\
306 Unitas	& \bf49 p \bf5	& \bf79		& \bf-35	& 8.73875	& 47 p 2	& 1.4	& 2.7	& \citet{Dur.ea:09}\\
		& 53 p 5	& 253		& -17		&		&		& 1.6	& 3.2	& \\
372 Palma	& 187 p 20	& 221		& -47		& 8.58189	& 189 p 3	& 10.1	& 10.6	& \citet{Han.ea:10} \\
		& 198 p 26	& 44		& 17		& 8.58191	&		& 13.2	& 13.8	& \\	  
409 Aspasia	& 173 p 17	& 3		& 30		& 9.02145	& 162 p 7	& 8.7	& 10.5	& \citet{War.ea:08}\\
471 Papagena	& 137 p 25	& 223		& 67		& 7.11539	& 134 p 5	& 12.4	& 15.5	& \citet{Han.ea:10} \\
747 Winchester	& 171 p 15	& 304		& -60		& 9.41480	& 172 p 3	& 7.6	& 8.3	& \citet{Mar.ea:09} \\
		& 		&		&		&		&		&	&	& this work\\		
849 Ara		& 76 p 14	& 223		& -40		& 4.116391	& 62 p 3	& 7.0	& 17.8	& \citet{Mar.ea:09}\\
925 Alphonsina	& 58 p 16	& 294		& 41		& 7.87754	& 54 p 3	& 8.0	& 11.5	& \citet{Han.ea:10}\\
1263 Varsavia	& 41 p 8	& 341		& -14		& 7.16495	& 49 p 1	& 3.9	& 4.0	& this work\\
\hline
\caption{The table lists the equivalent diameter $D$ with its estimated uncertainty, 
the spin vector ecliptic coordinates $\lambda_\mathrm{p}$ and $\beta_\mathrm{p}$, the sidereal rotation period $P$, the diameter 
$D_\mathrm{IRAS}$ derived from the IRAS infrared measurements by \cite{Ted.ea:04}, the rms residual of the fit with visual chords allowed to move
along the relative velocity vector (rms$_1$) and with visual chords fixed (rms$_2$), and
the reference to the paper, where the original shape model was published. If one of two possible pole directions and the corresponding
shape model is preferred, it is printed in bold font.}
\label{table2}
\end{longtable}

\clearpage

\begin{longtable}{cccrcrc}
\multicolumn{7}{c}{\textbf{Aspect data for new observations}}\\
\hline
\hline
Date        	& $r$ 	& $\Delta$	& \multicolumn{1}{c}{$\alpha$}	& $\lambda$	& \multicolumn{1}{c}{$\beta$}	& Observer	\\
		& [AU] 	& [AU]          & \multicolumn{1}{c}{[deg]}	& [deg]		& \multicolumn{1}{c}{[deg]}	&	\\
\hline
\endhead
\hline
\endfoot
\multicolumn{7}{c}{\textbf{(28) Bellona}}\\
2007 04 28.2  & 2.635    & 1.684  &  9.0     & 196.8     & $12.8$  &   Warner   \\
2007 04 29.3  & 2.637    & 1.690  &  9.4     & 196.6     & $12.8$  &   Warner   \\
2007 05 04.3  & 2.644    & 1.723  & 11.0     & 195.8     & $12.8$  &   Warner   \\
2007 05 10.2  & 2.653    & 1.770  & 13.0     & 195.0     & $12.6$  &   Warner   \\
2008 06 06.3  & 3.143    & 2.297  & 12.0     & 295.0     & $ 8.5$  &   Pilcher   \\
2008 06 07.3  & 3.144    & 2.288  & 11.7     & 294.9     & $ 8.5$  &   Pilcher   \\
2008 06 10.3  & 3.146    & 2.265  & 10.9     & 294.5     & $ 8.6$  &   Pilcher   \\
2008 06 14.3  & 3.148    & 2.237  &  9.7     & 293.9     & $ 8.6$  &   Pilcher   \\
2008 06 24.3  & 3.154    & 2.184  &  6.6     & 292.2     & $ 8.5$  &   Pilcher   \\
2008 07 02.3  & 3.159    & 2.160  &  4.2     & 290.6     & $ 8.3$  &   Pilcher   \\
2008 07 06.3  & 3.161    & 2.154  &  3.1     & 289.8     & $ 8.2$  &   Pilcher   \\
2008 08 02.2  & 3.174    & 2.234  &  8.3     & 284.4     & $ 7.1$  &   Pilcher   \\
2010 12 16.4  & 2.385    & 1.517  & 14.0     & 118.9     & $-9.4$  &   Pilcher   \\
2010 12 19.3  & 2.383    & 1.495  & 12.9     & 118.5     & $-9.4$  &   Pilcher   \\
2010 12 22.4  & 2.382    & 1.474  & 11.6     & 118.1     & $-9.3$  &   Pilcher   \\
\multicolumn{7}{c}{\textbf{(747) Winchester}}\\
2008 05 09.3  & 3.949    & 3.031  &  7.0     & 246.7     & $22.3$  &   Pilcher   \\
2008 05 12.3  & 3.946    & 3.015  &  6.5     & 246.1     & $22.4$  &   Pilcher   \\
2008 05 13.3  & 3.945    & 3.011  &  6.4     & 245.9     & $22.4$  &   Pilcher   \\
2008 05 26.3  & 3.932    & 2.976  &  5.6     & 243.3     & $22.4$  &   Pilcher   \\
2008 05 27.9  & 3.930    & 2.975  &  5.7     & 243.0     & $22.3$  &   Gal\'ad   \\
2008 05 28.9  & 3.929    & 2.975  &  5.7     & 242.8     & $22.3$  &   Gal\'ad   \\
2008 05 29.9  & 3.928    & 2.975  &  5.8     & 242.6     & $22.3$  &   Gal\'ad   \\
2008 05 30.9  & 3.927    & 2.975  &  5.9     & 242.4     & $22.2$  &   Gal\'ad   \\
\hline
\caption{The table lists asteroid distance from the Sun~$r$, from the Earth~$\Delta$, the solar phase angle~$\alpha$, 
and the geocentric ecliptic coordinates of the asteroid $(\lambda, \beta)$.}
\label{table_aspects}
\end{longtable}
\twocolumn

\clearpage

\onecolumn

\clearpage

\clearpage

\begin{thebibliography}{54}
\expandafter\ifx\csname natexlab\endcsname\relax\def\natexlab#1{#1}\fi
\expandafter\ifx\csname url\endcsname\relax
  \def\url#1{\texttt{#1}}\fi
\expandafter\ifx\csname urlprefix\endcsname\relax\def\urlprefix{URL }\fi

\bibitem[{{Carry} et~al.(2010){Carry}, {Dumas}, {Kaasalainen}, {Berthier},
  {Merline}, {Erard}, {Conrad}, {Drummond}, {Hestroffer}, {Fulchignoni}, and
  {Fusco}}]{Car.ea:10}
{Carry}, B., {Dumas}, C., {Kaasalainen}, M., {Berthier}, J., {Merline}, W.~J.,
  {Erard}, S., {Conrad}, A., {Drummond}, J.~D., {Hestroffer}, D.,
  {Fulchignoni}, M., {Fusco}, T., Feb. 2010. {Physical properties of (2)
  Pallas}. Icarus 205, 460--472.

\bibitem[{{Degenhardt}(2009)}]{Deg:09}
{Degenhardt}, S., May 2009. {High resolution asteroid profile by multi chord
  occultation observations}. Society for Astronomical Sciences Annual Symposium
  28, 19--21.

\bibitem[{{Delbo} and {Tanga}(2009)}]{Del.Tan:09}
{Delbo}, M., {Tanga}, P., Feb. 2009. {Thermal inertia of main belt asteroids
  smaller than 100 km from IRAS data}. \planss 57, 259--265.

\bibitem[{{Descamps} et~al.(2008){Descamps}, {Marchis}, {Pollock}, {Berthier},
  {Birlan}, {Vachier}, and {Colas}}]{Des.ea:08}
{Descamps}, P., {Marchis}, F., {Pollock}, J., {Berthier}, J., {Birlan}, M.,
  {Vachier}, F., {Colas}, F., Nov. 2008. {2007 Mutual events within the binary
  system of (22)~Kalliope}. \planss 56, 1851--1856.

\bibitem[{{Drummond} and {Cocke}(1989)}]{Dru.Coc:89}
{Drummond}, J.~D., {Cocke}, W.~J., Apr. 1989. {Triaxial ellipsoid dimensions
  and rotational pole of 2 Pallas from two stellar occultations}. Icarus 78,
  323--329.

\bibitem[{{Dunham}(1999)}]{Dun:99}
{Dunham}, D., Feb. 1999. {Planetary occulations for 1999}. \skytel 97~(2), 106.

\bibitem[{{Dunham}(1998)}]{Dun:98}
{Dunham}, D.~W., Feb. 1998. {Planetary occultations for 1998}. \skytel 95~(2),
  86.

\bibitem[{{Dunham}(2002)}]{Dun:02}
{Dunham}, D.~W., Mar. 2002. {Planetary occultations for 2002}. \skytel 103~(3),
  92--97.

\bibitem[{{Dunham}(2005)}]{Dun:05}
{Dunham}, D.~W., Mar. 2005. {Asteroid occultations for March-July 2005}.
  \skytel 109~(3), 70--72.

\bibitem[{{Dunham}(2006)}]{Dun:06}
{Dunham}, D.~W., Jun. 2006. {Upcoming asteroid occultations}. \skytel 111~(6),
  63--64.

\bibitem[{{Dunham} et~al.(1990){Dunham}, {Dunham}, {Binzel}, {Evans}, {Freuh},
  {Henry}, {A'Hearn}, {Schnurr}, {Betts}, {Haynes}, {Orcutt}, {Bowell},
  {Wasserman}, {Nye}, {Giclas}, {Chapman}, {Dietz}, {Moncivais}, {Douglass},
  {Parker}, {Beish}, {Martin}, {Monger}, {Hubbard}, {Reitsema}, {Klemola},
  {Lee}, {McNamara}, {Maley}, {Manly}, {Markworth}, {Nolthenius}, {Oswalt},
  {Smith}, {Strother}, {Povenmire}, {Purrington}, {Trenary}, {Schneider},
  {Schuster}, {Moreno}, {Guichard}, {Sanchez}, {Taylor}, {Upgren}, and {von
  Flandern}}]{Dun.ea:90}
{Dunham}, D.~W., {Dunham}, J.~B., {Binzel}, R.~P., {Evans}, D.~S., {Freuh}, M.,
  {Henry}, G.~W., {A'Hearn}, M.~F., {Schnurr}, R.~G., {Betts}, R., {Haynes},
  H., {Orcutt}, R., {Bowell}, E., {Wasserman}, L.~H., {Nye}, R.~A., {Giclas},
  H.~L., {Chapman}, C.~R., {Dietz}, R.~D., {Moncivais}, C., {Douglass}, W.~T.,
  {Parker}, D.~C., {Beish}, J.~D., {Martin}, J.~O., {Monger}, D.~R., {Hubbard},
  W.~B., {Reitsema}, H.~J., {Klemola}, A.~R., {Lee}, P.~D., {McNamara}, B.~R.,
  {Maley}, P.~D., {Manly}, P., {Markworth}, N.~L., {Nolthenius}, R., {Oswalt},
  T.~D., {Smith}, J.~A., {Strother}, E.~F., {Povenmire}, H.~R., {Purrington},
  R.~D., {Trenary}, C., {Schneider}, G.~H., {Schuster}, W.~J., {Moreno}, M.~A.,
  {Guichard}, J., {Sanchez}, G.~R., {Taylor}, G.~E., {Upgren}, A.~R., {von
  Flandern}, T.~C., May 1990. {The size and shape of (2) Pallas from the 1983
  occultation of 1 Vulpeculae}. \aj 99, 1636--1662.

\bibitem[{{Dunham} et~al.(2002){Dunham}, {Goffin}, {Manek}, {Federspiel},
  {Stone}, and {Owen}}]{Dun.ea:02}
{Dunham}, D.~W., {Goffin}, E., {Manek}, J., {Federspiel}, M., {Stone}, R.,
  {Owen}, W., Sep. 2002. {Asteroidal occultation results multiply helped by
  Hipparcos}. Memorie della Societa Astronomica Italiana 73, 662.

\bibitem[{{Dunham} and {Herald}(2009)}]{Dun.Her:09}
{Dunham}, D.~W., {Herald}, D., Jul. 2009. {Asteroid Occultations V7.0}. NASA
  Planetary Data System 111.

\bibitem[{{\SortNoop{Durech03}{\v D}urech} and
  {Kaasalainen}(2003)}]{Dur.Kaa:03}
{\SortNoop{Durech03}{\v D}urech}, J., {Kaasalainen}, M., Jun. 2003.
  {Photometric signatures of highly nonconvex and binary asteroids}. \aap 404,
  709--714.

\bibitem[{{\SortNoop{Durech06}\v{D}urech}
  et~al.(2007){\SortNoop{Durech06}\v{D}urech}, {Scheirich}, {Kaasalainen},
  {Grav}, {Jedicke}, and {Denneau}}]{Dur.ea:07}
{\SortNoop{Durech06}\v{D}urech}, J., {Scheirich}, P., {Kaasalainen}, M.,
  {Grav}, T., {Jedicke}, R., {Denneau}, L., 2007. {Physical models of asteroid
  from sparse photometric data}. In: {Milani}, A., {Valsecchi}, G.~B.,
  {Vokrouhlick\'y}, D. (Eds.), {Near Earth Objects, our Celestial Neighbors:
  Opportunity and Risk}. {Cambridge University Press}, Cambridge, p. 191.

\bibitem[{{\SortNoop{Durech10}{\v D}urech} et~al.(2009){\SortNoop{Durech10}{\v
  D}urech}, {Kaasalainen}, {Warner}, {Fauerbach}, {Marks}, {Fauvaud},
  {Fauvaud}, {Vugnon}, {Pilcher}, {Bernasconi}, and {Behrend}}]{Dur.ea:09}
{\SortNoop{Durech10}{\v D}urech}, J., {Kaasalainen}, M., {Warner}, B.~D.,
  {Fauerbach}, M., {Marks}, S.~A., {Fauvaud}, S., {Fauvaud}, M., {Vugnon},
  J.-M., {Pilcher}, F., {Bernasconi}, L., {Behrend}, R., Jan. 2009. {Asteroid
  models from combined sparse and dense photometric data}. \aap 493, 291--297.

\bibitem[{{\SortNoop{Durech10}\v{D}urech}
  et~al.(2010){\SortNoop{Durech10}\v{D}urech}, {Sidorin}, and
  {Kaasalainen}}]{Dur.ea:10}
{\SortNoop{Durech10}\v{D}urech}, J., {Sidorin}, V., {Kaasalainen}, M., Apr.
  2010. {DAMIT: a database of asteroid models}. \aap 513, A46+.

\bibitem[{{Hanu{\v s}} and {\v{D}urech}(2010)}]{Han.ea:10}
{Hanu{\v s}}, J., {\v{D}urech}, J., Oct. 2010. {New asteroid models based on
  combined dense and sparse photometry}. In: Bulletin of the American
  Astronomical Society. Vol.~42 of Bulletin of the American Astronomical
  Society. p. 1035.

\bibitem[{{Kaasalainen}(2003)}]{Kaa:03}
{Kaasalainen}, M., Dec. 2003. {Unveiling asteroids: International observing
  project and amateur-professional connection}. \jrasc 97, 283.

\bibitem[{{Kaasalainen} and {Torppa}(2001)}]{Kaa.Tor:01}
{Kaasalainen}, M., {Torppa}, J., Sep. 2001. {Optimization methods for asteroid
  lightcurve inversion. I. Shape determination}. Icarus 153, 24--36.

\bibitem[{{Kaasalainen} et~al.(2001){Kaasalainen}, {Torppa}, and
  {Muinonen}}]{Kaa.ea:01}
{Kaasalainen}, M., {Torppa}, J., {Muinonen}, K., Sep. 2001. {Optimization
  methods for asteroid lightcurve inversion. II. The complete inverse problem}.
  Icarus 153, 37--51.

\bibitem[{{Kaasalainen} et~al.(2002{\natexlab{a}}){Kaasalainen}, {Torppa}, and
  {Piironen}}]{Kaa.ea:02b}
{Kaasalainen}, M., {Torppa}, J., {Piironen}, J., Mar. 2002{\natexlab{a}}.
  {Binary structures among large asteroids}. Astron. Astrophys. 383, L19--L22.

\bibitem[{{Kaasalainen} et~al.(2002{\natexlab{b}}){Kaasalainen}, {Torppa}, and
  {Piironen}}]{Kaa.ea:02}
{Kaasalainen}, M., {Torppa}, J., {Piironen}, J., Oct. 2002{\natexlab{b}}.
  {Models of twenty asteroids from photometric data}. Icarus 159, 369--395.

\bibitem[{{Kaasalainen} et~al.(2005){Kaasalainen}, {Kaasalainen}, and
  {Piironen}}]{KaaS.ea:05}
{Kaasalainen}, S., {Kaasalainen}, M., {Piironen}, J., Sep. 2005. {Ground
  reference for space remote sensing. Laboratory photometry of an asteroid
  model}. \aap 440, 1177--1182.

\bibitem[{{Koschny} et~al.(2009){Koschny}, {Drolshagen}, and
  {Bobrinky}}]{Kos.ea:09}
{Koschny}, D., {Drolshagen}, G., {Bobrinky}, N., 2009. {The relevance of
  asteroid occultation measurements to near-Earth objects}. In: Hazards of
  Near-Earth Objects, in press.

\bibitem[{{Kristensen}(1984)}]{Kri:84}
{Kristensen}, L.~K., 1984. {(9) Metis Okkultationen den 19. Februar 1984}.
  Astronomie \& Raumfahrt 44, 76.

\bibitem[{{Lagerkvist} et~al.(2001){Lagerkvist}, {Piironen}, and
  {Erikson}}]{Lag.ea:01b}
{Lagerkvist}, C.-I., {Piironen}, J., {Erikson}, A., 2001. {Asteroid photometric
  catalogue, fifth update}. Uppsala Astronomical Observatory.

\bibitem[{{Marchis} et~al.(2006){Marchis}, {Kaasalainen}, {Hom}, {Berthier},
  {Enriquez}, {Hestroffer}, {Le Mignant}, and {de Pater}}]{Mar.ea:06}
{Marchis}, F., {Kaasalainen}, M., {Hom}, E.~F.~Y., {Berthier}, J., {Enriquez},
  J., {Hestroffer}, D., {Le Mignant}, D., {de Pater}, I., Nov. 2006. {Shape,
  size and multiplicity of main-belt asteroids}. Icarus 185, 39--63.

\bibitem[{{Marciniak} et~al.(2009){Marciniak}, {Micha{\l}owski}, {Hirsch},
  {Poli{\'n}ska}, {Kami{\'n}ski}, {Kwiatkowski}, {Kryszczy{\'n}ska}, {Behrend},
  {Bernasconi}, {Micha{\l}owski}, {Starczewski}, {Fagas}, and
  {Sobkowiak}}]{Mar.ea:09}
{Marciniak}, A., {Micha{\l}owski}, T., {Hirsch}, R., {Poli{\'n}ska}, M.,
  {Kami{\'n}ski}, K., {Kwiatkowski}, T., {Kryszczy{\'n}ska}, A., {Behrend}, R.,
  {Bernasconi}, L., {Micha{\l}owski}, J., {Starczewski}, S., {Fagas}, M.,
  {Sobkowiak}, K., Apr. 2009. {Photometry and models of selected main belt
  asteroids. VI. 160 Una, 747 Winchester, and 849 Ara}. \aap 498, 313--320.

\bibitem[{{Marciniak} et~al.(2007){Marciniak}, {Micha{\l}owski}, {Kaasalainen},
  {\v{D}urech}, {Poli{\'n}ska}, {Kwiatkowski}, {Kryszczy{\'n}ska}, {Hirsch},
  {Kami{\'n}ski}, {Fagas}, {Colas}, {Fauvaud}, {Santacana}, {Behrend}, and
  {Roy}}]{Mar.ea:07}
{Marciniak}, A., {Micha{\l}owski}, T., {Kaasalainen}, M., {\v{D}urech}, J.,
  {Poli{\'n}ska}, M., {Kwiatkowski}, T., {Kryszczy{\'n}ska}, K., {Hirsch}, R.,
  {Kami{\'n}ski}, K., {Fagas}, M., {Colas}, F., {Fauvaud}, S., {Santacana}, G.,
  {Behrend}, R., {Roy}, R., Oct. 2007. {Photometry and models of selected main
  belt asteroids. IV. 184 Dejopeja, 276 Adelheid, 556 Phyllis}. \aap 473,
  633--639.

\bibitem[{{Micha{\l}owski} et~al.(2004){Micha{\l}owski}, {Kwiatkowski},
  {Kaasalainen}, {Pych}, {Kryszczy{\' n}ska}, {Dybczy{\' n}ski}, {Velichko},
  {Erikson}, {Denchev}, {Fauvaud}, and {Szab{\' o}}}]{Mic.ea:04}
{Micha{\l}owski}, T., {Kwiatkowski}, T., {Kaasalainen}, M., {Pych}, W.,
  {Kryszczy{\' n}ska}, A., {Dybczy{\' n}ski}, P.~A., {Velichko}, F.~P.,
  {Erikson}, A., {Denchev}, P., {Fauvaud}, S., {Szab{\' o}}, G.~M., Mar. 2004.
  {Photometry and models of selected main belt asteroids I. 52 Europa, 115
  Thyra, and 382 Dodona}. \aap 416, 353--366.

\bibitem[{{Millis} and {Elliot}(1979)}]{Mil.Ell:79}
{Millis}, R.~L., {Elliot}, J.~L., 1979. {Direct determination of asteroid
  diameters from occultation observations}. In: {Gehrels}, T., {Matthews},
  M.~S. (Eds.), {Asteroids}. {University of Arizona Press}, Tucson, pp.
  98--118.

\bibitem[{{Millis} et~al.(1981){Millis}, {Wasserman}, {Bowell}, {Franz},
  {White}, {Lockwood}, {Nye}, {Bertram}, {Klemola}, {Dunham}, and
  {Morrison}}]{Mil.ea:81}
{Millis}, R.~L., {Wasserman}, L.~H., {Bowell}, E., {Franz}, O.~G., {White},
  N.~M., {Lockwood}, G.~W., {Nye}, R., {Bertram}, R., {Klemola}, A., {Dunham},
  E., {Morrison}, D., Feb. 1981. {The diameter of Juno from its occultation of
  AG+0$^\circ$1022}. \aj 86, 306--313.

\bibitem[{{Millis} et~al.(1983){Millis}, {Wasserman}, {Franz}, {White},
  {Bowell}, {Klemola}, {Elliott}, {Smethells}, {Price}, {McKay}, {Steel},
  {Everhart}, and {Everhart}}]{Mil.ea:83}
{Millis}, R.~L., {Wasserman}, L.~H., {Franz}, O.~G., {White}, N.~M., {Bowell},
  E., {Klemola}, A., {Elliott}, R.~C., {Smethells}, W.~G., {Price}, P.~M.,
  {McKay}, C.~P., {Steel}, D.~I., {Everhart}, E., {Everhart}, E.~M., Feb. 1983.
  {The diameter of 88 Thisbe from its occultation of SAO 187124}. \aj 88,
  229--235.

\bibitem[{{Ostro} et~al.(2010){Ostro}, {Magri}, {Benner}, {Giorgini}, {Nolan},
  {Hine}, {Busch}, and {Margot}}]{Ost.ea:10}
{Ostro}, S.~J., {Magri}, C., {Benner}, L.~A.~M., {Giorgini}, J.~D., {Nolan},
  M.~C., {Hine}, A.~A., {Busch}, M.~W., {Margot}, J.~L., May 2010. {Radar
  imaging of Asteroid 7 Iris}. Icarus 207, 285--294.

\bibitem[{{Sato} et~al.(1993){Sato}, {Soma}, and {Hirose}}]{Sat.ea:93}
{Sato}, I., {Soma}, M., {Hirose}, T., Apr. 1993. {The occultation of gamma
  Geminorum by the asteroid 381 Myrrha}. \aj 105, 1553--1561.

\bibitem[{{Sato} et~al.(2000){Sato}, {{\v S}arounov{\' a}}, and
  {Fukushima}}]{Sat.ea:00}
{Sato}, I., {{\v S}arounov{\' a}}, L., {Fukushima}, H., May 2000. {Size and
  shape of Trojan asteroid Diomedes from its occultation and photometry}.
  Icarus 145, 25--32.

\bibitem[{{Schmidt} et~al.(2009){Schmidt}, {Thomas}, {Bauer}, {Li}, {McFadden},
  {Mutchler}, {Radcliffe}, {Rivkin}, {Russell}, {Parker}, and
  {Stern}}]{Sch.ea:09}
{Schmidt}, B.~E., {Thomas}, P.~C., {Bauer}, J.~M., {Li}, J., {McFadden}, L.~A.,
  {Mutchler}, M.~J., {Radcliffe}, S.~C., {Rivkin}, A.~S., {Russell}, C.~T.,
  {Parker}, J.~W., {Stern}, S.~A., Oct. 2009. {The shape and surface variation
  of 2 Pallas from the Hubble Space Telescope}. Science 326, 275--278.

\bibitem[{{Shevchenko} and {Tedesco}(2006)}]{She.Ted:06}
{Shevchenko}, V.~G., {Tedesco}, E.~F., Sep. 2006. {Asteroid albedos deduced
  from stellar occultations}. Icarus 184, 211--220.

\bibitem[{{Slivan} et~al.(2003){Slivan}, {Binzel}, {Crespo da Silva},
  {Kaasalainen}, {Lyndaker}, and {Kr{\v c}o}}]{Sli.ea:03}
{Slivan}, S.~M., {Binzel}, R.~P., {Crespo da Silva}, L.~D., {Kaasalainen}, M.,
  {Lyndaker}, M.~M., {Kr{\v c}o}, M., Apr. 2003. {Spin vectors in the Koronis
  family: comprehensive results from two independent analyses of 213 rotation
  lightcurves}. Icarus 162, 285--307.

\bibitem[{{S\^{o}ma} et~al.(2007){S\^{o}ma}, {Hayamizu}, {Miyashita},
  {Setoguchi}, and {Hirose}}]{Som.ea:07}
{S\^{o}ma}, M., {Hayamizu}, T., {Miyashita}, K., {Setoguchi}, T., {Hirose}, T.,
  2007. {Occultation by (22)~Kalliope and its satellite Linus}. In: Proc.
  International Astronomical Union. Vol.~3. pp. 130--131.

\bibitem[{{Stamm}(1985)}]{Sta:85}
{Stamm}, J., 1985. {Asteroidal appulse and occultation observations.} Occ.
  Newsl. 3, 296--299.

\bibitem[{{Stamm}(1989)}]{Sta:89}
{Stamm}, J., 1989. {Reports of asteroidal appulses and occultations}. Occ.
  Newsl. 5, 327--328.

\bibitem[{{Tanga} and {Delbo}(2007)}]{Tan.Del:07}
{Tanga}, P., {Delbo}, M., Nov. 2007. {Asteroid occultations today and tomorrow:
  toward the GAIA era}. \aap 474, 1015--1022.

\bibitem[{{Tanga} et~al.(2003){Tanga}, {Hestroffer}, {Cellino}, {Lattanzi}, {Di
  Martino}, and {Zappal{\` a}}}]{Tan.ea:03}
{Tanga}, P., {Hestroffer}, D., {Cellino}, A., {Lattanzi}, M., {Di Martino}, M.,
  {Zappal{\` a}}, V., Apr. 2003. {Asteroid observations with the Hubble Space
  Telescope. II. Duplicity search and size measurements for 6 asteroids}.
  Astron. Astrophys. 401, 733--741.

\bibitem[{{Taylor} and {Dunham}(1978)}]{Tay.Dun:78}
{Taylor}, G.~E., {Dunham}, D.~W., Apr. 1978. {The size of minor planet 6 Hebe}.
  Icarus 34, 89--92.

\bibitem[{{Tedesco} et~al.(2004){Tedesco}, {Noah}, {Noah}, and
  {Price}}]{Ted.ea:04}
{Tedesco}, E.~F., {Noah}, P.~V., {Noah}, M., {Price}, S.~D., Oct. 2004. {IRAS
  Minor Planet Survey V6.0}. NASA Planetary Data System 12.

\bibitem[{{Thompson} and {Yeelin}(2006)}]{Tho.Yee:06}
{Thompson}, B., {Yeelin}, T., Dec. 2006. {Duplicity in 16 Piscium confirmed
  from its occultation by 7 Iris on 2006 May 5}. \pasp 118, 1648--1655.

\bibitem[{{Timerson} et~al.(2009){Timerson}, {\v{D}urech}, {Aguirre}, {Benner},
  {Blacnhette}, {Breit}, {Campbell}, {Campbell}, {Carlisle}, {Castro}, {Clark},
  {Clark}, {Correa}, {Coughlin}, {Degenhardt}, {Dunham}, {Fleishman},
  {Frankenberger}, {Gabriel}, {Harris}, {Herald}, {Hicks}, {Hofler}, {Holmes},
  {Jones}, {Lambert}, {Lucas}, {Lyzenga}, {Macdougal}, {Maley}, {Morgan},
  {Mroz}, {Nolthenius}, {Nugent}, {Preston}, {Rodriguez}, {Royer}, {Sada},
  {Sanchez}, {Sanford}, {Sorensen}, {Stanton}, {Venable}, {Vincent}, {Wasson},
  and {Wilson}}]{Tim.ea:09}
{Timerson}, B., {\v{D}urech}, J., {Aguirre}, S., {Benner}, L., {Blacnhette},
  D., {Breit}, D., {Campbell}, S., {Campbell}, R., {Carlisle}, R., {Castro},
  E., {Clark}, D., {Clark}, J., {Correa}, A., {Coughlin}, K., {Degenhardt}, S.,
  {Dunham}, D., {Fleishman}, R., {Frankenberger}, R., {Gabriel}, P., {Harris},
  B., {Herald}, D., {Hicks}, M., {Hofler}, G., {Holmes}, A., {Jones}, R.,
  {Lambert}, R., {Lucas}, G., {Lyzenga}, G., {Macdougal}, C., {Maley}, P.,
  {Morgan}, W., {Mroz}, G., {Nolthenius}, R., {Nugent}, R., {Preston}, S.,
  {Rodriguez}, C., {Royer}, R., {Sada}, P., {Sanchez}, E., {Sanford}, B.,
  {Sorensen}, R., {Stanton}, R., {Venable}, R., {Vincent}, M., {Wasson}, R.,
  {Wilson}, E., Jul. 2009. {A trio of well-observed asteroid occultations in
  2008}. Minor Planet Bulletin 36, 98--100.

\bibitem[{{Torppa} et~al.(2003){Torppa}, {Kaasalainen}, {Michalowski},
  {Kwiatkowski}, {Kryszczy{\' n}ska}, {Denchev}, and {Kowalski}}]{Tor.ea:03}
{Torppa}, J., {Kaasalainen}, M., {Michalowski}, T., {Kwiatkowski}, T.,
  {Kryszczy{\' n}ska}, A., {Denchev}, P., {Kowalski}, R., Aug. 2003. {Shapes
  and rotational properties of thirty asteroids from photometric data}. Icarus
  164, 346--383.

\bibitem[{{Warner} et~al.(2008){Warner}, {\v{D}urech}, {Fauerbach}, and
  {Marks}}]{War.ea:08}
{Warner}, B.~D., {\v{D}urech}, J., {Fauerbach}, M., {Marks}, S., Oct. 2008.
  {Shape and spin models for four asteroids}. Minor Planet Bulletin 35,
  167--171.

\bibitem[{{Warner} et~al.(2009){Warner}, {Harris}, and {Pravec}}]{War.ea:09}
{Warner}, B.~D., {Harris}, A.~W., {Pravec}, P., Jul. 2009. {The asteroid
  lightcurve database}. Icarus 202, 134--146.

\bibitem[{{Wasserman} et~al.(1986){Wasserman}, {Millis}, and
  {Franz}}]{Was.ea:86}
{Wasserman}, L.~H., {Millis}, R.~L., {Franz}, O., Jun. 1986. {The occultation
  of AG+20$^\circ$1138 by 129 Antigone on 11 April 1985}. \baas 18, 797.

\bibitem[{{Wasserman} et~al.(1979){Wasserman}, {Millis}, {Franz}, {Bowell},
  {White}, {Giclas}, {Martin}, {Elliot}, {Dunham}, {Mink}, {Baron},
  {Honeycutt}, {Henden}, {Kephart}, {A'Hearn}, {Reitsema}, {Radick}, and
  {Taylor}}]{Was.ea:79}
{Wasserman}, L.~H., {Millis}, R.~L., {Franz}, O.~G., {Bowell}, E., {White},
  N.~M., {Giclas}, H.~L., {Martin}, L.~J., {Elliot}, J.~L., {Dunham}, E.,
  {Mink}, D., {Baron}, R., {Honeycutt}, R.~K., {Henden}, A.~A., {Kephart},
  J.~E., {A'Hearn}, M.~F., {Reitsema}, H.~J., {Radick}, R., {Taylor}, G.~E.,
  Feb. 1979. {The diameter of Pallas from its occultation of SAO 85009}. \aj
  84, 259--268.

\end{thebibliography}
\end{document}